\documentclass[10pt,journal,compsoc]{IEEEtran}

\usepackage[misc]{ifsym}

\usepackage{xspace}
\usepackage{enumitem}

\usepackage{booktabs} %
\usepackage{multirow}

\usepackage{hyperref}
\hypersetup{
  colorlinks=true,breaklinks,
  linktoc=section,
  linkcolor=red,
  citecolor=blue,
  urlcolor=blue,
  }

\usepackage[export]{adjustbox}
\usepackage[table,dvipsnames]{xcolor}
\usepackage{tcolorbox}

\usepackage{graphicx}

\usepackage[caption=false,font=footnotesize,textfont=sf]{subfig}
\usepackage{lmodern}
\usepackage[T1]{fontenc}
\usepackage{mathtools}
\usepackage{amsthm}
\usepackage{amssymb}
\usepackage{amsfonts}
\usepackage{bbm}

\usepackage{algorithm}
\usepackage{algpseudocode}

\newcommand{\overbar}[1] {\mkern 1.5mu\overline{\mkern-3.5mu#1\mkern-1.0mu}\mkern 1.5mu}

\newcommand{\Sec}[1]		{Sec.\,\ref{#1}}
\newcommand{\Fig}[1]		{Fig.\,\ref{#1}}

\newcommand{\Eq}[1]			{Eq.\,\ref{#1}}
\newcommand{\Tab}[1]		{Tab.\,\ref{#1}}
\newcommand{\Alg}[1]		{Alg.\,\ref{#1}}
\newcommand{\Problem}[1]		{Problem\,\ref{#1}}

\newcommand{\ie}   			{i.e.\@\xspace}
\newcommand{\eg}   			{e.g.\@\xspace}
\newcommand{\etc}   		{etc.\xspace}

\newcommand{\Id}        {I}

\newcommand{\notX}[1][] {\overbar{X} \ifthenelse{\isempty{#1}}{}{_{\!#1}}} %

\renewcommand*{\top}{{\mkern-1.5mu\mathsf{T}}}

\newcommand{\inlinetitle}[2]  {\smallskip\noindent\textbf{\emph{#1}{#2}}}

\newcommand{\vbar}{\,|\,}

\usepackage[normalem]{ulem}

\newcounter{marginNoteCounter}

\DeclareMathAlphabet{\altmathcal}{OMS}{cmsy}{m}{n}
\renewcommand{\mathcal}[1]  {\altmathcal{#1}}

\makeatletter
\DeclareRobustCommand*{\escapeus}[1]{%
    \begingroup\@activeus\scantokens{#1\endinput}\endgroup}
\begingroup\lccode`\~=`\_\relax
    \lowercase{\endgroup\def\@activeus{\catcode`\_=\active \let~\_}}
\makeatother

\newcommand{\mySqBullet}		{\raisebox{0.25em}{{\scriptsize$_\blacksquare$}}}

\usepackage{scalerel}
\usepackage{tikz}
\usetikzlibrary{svg.path}

\definecolor{orcidlogocol}{HTML}{A6CE39}
\tikzset{
  orcidlogo/.pic={
    \fill[orcidlogocol] svg{M256,128c0,70.7-57.3,128-128,128C57.3,256,0,198.7,0,128C0,57.3,57.3,0,128,0C198.7,0,256,57.3,256,128z};
    \fill[white] svg{M86.3,186.2H70.9V79.1h15.4v48.4V186.2z}
                 svg{M108.9,79.1h41.6c39.6,0,57,28.3,57,53.6c0,27.5-21.5,53.6-56.8,53.6h-41.8V79.1z M124.3,172.4h24.5c34.9,0,42.9-26.5,42.9-39.7c0-21.5-13.7-39.7-43.7-39.7h-23.7V172.4z}
                 svg{M88.7,56.8c0,5.5-4.5,10.1-10.1,10.1c-5.6,0-10.1-4.6-10.1-10.1c0-5.6,4.5-10.1,10.1-10.1C84.2,46.7,88.7,51.3,88.7,56.8z};
  }
}

\newcommand\orcidicon[1]{\href{https://orcid.org/#1}{\mbox{\scalerel*{
\begin{tikzpicture}[yscale=-1,transform shape]
\pic{orcidlogo};
\end{tikzpicture}
}{|}}}}

\newcommand{\methodName}{Mixture of Interacting Cascades\xspace}
\newcommand{\MIC}{\texttt{MIC}\xspace}
\newcommand{\MIClin}{\texttt{linMIC}\xspace}
\newcommand{\IC}{\texttt{IC}\xspace}
\newcommand{\CC}{\texttt{CC}\xspace}

\newcommand{\Nu}{N_{u}} %
\newcommand{\Ne}{N_{e}} %
\newcommand{\Nc}{N_{c}} %

\newcommand{\Weight}{{W}}%
\newcommand{\weight}{{w}}%

\newcommand{\convprod}{\,\operatorname{\star}\,}
\newcommand{\SF}[2]{S(#1,#2)}
\newcommand{\LogSF}[2]{\log (S(#1,#2))}
\ifCLASSOPTIONcompsoc
  \usepackage[nocompress]{cite}
\else
  \usepackage{cite}
\fi

\ifCLASSINFOpdf

\else
\fi
\begin{document}

\title{Uncovering Social Network Activity \\Using Joint User and Topic Interaction}

\author{Gaspard~Abel$^{\star \times}$,~
        Argyris Kalogeratos$^{\star}$,
        Jean-Pierre Nadal$^{\times\dagger}$,
        Julien Randon-Furling$^{\star +}$
\IEEEcompsocitemizethanks{%
\IEEEcompsocthanksitem[${\star}$] Université Paris Saclay, Université Paris Cité, ENS Paris Saclay, CNRS, SSA, INSERM, Centre
Borelli, 91190 Gif-sur-Yvette, France
\IEEEcompsocthanksitem[${\times}$] Centre d'Analyse et de Mathématique Sociales, EHESS, CNRS, 75006 Paris, France.
\IEEEcompsocthanksitem[${\dagger}$] Laboratoire de Physique de l'École Normale Supérieure, ENS, Université PSL, CNRS, Sorbonne Université, Université Paris Cité, 75005 Paris, France
\IEEEcompsocthanksitem[${+}$] College of Computing, UM6P, Ben Guerir, Morocco.
\IEEEcompsocthanksitem[{\scriptsize \Letter}] Corresponding author: Argyris Kalogeratos.
\\E-mail: \texttt{name.surname@ens-paris-saclay.fr}.}%
\thanks{Manuscript received December XX, XXXX; revised XXXX XX, XXXX.}
}

%
%
\IEEEtitleabstractindextext{
\begin{abstract}
The emergence of online social platforms, such as social networks and social media, has drastically affected the way people apprehend the information flows to which they are exposed. %
In such platforms, various information cascades spreading among users is the main force creating complex dynamics of opinion formation, each user being characterized by their own behavior adoption mechanism.
Moreover, the spread of multiple pieces of information or beliefs in a networked population is rarely uncorrelated. In this paper, we introduce the \emph{\methodName} (\MIC), a model of marked multidimensional Hawkes processes with the capacity to model jointly non-trivial interaction between cascades and users. We emphasize on the interplay between information cascades and user activity, and use a mixture of temporal point processes to build a coupled user/cascade point process model.
Experiments on synthetic and real data highlight the benefits of this approach and demonstrate that \MIC achieves superior performance to existing methods in modeling the spread of information cascades. Finally, we demonstrate how \MIC can provide, through its learned parameters, insightful bi-layered visualizations of real social network activity data. %
\end{abstract}

}

\maketitle

\IEEEdisplaynontitleabstractindextext
\IEEEpeerreviewmaketitle

\IEEEraisesectionheading{\section{Introduction}\label{sec:introduction}}
\IEEEPARstart{T}{he} advent of wireless communication technologies and online social platforms, coupled with the miniaturization and accessibility of mobile devices, has %
transformed the way individuals are exposed to information, as well as how they interact with each other in digital diffusion networks. The growing role of online social platforms, namely online social media and even more typically online social networks (OSNs) %
in public debate and opinion formation has attracted research interest in the last decades, trying to understand how the information diffusion shapes the beliefs of a population; %
this is investigated in the field of collective phenomena, among other fields. The associated class of problems is relevant in various domains, such as epidemiology \cite{yangBiVirusCompetingSpreading2018,hethcoteMathematicsInfectiousDiseases}, econometrics \cite{bacryModellingMicrostructureNoise2013}, viral diffusion \cite{zhouLearningTriggeringKernels2013,wengCompetitionMemesWorld2012}, or agenda setting \cite{luceriUnmaskingWebDeceit2024,nianInfluenceOpinionLeaders2025}. Additionally, studies on social network activity have considered an information cascades approach. This formulation is supported by the temporal property of information propagation \cite{myersBurstyDynamicsTwitter2014} and the universal properties of information sharing \cite{muchnikOriginsPowerlawDegree2013,sadriExploringNetworkProperties2020,notarmuziUniversalityCriticalityComplexity2022}.

When users share content in a social network, they express themselves on a specific topic and at the same time assist in spreading the associated information cascade to their vicinity. Despite being initiated locally, this phenomenon may extend to a non-negligible fraction of the population, up to a point where consensus is reached. On another level, OSNs are also characterized by the simultaneous spread of a multitude of topics, which may be related in a way that affects user behavior as well. %
As several studies show, the joint dynamics of multiple topics and user activity generate socio-semantic coupling --entanglement between information polarization and community formation-- and contribute to echo-chamber or filter-bubble effects \cite{aielloFriendshipPredictionHomophily2012,barberaTweetingLeftRight2015,rothSociosemanticConfigurationOnline2023a}. Yet %
limited attention has been paid to modeling the interplay between cascade diffusion and user-level behavior. Real OSNs exhibit also strong heterogeneity in user and topic activity, which creates hierarchical and uneven information pathways \cite{muchnikOriginsPowerlawDegree2013,sadriExploringNetworkProperties2020} that are often overlooked by macroscopic models. Finally, recent diffusion models prioritize predictive accuracy over social interpretability, offering limited insight into the mechanisms that the drive observed patterns. A review of the related literature can be found in \Sec{sec:rev}.

Motivated by these challenges, we first argue that capturing non-trivial interactions between information cascades and user behaviors is essential for modeling realistic network activity. Then, we provide conceptual tools toward bridging theoretical modeling and data-driven analysis, improving both explanatory power and interpretability.
A summary of the contribution of this article is as follows:\\
\mySqBullet~We introduce the \textit{\methodName} (\MIC) model (\Sec{sec:model-def}), a two-layer temporal point process, namely multidimensional marked Hawkes processes, with the capacity to model complex patterns of social network activity generated by the interplay between users and cascades. %
\MIC models jointly cascade-to-cascade, cascade-to-user, and finally user-to-user interactions, where the interactions between cascades are implicit in the event generation process driven by users (\Fig{fig:adoption_scheme}).
To achieve that, \MIC encodes the pairwise influence between cascades in a \emph{cascade interaction matrix}. Moreover, \MIC encompasses existing models in the literature, namely the \emph{Independent Cascades} (\IC) \cite{goldenbergTalkNetworkComplex,hodasSimpleRulesSocial2014} and \emph{Correlated Cascades} (\CC) \cite{zarezadeCorrelatedCascadesCompete2017} models, %
while its gains have a reasonable complexity overhead, given that the cascading topics are usually substantially less in number than users \cite{myersClashContagionsCooperation2012}.\\
\mySqBullet~We derive closed-form expressions for the conditional intensity and the number of events, which we validate with numerical simulations (\Sec{sec:th_derivation}). Providing this type of theoretical results is not typical in the related literature.\\
\mySqBullet~We provide a simple yet effective Maximum Likelihood Estimation scheme for learning the \MIC %
parameters using input data (\Sec{sec_inference}), and we validate
\MIC's advantage in capturing the dynamics of multiple interacting cascades in synthetic experiments comparing to existing Hawkes-process-based models.\\
\mySqBullet~In comparative experiments on real data (\Sec{sec:experiments}), we show that \MIC %
captures more accurately information spreading dynamics in several different datasets. %
Moreover, results on real data highlight the inherent limitation of all the compared temporal point processes methods in
handling the heterogeneity of volume of user (and cascade) activity. This panel of evaluation measures suggest that \MIC is the best candidate for encompassing such properties.\\
\mySqBullet~We demonstrate that \MIC can also offer qualitative insights on the structure of information pathways by visualizing social network activity,
as portrayed in the schematic view of \Fig{fig:adoption_scheme}.

\section{Related work}\label{sec:rev}

\subsection{Information diffusion in online social networks}
The spread of information in OSNs is studied at different granularity levels: from local user-based dynamics, to global network analysis aiming at predicting the popularity of contents. Considering the diffusion of content, \textit{Information Cascades} is a popular formulation \cite{zhouSurveyInformationCascade2022}. Although the network structure is assumed to be known, the complexity of these spreading processes lies in the heterogeneous social influence between users. In fact, learning the underlying influence weights of %
the associated graph %
has attracted significant attention: NetInf \cite{rodriguezUncoveringTemporalDynamics2011} and \cite{zhouLearningTriggeringKernels2013} propose optimization methods to learn the local infectivity of a graph that is assumed to be low rank and sparse, while \cite{yangMixtureMutuallyExciting2013} tackles the joint problem of diffusion network inference and meme tracking to highlight the importance of characterizing topics for the detection of viral content.

OSN dynamics are also investigated through %
of influence estimation and dynamic graph evolution \cite{farajtabarCOEVOLVEJointPoint2017}. Unlike traditional broadcast media, the information cascades in OSNs are characterized by user interactions and content sharing. \textit{Independent Cascades} (\IC) \cite{goldenbergTalkNetworkComplex} was among the first models for information spread in networks. More refined models were introduced \cite{iwataDiscoveringLatentInfluence2013,lindermanDiscoveringLatentNetwork2014,valeraModelingAdoptionUsage2015,xuLearningGrangerCausality2016a,hosseiniRecurrentPoissonFactorization2017,isikFlexibleTriggeringKernels2022}, also the \textit{Correlated Cascades} (\CC) model \cite{zarezadeCorrelatedCascadesCompete2017} that include correlations between products and user influence, %
yet without encompassing the complex
interplay between users and information cascades.

\begin{figure}[t]
  \centering
  \includegraphics[width= 0.95\columnwidth, viewport=90 20 630 410,clip]{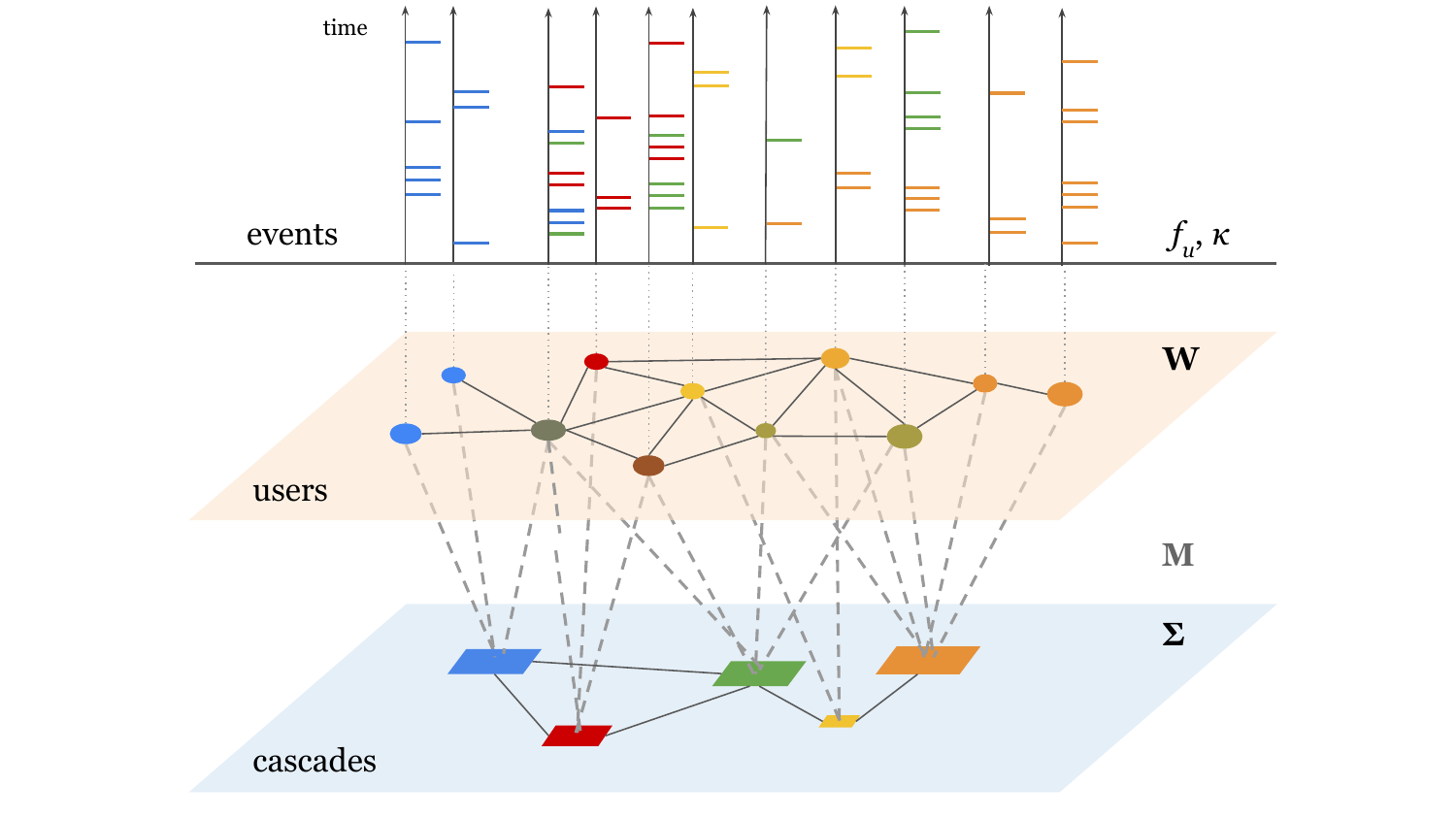}%
  \caption{Bi-layer scheme for the \MIC model. Users interact on the top of a layer of interacting cascades. Node size is proportional to the associated volume of events, and node color depicts the mixture of cascades. \MIC encompasses complex patterns of social network activity related to both layers and their interplay, namely by modeling jointly cascade-to-cascade ($\mathbf{\Sigma}$), cascade-to-user ($\mathbf{M}$), and user-to-user ($\mathbf{\Weight}$) interactions. Cascade(-to-cascade) interactions are implicit through the event generation process driven by users. Model parameters $\mathbf{\Sigma}$,\,$\mathbf{\Weight}$,\,$\mathbf{M}$, and the design components $f_u(\cdot)$, $\kappa(\cdot)$ (see \Sec{sec:model-def}) appear on the right.}
  \label{fig:adoption_scheme}
\end{figure}

\subsection{Opinion formation in social networks}
Another important line of work studies the opinion evolution in social networks. Over the last decades, opinion dynamics and theoretical modeling are being also intensively researched, from the prototypical DeGroot model \cite{degrootReachingConsensus1974} to more complex models that account for the influence of the network structure, the presence of stubborn agents \cite{vermaImpactCompetingZealots2014}, an acceptance threshold \cite{deffuantMixingBeliefsInteracting2000,rainerOpinionDynamicsBounded2002}, multiple topics \cite{baumannEmergencePolarizedIdeological2021}, or the influence of a global steering mechanism \cite{conjeaudDeGrootBasedOpinionFormation2024}. Beyond theoretical modeling that makes the unrealistic assumption of having access to the true opinion of agents, studying opinion formation in OSNs can be enabled by estimating user opinions based on their observed online activity, \eg via sentiment analysis \cite{tanUserlevelSentimentAnalysis2011,westExploitingSocialNetwork2014,royHatefulSentimentDetection2024}, or by looking at retweet dynamics \cite{liuCTLISLearning2019a,chavalariasCanSingleLine2024,vendevilleVoterModelCan2025}. Notably, the SLANT model \cite{deLearningForecastingOpinion2016} comprises a set of marked jumped stochastic differential equations to capture both asynchronous event dynamics and user opinions, while in \cite{wangStochasticDifferentialEquation} an optimal control framework is developed. The limitation of these models is that they only consider that each user has a varying opinion with respect to a single topic, and not on multiple interacting topics that is the case in the content shared in real OSNs. %
In this context, the implicit interactions between topics take roots into their contextual links, as well as user bias \cite{wengCompetitionMemesWorld2012}. %

\subsection{Temporal point processes for social network activity}
Recurrent temporal point processes %
have been used for capturing the patterns of information diffusion in social networks. In particular, starting from the original Hawkes processes \cite{hawkesSpectraSelfexcitingMutually1971}, the extension of multidimensional Hawkes processes \cite{thomaslinigerMultivariateHawkesProcesses2009} has been considered for capturing microscopic information diffusion pathways \cite{farajtabarCOEVOLVEJointPoint2017,valeraModelingAdoptionUsage2015,hosseiniRecurrentPoissonFactorization2017}. This formulation is convenient for accounting for both social and self-exciting collective phenomena, as well for considering the multitude of cascades propagating through an OSN. There has been a growing interest in %
efficient parameter estimation methods for %
temporal point processes \cite{yangMixtureMutuallyExciting2013,zhouLearningTriggeringKernels2013,lemonnierMultivariateHawkesProcesses2017,isikFlexibleTriggeringKernels2022,Limnios2025}, as well as in hybrid variations %
combining deep learning architectures \cite{meiNeuralHawkesProcess2017,caoDeepHawkesBridgingGap2017, meiTransformerEmbeddingsIrregularly2021,liPublicOpinionField2025}.
Gathered and compared in \cite{xueEasyTPPOpenBenchmarking2023a}, such works primarily focus on a macroscopic description of social network activity, \ie %
the description of the spread of a set of event types. Furthermore, these models rely on input data composed of independent event sequences, which limits their capacity to describe user interactions. %
Far less effort has been invested in providing %
modeling frameworks to tackle analytically and numerically complex information diffusion patterns at a microscopic scale, both at the cascade and the user level. %

\section{%
Background on Hawkes processes}%
\label{sec:background}

\inlinetitle{Preliminaries}{.}~%
A directed graph $\mathcal{G} = \left(\mathcal{V},E \right)$ represents a network, with $\mathcal{V}$ being a set of $\Nu$ nodes (users) and $E$ a set of edges, together inducing an adjacency matrix $\mathbf{A}$. Let $\mathbf{\Weight} = \{\weight_{vu}\}_{u,v}$ be the associated weight matrix quantifying the \textit{social influence} exerted by user $v$ on user $u$, for each edge $(v,u)\in E$. Let also $\mathcal{F}_u$ be the set of `followers' of $u$, \ie nodes with directed edges to $u$.
Suppose a set $\mathcal{C} =\{c_i\}_{i=1}^{\Nc}$ of \textit{information cascades} (tweets/posts that represent topics, opinions, products). User activity
is a sequence of events, represented by the order set $\mathcal{H} = \{e_i = (u_i,t_i,c_i)\}_{i=1}^{N_e}$, and each event refers to a user $u_i$, a time of occurrence $t_i$%
, and a cascade $c_i$ to which it contributes. %
The event history $\mathcal{H}(t) = \{e_i= (u_i,t_i,c_i)  \vbar  t_i<t\}$ is the set of events occurring in the time interval $[0,t]$ related to any cascade;
$\mathcal{H}_u(t)$ refers to the %
events of user $u$; $\mathcal{H}^{(c)}(t)$ refers to cascade $c$, and $\mathcal{H}_u^{(c)}(t)$ is only about $u$ and $c$. We omit the time index when referring to the total history of length $T$, $\mathcal{H} := \mathcal{H}(T)$, same %
for other time-dependent notation. The full symbol list is in \Tab{tab:symbols}.

\inlinetitle{Multidimensional Hawkes processes (MHPs)}{.}~%
Hawkes processes \cite{hawkesSpectraSelfexcitingMutually1971} is a class of self-exciting temporal point processes suitable for describing recurrent %
discrete asynchronous events. They are widely used in modeling collective phenomena characterized by activity bursts, like those observed in OSNs. MHPs can be formulated via a counting process vector $\mathbf n(t) = (n_u(t))_{u} \in \mathbb{N}^{\Nu}$ %
, which counts the amount of events up to time $t$ for each user, and $d \mathbf n(t)$ denotes the infinitesimal increase of $\mathbf n(t)$ at $t$. %
Given a history $\mathcal{H}(t)$, the occurrence rate for an event created by any user in the interval $[t, t+dt]$ is expressed by the conditional \textit{intensity function} $\boldsymbol{\lambda}(t)$:
\begin{equation}
    \boldsymbol\lambda(t) dt := \mathbb{E} [d \mathbf{n}(t)  \vbar  \mathcal{H}(t)].
\end{equation}
The intensity for a user $u$ is defined as (see \cite{hawkesSpectraSelfexcitingMutually1971}):
\begin{equation}
    \lambda_u(t) = \mu_u(t) + \sum_{v \in \mathcal{F}_u} \weight_{vu}(t)\convprod dn_v(t),
\end{equation}
where $\mu_u(t)$ is the \textit{background intensity}, $\mathcal F_u$ is the set of users that $u$ follows, $\weight_{vu}(t)$ is a temporal kernel for the \textit{social influence} exerted by $v$ on $u$, and
$\star$ is the convolution product. Note that %
in many existing works on social network analysis with practical considerations \cite{valeraModelingAdoptionUsage2015,zarezadeCorrelatedCascadesCompete2017} the background intensity $\mu_u$ is set to be stationary, and the social influence of $v$ to $u$ induced by an event of $v$ that occurred at time $t'$ is expressed as $\weight_{vu}(t) := \weight_{vu}\kappa(t-t')$, where $\kappa(\cdot)$ is a time-decaying kernel. Thus, the simplified intensity function is:
\begin{equation}
  \lambda_u(t) = \mu_u + \sum_{v \in \mathcal{F}_u} \weight_{vu} \!\!\sum_{e_i \in \mathcal{H}_{v}(t)} \kappa (t-t_i).
\end{equation}
Unlike synchronous and social averaging models, %
a time-decaying kernel allows for more refined user attention mechanisms. %
In this work, we use a typical exponential kernel: $\kappa(t) := \mathbbm{1}\{t>0\} \exp (- t/\tau)$, where smaller values of $\tau \in \mathbb{R}_+$ accelerate the fade-out effect, though our analysis remains valid for other types of time-decaying kernels.
As a literature note, MHPs have been generalized to multiple cascades or event types \cite{poux-medardMultivariatePoweredDirichletHawkes2023,zhuSpatiotemporaltextualPointProcesses2022}, and learning non-parametric triggering kernels for Hawkes processes is of broad interest \cite{zhouLearningTriggeringKernels2013}, \cite{lemonnierMultivariateHawkesProcesses2017}.

\inlinetitle{Marked MHPs (MMHPs)}{.}~%
Marked Hawkes processes offer one way to generalize MHPs for a network of users by accommodating multiple cascading topics. Each mark $c$ corresponds to a cascade and is assigned to a user $u$ via the user's \textit{conditional mark probability density function} $f_u(c \vbar t)$. The marked intensity of user $u$ for cascade $c$ is written as (see \cite{daleyIntroductionTheoryPoint2003}):
\begin{equation}\label{eq:marked-intensity-u}
    \lambda_u^{(c)}(t) = \lambda_u (t) f_u(c \vbar t),
\end{equation}
which is time-dependent, hence depends on the event history. The total marked intensity of cascade $c$ is:
$\lambda^{(c)}(t) = \sum_{u\in\mathcal{V}}\lambda_u (t) f_u(c \vbar t)$. %
An interesting link with opinion formation is that $f_u$ can be seen as a representation of $u$'s relative opinion on the $N_c$ topics, and $\lambda_u^{(c)}(t)$ as the expressed opinion through the marked user intensity.
Most importantly, $f_u$ allows to formalize joint user-cascade dynamics. Probabilistic models have been defined for $f_u$ (\eg in \cite{poux-medardMultivariatePoweredDirichletHawkes2023}, a Dirichlet distribution has been used for cascade dynamics, yet user dynamics were not considered), while the \CC model \cite{zarezadeCorrelatedCascadesCompete2017} proposed the use of the \emph{Boltzmann distribution} (\ie the \emph{soft-max} function) for reweighting $f_u$'s components and this way allowing for cascade competition. Finally, note that the \IC model can be expressed as an MMHP by considering $f_u$ to be a simple linear normalization of the user intensities $f_u(\cdot \vbar t)$ %
\cite{zarezadeCorrelatedCascadesCompete2017}. More light will be shed on this discussion in \Sec{sec:model-def}.

\section{\methodName (\MIC)}\label{sec:model-def}

\begin{tcolorbox}%
\begin{center}{\textbf{The MIC model}}\end{center}%
\vspace{-0.8em}
\begin{align}
    &\!\!\!\!\!\mySqBullet~\text{Independent user intensity for a cascade:}\nonumber\\
		&\quad \nu_u^{(c)} (t) = \mu_u^{(c)} + \sum_{v \in \mathcal{F}_u} \weight_{vu} \!\!\!\sum_{e_i \in \mathcal{H}_{v}^{(c)}(t)} \!\!\!\!\!\kappa (t-t_i)  \label{eq:ind-user-intensity}\\
		&\!\!\!\!\!\mySqBullet~\text{Context-sensitive indep. user intensity for a cascade:}\nonumber\\
    & \quad \nu_u^{*(c)} = \mu_u^{(c)} + \sum_s \sigma_{s c} \sum_{v \in \mathcal{F}_u} \weight_{vu} \!\!\!\sum_{e_i \in \mathcal{H}_{v}^{(s)}(t)} \!\!\!\!\!\kappa (t-t_i) \\
		&\!\!\!\!\!\mySqBullet~\text{Cascade mixing function:}\nonumber\\
    &\quad f_u(c \vbar t;\phi) = \frac{\phi \big(\nu_u^{*(c)}(t) \big)}{\sum_s \phi\big( \nu_u^{*(s)}(t)\big)} %
		\label{eq:condmarkprob}\\
		&\!\!\!\!\!\mySqBullet~\text{Global user intensity:}\nonumber\\
    &\quad \!\!\!\!\!\!\lambda_u(t) = \sum_{c\in \mathcal{C}} \nu_u^{(c)}(t)= \mu_u + \sum_{v \in \mathcal{F}_u} \weight_{vu} \!\!\!\sum_{e_i \in \mathcal{H}_{v}(t)} \!\!\!\!\!\kappa (t-t_i)\!\!\!\!\!\!\\
    &\!\!\!\!\!\mySqBullet~\text{Marked user intensity for a cascade:}\nonumber\\
		&\quad \lambda_u^{(c)} (t) = \lambda_u(t) f_u(c \vbar t) \label{eq:mui-for-a-cascade}
\end{align}%
\end{tcolorbox}

\noindent The \MIC model is defined over the MMHPs framework %
and %
builds upon the idea that user interactions take place over a landscape of contextually interacting topics (see the conceptual illustration in \Fig{fig:adoption_scheme}). More specifically, the cascade interactions are represented as \emph{positive pairwise contextual links}, which are taken into account for defining a context-sensitive independent user intensity for each cascade. These are then fed to a mixing function that shapes the final conditional mark probability density $f_u$.

The main formulas of the model appear in Eqs.\,\ref{eq:ind-user-intensity}-\ref{eq:mui-for-a-cascade}. %
$\mu_u^{(c)}$ represents the baseline user interest for cascade $c$, where $\forall u$, $\mu_u = \sum_{c\in \mathcal{C}} \mu_u^{(c)}$; the context-sensitive independent user intensity for a cascade, $\nu_u^{*(c)}$, %
combines all the associated contextual links of the \emph{cascade interaction matrix} $\mathbf{\Sigma} =\left[\sigma_{sc}\right]_{sc}$ where $\sigma_{sc} \in [0,1]$ is the contextual influence that a cascade $s$ exerts on cascade $c$, and by convention $\forall c$, $\sum_s \sigma_{sc} = 1$. The contextual links are generally asymmetric, \ie $\sigma_{sc} \neq \sigma_{cs}$; also, positivity implies interactions between two cascades ranging from being neutral to being reinforcing. Modeling-wise, contextual links are abstract, hence able to describe a wide range of possible cascade interactions: semantic similarity, co-occurrence, topic/opinion alignment, \etc %

The \emph{cascade mixing function} (\Eq{eq:condmarkprob})
maps each $\nu_u^{*(s)}$ through $\phi$. For instance, with $\phi(x) := x$, $f_u$ performs a simple normalization, while for $\phi(x) := \exp(\beta x)$ (with $\beta \geq 0$ being a hyperparameter), $f_u$ becomes the Boltzmann distribution that reshapes the density.
Regardless the choice of $\phi$, $f_u$
introduces one-vs-all cascade competition. Especially for the Boltzmann distribution, the mixing is modulated by %
$\beta$ and the magnitudes of the set $(\nu_u^{(c)*}(t))_{c\in [1,\Nc]}$ into a softmax %
function, inducing both reinforcing and inhibitory effects: as $\beta \to \infty$, $f_u$ assigns all the activity to the most popular cascade, \ie $c = \arg\max_s f_u(s \vbar t)\}\,:\,f_u(c \vbar t) \approx 1$, otherwise $f_u(c \vbar t) \approx 0$. For $\beta \to 0$, $f_u$ converges to a uniform distribution, \ie $f_u(c \vbar t) \approx 1/N_c$. Note that from a statistical mechanics point-of-view, the denominator of the Boltzmann distribution is the \emph{partition function} of the conditional mark probability $f_u$, denoted by $Z_u(t) = \sum_k \exp \big( \beta \nu_u^{*(s)}(t)\big)$.

The complete \MIC model is defined by the parameters set $\mathbf{\Theta} = \big( [\mu_u^{(c)}]_{uc},  [\sigma_{sc}]_{sc}, [\weight_{vu}]_{vu} \big) = ( \mathbf{M} , \mathbf{\Sigma}, \mathbf{\Weight} )$, which gives a total of $\Nu \times \Nc + \Nc^2 + \Nu^2 $ parameters to be inferred, and %
$\beta$ can be estimated by cross-validation. Given that usually $\Nc \ll \Nu$, inferring the additional $\Nc^2$ parameters of the model does not change the order of the computational cost.

\inlinetitle{Comparison with other closely related models}{.}~For $\boldsymbol{\Sigma} = \mathbf{\Id}$ and linear $f_u$ mixing ($\phi(x):=x$), \MIC recovers the \IC model \cite{hodasSimpleRulesSocial2014} (hence, it holds $\lambda_u^{(c)}(t) = \nu_u^{(c)}(t)$). Otherwise, using the $f_u$ with exponential mixing (Boltzmann distribution), \MIC recovers the \CC model \cite{zarezadeCorrelatedCascadesCompete2017}. As mentioned earlier, $f_u$ introduces mere one-vs-all competition, which is included in all the \IC, \CC, and \MIC models. However, our formulation is more flexible since it allows pairwise asymmetric contextual links between cascades, which in turn help us define contextual user intensities, and this way extend the notion of cascade competition.
\Tab{tab:boltzmann-regimes} summarizes how the compared models encompass the different regimes of cascade interactions.

\section{Learning a MIC model}%
 \label{sec_inference}

\MIC proposes the addition of $\Nc^2$ parameters to capture pairwise cascade interactions (see \Sec{sec:model-def}). %
Here, our intention is show that those parameters can be efficiently learned by a comprehensive and not complex Maximum Likelihood Estimation (MLE) approach. %
Moreover, when the likelihood can be factorized into a product of user likelihoods, then thanks to the partial log-likelihood formulas, these optimization sub-problems can be solved in parallel. %
Although not in the scope of this paper, there is literature dedicated to parameter inference by developing a highly competitive optimization for Hawkes processes, which use methods such as kernel decomposition \cite{zhouLearningTriggeringKernels2013,lemonnierMultivariateHawkesProcesses2017,isikFlexibleTriggeringKernels2022} or Bayesian approaches \cite{zhangEfficientNonparametricBayesian2022,nayakSmartInformationSpreading2019}.

\begin{table}[!t]
  \caption{Regimes of pairwise cascade interaction for two cascades, $s$ and $c$, induced by the cascade mixing function $f_u$ (\Eq{eq:condmarkprob}) of the {\color{orange}\MIC} model. The characterization depends on the involved map $\phi$, the scaling hyperparameter $\beta$ of the Boltzmann distribution, and the influence $\sigma_{sc}$. \MIC encompasses its linear variant {\color{purple}\MIClin}, {\color{blue}\IC} \cite{goldenbergTalkNetworkComplex}, and {\color{Green}\CC} \cite{zarezadeCorrelatedCascadesCompete2017}.}
  \label{tab:boltzmann-regimes}
  \centering
	\renewcommand{\arraystretch}{1.3}
	\vspace{-2mm}
	\resizebox{\columnwidth}{!}{
	  \begin{tabular}{r c c c}
  \toprule
  \multicolumn{2}{c}{\textsc{Mixing Function}} &  \textsc{Neutrality} & \textsc{Reinforcement} \\
	 \multicolumn{2}{c}{$f_u$} & $\sigma_{sc} = 0$ & $0 < \sigma_{sc} \leq 1$\\
  \midrule
	\textsc{uniformity} & $\phi(x):=\exp(\beta x)$, $\beta\rightarrow 0$\phantom{0}  & \!\!{\phantom{\color{blue}\IC |}} {\color{Green}\CC} | {\color{orange}\MIC} & {\color{orange}\MIC}\\
	\textsc{linearity} & $\phi(x):=x$\phantom{xxxxxxxxxxxx}  & \!\phantom{xxxxxx}{\color{blue}\IC} | {\color{purple}\MIClin} & {\color{purple}\MIClin}\\
	\textsc{competition} & $\phi(x):=\exp(\beta x)$, $\beta \,>\, 0$\phantom{0} & \!\!\phantom{\IC | }{\color{Green}\CC} | {\color{orange}\MIC} & {\color{orange}\MIC}\\
	\bottomrule
  \end{tabular}}
  \end{table}

\subsection{Computation of the log-likelihood}

First, using the formulation introduced by \cite{aalenSurvivalEventHistory2008}, for a MMHP and given the model parameters $\mathbf{\Theta}$, the conditional density function %
for the event $e_i=(u_i,c_i,t_i)$ to occur at $t_i$, combined with no other event to occur before $t_{i+1}$, is written as:
\begin{equation}\label{eq:cond_intensity}
  l(e_i,t_{i+1} ; \mathbf{\Theta}) =
	\lambda_{u_i}^{(c_i)}(t_i)%
	\ \mathbb{P}(\text{no event in}\ (t_i,t_{i+1}])),
\end{equation}
where $\lambda_{u_i}^{(c_i)}(t_i)$ is the rate of occurrence of event $e_i$. %
The probability of \Eq{eq:cond_intensity} is known as \emph{exponential survival function}, which we write here for an arbitrary time interval $(t,t']$ as: %
\begin{align}\label{eq:survival-function}
  \begin{split}
	\SF{t}{t'} &
	= \mathbb{P}(\text{no event in}\ (t,t'])) = \exp\bigg(\!\!-\int_{t}^{t'}\!\!\! \sum_{u\in \mathcal V}\lambda_{u}(\tau) d\tau\bigg)\\
	&
	= \exp\bigg(\!\!- \sum_{u\in \mathcal V}\bigg[ (t'\!-t) \mu_{u}+ \sum_{v \in \mathcal F_u}\sum_{e_j \in \mathcal{H}_{v}(t,t')} \!\!\!\!\weight_{vu} K (t,t'\!-t_j)\bigg]\bigg)\!,%
  \end{split}
\end{align}
where $K(t,t'-t_j) = \int_t^{t'-t_j}\kappa (\tau) d\tau$, with $t<t_j\leq t'$ and $\kappa(\cdot)$ the exponential kernel.

Next, we express \MIC's full observation likelihood of the whole event history $\mathcal{H} = \{e_i = (u_i,t_i,c_i)\}_{i=1}^{N_e}$ %
as:
\begin{align}\label{eq:full-likelihood}
  \begin{split}
    \!\!\!\!\!\!L (\mathcal H ; \mathbf{\Theta}) &= \prod_{e_i \in \mathcal{H}} l(e_i,t_{i+1} ; \mathbf{\Theta}) %
	= \prod_{e_i \in \mathcal{H}} \left[\lambda_{u_i}^{(c_i)}(t_i) \ \SF{t_i}{t_{i+1}} \right]\\
   &= \bigg[ \prod_{e_i \in \mathcal{H}} \lambda_{u_i}^{(c_i)}(t_i) \bigg]
	\, \SF{0}{T},
  \end{split}
\end{align}
where $\prod_{e_i \in \mathcal{H}} \SF{t_i}{t_{i+1}} = \SF{0}{T}$, given \Eq{eq:survival-function}. As mentioned earlier, the full likelihood can be factorized by the partial likelihoods for each user $u$ as: $L (\mathcal H ; \mathbf{\Theta}) = \prod_{u\in \mathcal{V}} L_u (\mathcal H ; \mathbf{\Theta})$, where, using the marked user intensity of \Eq{eq:marked-intensity-u}, we can write:
\begin{align}
  \begin{split}
    \!\!\!\!\!\!\!\!\!\!\!L_u (\mathcal H ; \mathbf{\Theta}) &=
		\bigg[\prod_{e_i \in \mathcal{H}_u} \!\!\lambda_u^{(c_i)}(t_i) \bigg] \, \SF{0}{T} \\
     &= \bigg[\prod_{e_i \in \mathcal{H}_u} \!\!\lambda_u(t_i) f_u(c_i \vbar t_i) \bigg]
		\ \SF{0}{T}.\!\!\!\!
  \end{split}
\end{align}
Next step is to give the expression for the log-likelihood:
\begin{allowdisplaybreaks}
\begin{align}\label{eq:loglikelihood}
  \begin{split}
    \mathcal{L}(\mathcal H ; \mathbf{\Theta}) &= \bigg[\sum_{e_i \in \mathcal{H}} \log(\lambda_{u_i}(t_i) f_{u_i}(c_i \vbar t_i)) \bigg] %
+ \LogSF{0}{T}%
		\\
     &= \bigg[\sum_{e_i \in \mathcal{H}}\log(\lambda_{u_i}(t_i)) + \beta\nu_{u_i}^{*(c_i)}(t_i)  \\
    & \ \ \ \ \ \ \  - \sum_{e_i \in \mathcal{H}}\log \bigg(\sum_{s\in \mathcal C} \exp \Big( \beta \nu_{u_i}^{*(s)}(t_i)\Big) \bigg) \bigg]\\
    & \ \ \ \ %
		+ \LogSF{0}{T}%
		\\
     &= \bigg[\sum_{e_i \in \mathcal{H}} \log(\lambda_{u_i}(t_i)) + \beta \nu_{u_i}^{*(c_i)}(t_i) \\
    & \ \ \ \ \ \ \ - \sum_{e_i \in \mathcal{H}}\log \left(Z_{u_i}(t_i) \right)\bigg]\\
    & \ \ \ \   - \sum_{u\in \mathcal V}\bigg[ T\mu_{u} + \sum_{v \in \mathcal F_u}\sum_{e_j \in \mathcal{H}_{v}} \weight_{vu}
		K (0,T-t_j)
		\bigg],
  \end{split}
\end{align}
\end{allowdisplaybreaks}%
where %
$Z_{u_i}(t_i)$ is the partition function defined in \Sec{sec:model-def}.
The partial log-likelihood is therefore expressed as follows:
\begin{align}\label{eq:partial_loglikelihood}
  \begin{split}
    \mathcal{L}_u(\mathcal H_u ; \mathbf{\Theta}) &= \bigg[\sum_{e_i \in \mathcal{H}_u} \log(\lambda_{u}(t_i)) + \beta \nu_{u}^{*(c_i)}(t_i)\bigg]\\
      & \ \ \ \   - \sum_{e_i \in \mathcal{H}_u}\log (Z_{u}(t_i) )\\
      & \ \ \ \   -  T \mu_{u} - \sum_{v \in \mathcal F_u}\sum_{e_j \in \mathcal{H}_{v}} \weight_{vu}
			K (0,T-t_j).
  \end{split}
  \end{align}
\noindent \textbf{Proposition 1}: The negative log-likelihood $\mathcal{L}$ is a convex function of the parameters $\mathbf{\Theta}$.
\begin{proof} The convexity of the negative partial log-likelihood is a direct consequence of the convexity of the $\operatorname{Log Sum Exp}$ function $\varphi:\mathbb R^{\Nc} \to \mathbb R$ \cite{boydConvexOptimization2004}, defined as:
\begin{equation}
    \varphi(\mathbf{z_i}) =  \log \bigg(\sum_{s=1}^{\Nc} \exp(z_i^{(s)})\bigg),
\end{equation}
where $\forall s \in \mathcal{C}$, $z_i^{(s)} = \beta \nu_{u_i}^{*(s)}(t_i)$. For \MIC, the set of parameters $\mathbf{\Theta} = \left( \mathbf{M} , \mathbf{\Sigma}, \mathbf{\Weight} \right)$ are positive, ensuring the preservation of convexity. Similarly, the other terms of the negative log-likelihood are convex functions of the parameters, which proves the overall convexity of $\mathcal{L}_u$, and subsequently of $\mathcal{L}$. In the case of linear mixing, in \Eq{eq:full-likelihood} we replace $\lambda_{u_i}^{(c_i)}(t_i)$ by $\nu_{u_i}^{(c_i)}(t_i)$ and we perform the same reasoning, which also leads to a convex formulation of the log-likelihood.
\end{proof}

\subsection{Parameter inference algorithm}\label{sec:par-inferrence}

Relying on the expression of the partial and total log-likelihood (\Eq{eq:loglikelihood} and \Eq{eq:partial_loglikelihood}) we decompose the parameter inference scheme into two optimization steps, presented in \Alg{alg:optimize_theta}. \Problem{eq:probSigma} uses Maximum Likelihood Estimation (MLE) to find the cascade interaction matrix $\mathbf{\Sigma} =\left[\sigma_{cs}\right]_{cs}$ by minimizing $-\mathcal{L}(\mathcal{H} ; \mathbf{\Theta})$, with $(\mathbf{M},\mathbf{\Weight})$ fixed.
Then, \Problem{eq:probMB} uses Maximum Partial Likelihood Estimation (MPLE) to find the user parameters $\mathbf{M}_u = (\mu_u^{(c)})_c$ and $\mathbf{\Weight}_u = (\weight_{vu})_{v}$, with $\mathbf{\Sigma}$ fixed. Due to the per-user factorization of the log-likelihood, the $\Nu$ sub-problems can be solved %
in parallel. The two problems are individually convex and their global optimum can be estimated when taken in parts, %
using the alternating optimization scheme of \Alg{alg:optimize_theta} until the convergence of the total log-likelihood. As this work focuses on modeling the interplay of user and cascade interactions, we are mainly interested in showing that learning a \MIC model is feasible with a typical optimization scheme. In that sense, our proposition employs alternating optimization, parallelized problem solving, while the rest of the inference scheme comes from the prior line of %
work and up to the work in \cite{zarezadeCorrelatedCascadesCompete2017}. For larger datasets, efficient precomputation of the past events kernels (\eg $(\sum_{e_j \in \mathcal H_v(t)} \kappa(t-t_j))_v$) was performed to leverage the sparsity of the data and reduce the overall computational cost.
Note that \MIC's hyperparameters $\beta$ and $\tau$ are estimated by cross-validation, as we discuss in the Appendix, where we also report the running times recorded for parameter inference across different datasets.%

\begin{algorithm}[t]\small
  \caption{%
	Inference of \MIC parameters
	}\label{alg:optimize_theta}
	\hspace*{\algorithmicindent} \textbf{Input:}~history of train events $\mathcal{H}$, termination threshold $\epsilon$\\
  \hspace*{\algorithmicindent} \textbf{Output:}~the set of inferred parameters $\Tilde{\boldsymbol{\mathbf{\Theta}}} = (\Tilde{\mathbf{M}}, \Tilde{\mathbf{\Weight}}, \Tilde{\mathbf{\Sigma}})$%
  \vspace{0.05cm}%
	\hrule
	\vspace{0.05cm}%
	\begin{algorithmic}[1]
	    \State Initialize $\mathbf{\Theta} = (\mathbf{M}, \mathbf{\Weight}, \mathbf{\Sigma})$

			 \State Let $\Tilde{\mathbf{\Theta}} = \mathbf{\Theta}$

      \Repeat

          \State \!\!\mySqBullet~Step~1 -- Underlying cascade interaction
          \State %
					\vspace{-.9em}%
					\begin{equation}\label{eq:probSigma}
						\hspace{-7.2em}%
						\Tilde{\mathbf{\Sigma}} = {{\arg\min}_{\mathbf{\Sigma}}} -\mathcal{L}(\mathcal{H}; \mathbf{M}, \mathbf{\Weight})
					\vspace{-.6em}%
					\end{equation}
					\hspace{1.65cm}%
					s.t. $\forall c,s \in \mathcal{C}, \sigma_{sc} \in [0,1]$ and $ \sum_c \sigma_{sc} = 1$

					\smallskip
          \State \!\!\mySqBullet~Step~2 -- Social influence and background intensity
          \State \ \ \  \textbf{for} $u \in \mathcal{V}$ \ \textbf{do in parallel}
              \State %
							\vspace{-.9em}%
							\begin{equation}\label{eq:probMB}
							\hspace{-3.5em}\Tilde{\mathbf{M}}_u, \Tilde{\mathbf{\Weight}}_u = {\arg\min}_{\mathbf{M}_u, \mathbf{\Weight}_u} -\mathcal{L}_u(\mathcal{H}; \Tilde{\mathbf{\Sigma}})
							\vspace{-.6em}%
							\end{equation}
              \hspace{2.45cm} s.t. $\mathbf{M}_u, \mathbf{\Weight}_u \geq 0$

					\State \ \ \  \textbf{end for}

					\smallskip
					\State $\mathbf{\Theta} = \Tilde{\mathbf{\Theta}}$

			\Until{$\mathcal{L}(\mathbf{\Tilde{\Theta}}) - \mathcal{L}(\mathbf{\Tilde{\Theta}}) < \epsilon$}

			 \State \Return $\Tilde{\mathbf{\Theta}} = (\Tilde{\mathbf{M}}, \Tilde{\mathbf{\Weight}}, \Tilde{\mathbf{\Sigma}})$
  \end{algorithmic}
\end{algorithm}

\section{Analytic derivation of characteristic quantities %
}\label{sec:th_derivation}
In this section, we derive closed-form expressions for the conditional intensity and the number of events for the \MIC model. The derivation follows a rather simple method to obtain moments of Hawkes processes, based on \cite{cuiElementaryDerivationMoments2020}. It can be noted that the following derivation is valid for other exponential-decaying kernel types (mixture of exponentials, gamma kernels). These analytical results are compared and validated in \Sec{sec:experiments} with experiments both on synthetic (\Fig{fig:int_eve_grid_ex}) and real data (\Fig{fig:ints_url}).

\subsection{Generic derivation}

\textit{Step~1}:~Let us consider the %
quantity $\mathbb E\big[{n_v^{(s)}}^m(t){\nu_u^{(c)}}^q(t)\big]$, where $n_v^{(s)}(t)$ is the number of events of user $v$ for cascade $s$ and $\nu_u^{(c)}(t)$ is the independent intensity of user $u$ for cascade $c$, both at time $t$ and exponentiated by $m, q \in \mathbb{N}$.

\noindent\textit{Step~2}:~We calculate the probabilities of the absence or occurrence of an arbitrary event $e_i = (u_i,c_i,t_i)$ that is or is not associated to user $v$ and cascade $s$, at order $o(\Delta t$):
\begin{align}
  \begin{split}\label{eq:step2}
  &\mathbb P\left( \neg e_i \vbar  \mathcal H_t \right) =  1-\Delta t\sum_{v\in \mathcal{V}} \sum_{p \in \mathcal{C}}\lambda_v^{(p)}(t) + o(\Delta t) \\
  &\mathbb P\left(e_i \neq (v,s,t_i) \vbar  \mathcal H_t \right) =\Delta t \sum_{v \in \mathcal{V}} \sum_{\substack{p \in \mathcal{C}\\(v,p) \neq(v,s)}}\!\!\!\!\!\!\!\lambda_v^{(p)}(t) + o(\Delta t) \\
  &\mathbb P\left( e_i = (v,s,t_i) \vbar  \mathcal H_t \right) = \Delta t \lambda_v^{(s)}(t) + o(\Delta t).
  \end{split}
\end{align}
\noindent
\textit{Step~3}:~We express the evolution of the independent cascade intensity at time $t+\Delta t$ given the absence and the occurrence of an event $e_i = (u_i,c_i,t_i)$:
When no event occurs:
\begin{align}
  \begin{split}\label{eq:step3a}
    \nu_{u,0}^{(c)}(t+\Delta t) &= \mu_u^{(c)} + \sum_{v\in \mathcal{F}_u}\weight_{vu} \sum_{e_j \in \mathcal{H}_v^{(c)}(t)} \!\!e^{-(t-t_j+\Delta t)/\tau}\\
    &= \nu_u^{(c)}(t) - \Delta t (\nu_u^{(c)}(t)-\mu_u^{(c)})/\tau + o(\Delta t),
  \end{split}
\end{align}
where $\tau$ controls the scale of the temporal kernel $\kappa$ (see \Sec{sec:background}).
When there is an event $(u_i,c_i,t_i)$ with $t<t_i<t+\Delta t$:
\begin{align}
  \begin{split}\label{eq:step3b}
    \!\!\!\!\!\!\nu_{u,e_i}^{(c)}(t+\Delta t) &= \nu_{u,0}^{(c)}(t+\Delta t) + \weight_{u_i u}\delta_{c_i,c}(1-\Delta t') + o(\Delta t')/\tau,\!\!  \end{split}
\end{align}
where $\delta_{\cdot,\cdot}$ is the Kronecker %
delta function and $\Delta t' = t_i - t$ is the time passed after the occurrence of event $e_i$ (\ie $0<\Delta t'<\Delta t$). Here we extend the notation of influence: $\weight_{u_i u} = 0$ if $u_i \notin \mathcal{F}_u$.

\noindent\textit{Step~4}:~We calculate and simplify the conditional expectation $h(t+\Delta t) = \mathbb E[n_v^{(s)}(t+\Delta t)\nu_u^{(c)}(t+\Delta t) \vbar \mathcal H_t]$ given the two events:
\begin{align}
  \begin{split}
    \!\!\!\!\!\!\!\! h(t+\Delta t) =\  & {n_v^{(s)}}^m(t){(\nu_{u,0}^{(c)})}^{q}(t) \mathbb P\left( \neg e_i \vbar  \mathcal H_t \right) \\
    &+ n_v^{(s)}(t){(\nu_{u,e_i}^{(c)})}^{q}(t) \mathbb P\left(e_i \neq (w,s,t_i) \vbar  \mathcal H_t \right) \\
    &+ (n_v^{(s)}(t)+1)^m{(\nu_{u,e_i}^{(c)})}^{q}(t) \mathbb P\left( e_i = (w,s,t_i) \vbar  \mathcal H_t \right).\! \\
    \end{split}\!\!\!\!\!\!\!\!\!\!\!\!\!\!\!\!\!\!
\end{align}
With $\lim_{t\to 0}\frac{\mathbb E[h(t+\Delta t) - h(t)]}{\Delta t} = \frac{d \mathbb E[{n_v^{(s)}}^m(t){\nu_u^{(c)}}^q(t)]}{dt}$, using \Eq{eq:step2}, \Eq{eq:step3a} and \Eq{eq:step3b}, and taking the full expectation, we obtain the following differential equation for $\mathbb E[{n_v^{(s)}}^m(t){\nu_u^{(c)}}^q(t)]$:
\begin{align}
  \begin{split}\label{eq:diff_eq_Ng}
    \!\!\!\!\!\!\!\!\!\!\!\!\frac{d}{dt} \mathbb E[&{n_v^{(s)}}^m(t){\nu_u^{(c)}}^q(t)] = \\
    & -q \mathbb E[{n_v^{(s)}}^m(t){\nu_u^{(c)}}^{q-1}(t)(\nu_u^{(c)}(t)-\mu_u^{(c)})]/\tau\\
    &+ \sum_{j \in \mathcal{V}} \sum_{k=0}^{q-1} \binom{q}{k} \weight_{ju}^{q-k} \mathbb E[{n_v^{(s)}}^m(t)\lambda_j^{(c)}(t){\nu_u^{(c)}}^{k}(t)]\\
    &+ \sum_{l=0}^{m-1} \sum_{k=0}^{q} \binom{m}{l} \binom{q}{k} \weight_{vu}^{q-k}\delta_{s,c}^{q-k} \mathbb E[{n_v^{(s)}}^{l}(t) \lambda_v^{(s)}(t){\nu_u^{(c)}}^{k}(t)].\\
  \end{split}\!\!\!\!\!\!\!\!\!\!\!\!\!\!\!\!\!\!\!\!\!\!\!\!
\end{align}
Note that the derivation relies on the same assumptions of regularity of Hawkes processes as in \cite{cuiElementaryDerivationMoments2020}. The moments of the conditional intensity and the number of events $\mathbb E[{n_v^{(s)}}^m(t)\lambda_u^q(t)]$ can be derived in a similar way for the \MIC model:
\begin{align}
  \begin{split}\label{eq:diff_eq_Nlambda}
    \!\!\!\!\!\!\!\!\!\!\!\!\frac{d}{dt} \mathbb E[{n_v^{(s)}}^m(t)& \lambda_u^q(t)] = -q \mathbb E[{n_v^{(s)}}^m(t)\lambda_u^{q-1}(t)(\lambda_u(t)-\mu_u)]/\tau\\
    &+ \sum_{j \in \mathcal{V}} \sum_{k=0}^{q-1} \binom{q}{k} \weight_{ju}^{q-k} \mathbb E[{n_v^{(s)}}^m(t)\lambda_j(t)\lambda_u^{k}(t)]\\
    &+ \sum_{l=0}^{m-1} \sum_{k=0}^{q} \binom{m}{l} \binom{q}{k} \weight_{vu}^{q-k} \mathbb E[{n_v^{(s)}}^{l}(t) \lambda_v(t)\lambda_u^{k}(t)].
  \end{split}\!\!\!\!\!\!\!\!\!\!\!\!\!\!\!\!\!\!\!\!\!\!\!\!
\end{align}

\subsection{Application: computation of the global user intensity and the number of events}
The differential \Eq{eq:diff_eq_Ng} can be solved for specific values of $m$, $n$.\!\!

\inlinetitle{Expectation of the cascade intensity}{.}~For $m=0$ and $q=1$, we obtain the differential equation for the expectation of the independent cascade intensity:
\begin{align}
  \begin{split}\label{eq:diff_eq_g}
      \!\!\!\!\!\!\!\!\!\!\!\!\!\!\frac{d}{dt} \mathbb E[\nu_u^{(c)}(t)] &= -( \mathbb E[\nu_u^{(c)}(t)]- \mu_u^{(c)})/\tau
      + \sum_{j \in \mathcal{V}} \weight_{ju} \mathbb E[\lambda_j(t)f_j(c \vbar t)]\\
      &=-( \mathbb E[\nu_u^{(c)}(t)]- \mu_u^{(c)})/\tau+ \sum_{j \in \mathcal{V}} \weight_{ju} \mathbb E[\lambda_j^{(c)}(t)].
  \end{split}\!\!\!\!\!\!\!\!\!\!\!\!\!\!\!\!\!\!\!\!\!
\end{align}
In this form, the first term corresponds to the exponential decay that accounts for the recency of past events, while the second term is the contribution from every neighboring activity at time $t$ for cascade $c$.
Summing \Eq{eq:diff_eq_g} over $c$ (or %
taking $(m,n)=(0,1)$ in \Eq{eq:diff_eq_Nlambda}), we obtain the differential equation for the expectation of user intensity $\mathbb E[\lambda_u(t)]$:
\begin{align}\label{eq:diff_eq_lambda}
  \begin{split}
    \frac{d}{dt} \mathbb E[\lambda_u(t)] &= -(\mathbb E[\lambda_u(t)] - \mu_u)/\tau + \sum_{v \in \mathcal{V}} \weight_{vu} \mathbb E[\lambda_v(t)].
  \end{split}
\end{align}
\Eq{eq:diff_eq_lambda} can be expressed in vectorial form for all users $u$:
\begin{align}
  \begin{split}
    \frac{d}{dt} \mathbb E[\boldsymbol \lambda(t)] &= -(\mathbb E[\boldsymbol \lambda(t)] - \boldsymbol \mu)/\tau + \mathbf{\Weight}^\top \mathbb E[\boldsymbol \lambda(t)].
  \end{split}
\end{align}
This equation is analogous to the one derived in \cite{deLearningForecastingOpinion2016,farajtabarCOEVOLVEJointPoint2017} and can be analytically solved for the \MIC model, if we consider the initial condition $\mathbb E[\boldsymbol \lambda(0)] = \boldsymbol \mu$:
\begin{align}\label{eq:th_lambda}
  \begin{split}
    \mathbb E[\boldsymbol \lambda(t)] &= \left[\boldsymbol B + (\mathbf{\Id} - \boldsymbol B) e^{- \boldsymbol B^{-1} t/\tau}\right]\boldsymbol \mu,
  \end{split}
\end{align}
where $\boldsymbol B = (\mathbf{\Id} - \mathbf{\Weight}^\top\tau)^{-1}$. %
The expected intensity converges to the stationary value $\boldsymbol \lambda_{\infty} = \boldsymbol B \boldsymbol \mu$ if it holds for the spectral radius: $\rho(\mathbf{\Weight}^\top) < 1/\tau$. This condition can be interpreted as the balance between the amplitude of event social influence and its temporal decay. Then, after a negative exponential increase, the expected intensity tends to the stationary value $\boldsymbol B \boldsymbol \mu$.

\inlinetitle{Expected number of events}{.}~Next, the definition of the expectation of the number of events can be recovered in \Eq{eq:diff_eq_Ng} for $m=1$, $q=0$: $\frac{d \mathbb E[n_v^{(s)}(t)]}{dt} = \mathbb E[\lambda_v^{(s)}(t)]$. The expectation of the number of events can then be expressed as:
\begin{align}
  \begin{split}
    \mathbb{E}[\boldsymbol n(t)] &= \left[\boldsymbol B t + (\boldsymbol{I} - \boldsymbol{B}) \boldsymbol B \tau\left(\boldsymbol I - e^{- \boldsymbol B^{-1} t/\tau}\right)\right]\boldsymbol \mu,
  \end{split}
\end{align}
with the initial condition $\mathbb E[\boldsymbol N(0)] = 0$. Therefore, the expected number of events tends to a linear regime of slope $\boldsymbol B \boldsymbol \mu$.
Given the non-independence between the intensity $\lambda_u$ and the cascade mixing function $f_u$, the analytical expression of the expected cascade-related intensity is not straightforward and goes beyond our scope. Nevertheless, to validate the closeness between analytical derivation and numerical simulations in \Sec{sec:experiments}, we formulate the following \emph{stationary state} approximation for the cascade intensity:
\begin{align}\label{eq:approx_lambda}
  \mathbb E[\lambda^{(c)}(t)] \approx \sum_{u\in \Nu}\mathbb E[\lambda_u(t)] \frac{\exp(\beta (\boldsymbol B\boldsymbol \mu \mathbf{\Sigma})_{u,c})}{\sum_{s\in \mathcal C}\exp(\beta (\boldsymbol B\boldsymbol \mu \mathbf{\Sigma})_{u,s})}.
\end{align}
This approximation holds experimentally when $\Nu$ is large and the observation time-scale satisfies $t_{\text{obs}} \gg \tau$, allowing $\lambda_u$ and $f_u$ to be treated as separable quantities that rapidly converge to their stationary values.

\section{Experiments}\label{sec:experiments}

This section evaluates the performance of the \MIC model, both qualitatively and quantitatively. The compared methods and the evaluation measures used are presented in \Sec{sec:competitors-measures}.
In \Sec{synth_exp}, we start by designing synthetic experiments with data generated by simulating the \MIC model. The aim is to evaluate \MIC's capacity to encompass different information cascade dynamics, and therefore to show that \MIC generalizes %
over the %
existing models discussed earlier. In \Sec{realexp}, we test how the model handles the complexity of real data. %
We also provide an ablation study of the features of \MIC, so as to assess the impact of the cascade mixing function. %
Next, in \Sec{sec:heterogenous-activity} we analyze the models' behavior from a user-centric point of view, which is a frequently neglected side in similar studies that focus on the global behavior of the agents in a social network. Although this latter is of high significance \cite{muchnikOriginsPowerlawDegree2013,sadriExploringNetworkProperties2020}, neglecting the user-wise evaluation of user-based models may hinder understanding what happens at the level of individual users.
Finally, in \Sec{sec:vizualization} we demonstrate \MIC's potential for producing comprehensive data visualizations of social network activity landscapes, which provide a qualitative insight into the model's behavior and the underlying user-to-user, user-to-cascade, and cascade-to-cascade interactions. Note that an empirical scalability analysis, reporting running times for \MIC and all compared methods, is provided in the Appendix.

\subsection{Competitors and evaluation measures}\label{sec:competitors-measures}

We consider as competitors the {Independent Cascades} (\IC) model \cite{deLearningForecastingOpinion2016, iwataDiscoveringLatentInfluence2013,lindermanDiscoveringLatentNetwork2014}, and the {Correlated Cascades} (\CC) model \cite{zarezadeCorrelatedCascadesCompete2017}.
We also include the linear variant presented earlier, \MIClin. %
As discussed in \Sec{sec:model-def} (see also \Tab{tab:boltzmann-regimes}), all compared models can be mapped to special cases of \MIC, therefore, the results that follow can be also analyzed from an ablation study viewpoint for \MIC's features. Moreover, treating them as \MIC cases conveniently allowed us to perform the simulation of all models using the same code implementation\footnote{Implementation available at: \url{https://github.com/gas-abel/MIC}}\footnote{We do not include the {Competing Products} model \cite{valeraModelingAdoptionUsage2015} as it has been shown that it is clearly outperformed by the former competitors. Moreover, our results can be put in perspective to the slightly different context of opinion formation, as for $\Nc=2$ cascades, the SLANT model \cite{deLearningForecastingOpinion2016} with binary opinion values is analogous to the \IC model.}.

Each experiment takes as input the adjacency matrix of a user graph and an activity log containing events of the form: $\mathcal{H} = \{e_i=(\text{user: }u_i,\,\text{cascade: }c_i,\,\text{timestamp: }t_i)\}_{i=1}^{N_e}$.
This can be a real event log, or a synthetic one where the events are generated by a source model, \eg using Ogata's thinning algorithm \cite{ogataLewisSimulationMethod1981}. Then, the objective is to infer the model parameters, \eg for \MIC, these are $\mathbf{\Theta}=(\mathbf{\Sigma}, \mathbf{M}, \mathbf{\Weight})$.
We use a train-test split where the first $80\%$ of the events is used as training set, and the last $20\%$ is used for testing the models. %
We use two ways to evaluate and compare the models:\\ %
\mySqBullet~\emph{Comparing the goodness-of-fit}: we compare how well a probabilistic model fits to an input event log, namely, the \emph{test log-likelihoods} of the models computed on the test events of a given dataset $\mathcal{H}^{\text{test}}$. A high test log-likelihood, $\mathcal{L}^{\text{test}} < 0$ and as close as possible to zero is desired. %
  The test log-likelihood can be also computed for a varying fraction of the train dataset $\mathcal{H}^{\mathrm{train}}$ (\eg the last $20\%$ or $60\%$ train events before the test dataset $\mathcal{H}^{\mathrm{test}}$) to evaluate model efficiency as a function of data availability.\\
	\mySqBullet~\emph{Comparing the realized event activity}:~we compare the empirical event intensity (rate of events over time) of the events generated by a model against the activity that comes either from a reference model (when this is available), or by an input event log. Similarly, we compare the activity in terms of number of events generated. We use as evaluation measures to compare such quantities:
		i)~\emph{Inverse $l_1$-distance}; ii)~\emph{Pearson correlation}; iii)~\emph{Distribution of ranked user (resp. cascade) activity}:~this refers to the comparison of the distribution of the user (resp. cascade) activity with respect to their ranking according to their total volume of activity.

All the above mentioned measures can be used at a global level (for all users and cascades), user-wise, and/or cascade-wise, to evaluate model performance. Each experiment is run $10$ times to get a model's empirical average behavior.

\begin{figure}[!t]
\vspace{-1em}
  \centering
  \hspace{-1.em}\subfloat[$\mathcal{L}_{\IC}^{\text{test}}\,/\,\mathcal{L}_{\MIC}^{\text{test}}$]{\includegraphics[width=1.218in]{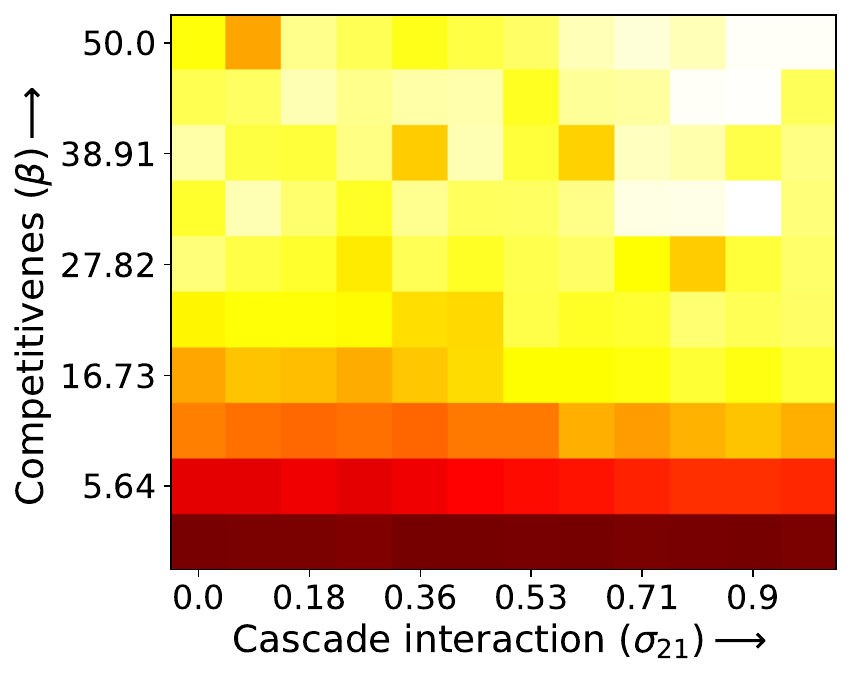}%
  \label{sub:grid_AUC_IC}}
  \hspace{-0.3em}
  \subfloat[$\mathcal{L}_{\MIClin}^{\text{test}}\,/\,\mathcal{L}_{\MIC}^{\text{test}}$]{\includegraphics[width=1.0in]{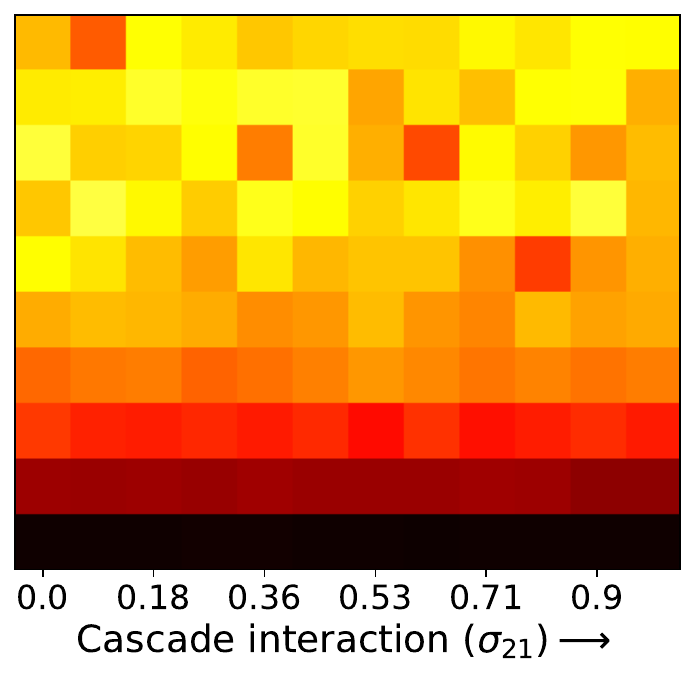}%
  \label{sub:grid_AUC_MIC_lin}}
  \hspace{-0.3em}
	\subfloat[$\mathcal{L}_{\CC}^{\text{test}}\,/\,\mathcal{L}_{\MIC}^{\text{test}}$]{\includegraphics[width=1.0in]{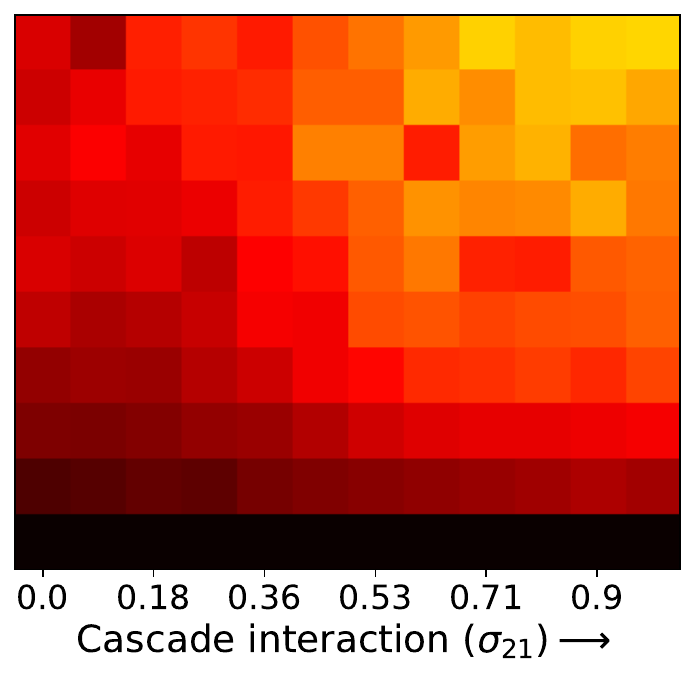}%
  \label{sub:grid_AUC_CC}}
  \hspace{-0.7em}
  \subfloat{
    \raisebox{0.1cm}{\includegraphics[width=0.4in]{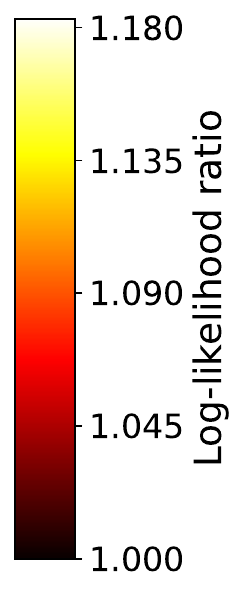}}}
  \caption{Heatmaps of the test log-likelihood ratios between the competitors and \MIC, with varying $\beta$ (y-axis) and the cascade interaction %
	$\sigma_{21}$ (x-axis), on synthetic event logs generated by \MIC. A ratio value larger than $1$ means that \MIC performs better. %
}
  \label{fig:grid_AUCs}
\end{figure}

\subsection{Experiments on synthetic data}\label{synth_exp}

\subsubsection{Data generation}\label{sec:data_gen}

Synthetic data are generated by the \MIC model, %
in a random Erd\H{o}s-Rényi network of $\Nu=50$ users and for $\Nc=3$ cascades. %
For each edge $(v,u)$ between users, we set the social influence at $\weight_{vu} \sim \mathcal{U}([0,1])$. For each user $u$ we set the baseline intensities for the three cascades at $\mu_u ^{(\cdot)}\sim \mathcal{U}([0,0.2])$. %
We vary the cascade dynamics along two dimensions: %
	\\--\,\,by varying the cascade competitiveness via $\beta \in [0.01, 100]$; \\
	--\,\,by varying the cascade interaction via $\mathbf{\Sigma}$: the reinforcement from $c_2$ to $c_1$ ($\sigma_{21}\in [0, 1]$) is gradually increasing as $\mathbf{\Sigma} = \big[\begin{smallmatrix}
  1 & 0 & 0\\
  0 & 1 & 0\\
  0 & 0 & 1
\end{smallmatrix}\big] \rightarrow %
\big[\begin{smallmatrix}
  1 & 0 & 0\\
  1 & 0 & 0\\
  0 & 0 & 1
\end{smallmatrix}\big]$. %
This explores a wide range of interactions between cascades, going gradually from pure independence to complex and asymmetric competition or cooperation. To get statistically sound results, synthetic event logs are generated $10$ times for each parameterization, and for times units $T=500$ and $\tau=3$.

\subsubsection{Goodness-of-fit evaluation%
}

The scenario challenges the compared models to learn from data generated by the \MIC model, as described in \Sec{sec:data_gen}. Note that \CC and \MIC are learned taking as input the true $\beta$ and $\tau$ values used to generate the train data. The models are then compared on how well they generalize to unseen synthetic event logs.
\Fig{fig:grid_AUCs} shows heatmaps of the test log-likelihood ratios $\mathcal{L}_{\texttt{<model>}}^{\text{test}}\,/\,\mathcal{L}_{\MIC}^{\text{test}}$. %
Starting from the uniform and independent cascades regime (low $\beta$), \MIC has on average a similar or slightly better performance to the three other methods. As $\beta$ increases, bringing nonlinear competitiveness, \MIC becomes %
more accurate than \IC (\Fig{sub:grid_AUC_IC}) and \MIClin (\Fig{sub:grid_AUC_MIC_lin}).
Those two first subfigures also indirectly compare \IC and \MIClin: the heatmap patterns are rather similar except from a difference in the scale of the ratio. It can further be observed that the performance of \IC decreases as $\sigma_{21}$ grows (along the x-axis), whereas, as expected by its design, \MIClin is unaffected by this variation, suggesting that overall it behaves similarly to \IC, \ie it is insensitive to both dimensions of variation, while having a bit better performance.
Additionally, once the interaction between cascades becomes more intricate, \MIC starts to outperform \CC (\Fig{sub:grid_AUC_CC}). %
The results suggest that \MIC is more flexible than the compared special cases, which is consistent with the regime summary given in \Tab{tab:boltzmann-regimes}.

\begin{figure}[!t]
  \centering
  \subfloat[Cascade $1$]{\hspace{-1em}\includegraphics[width=1.75in]{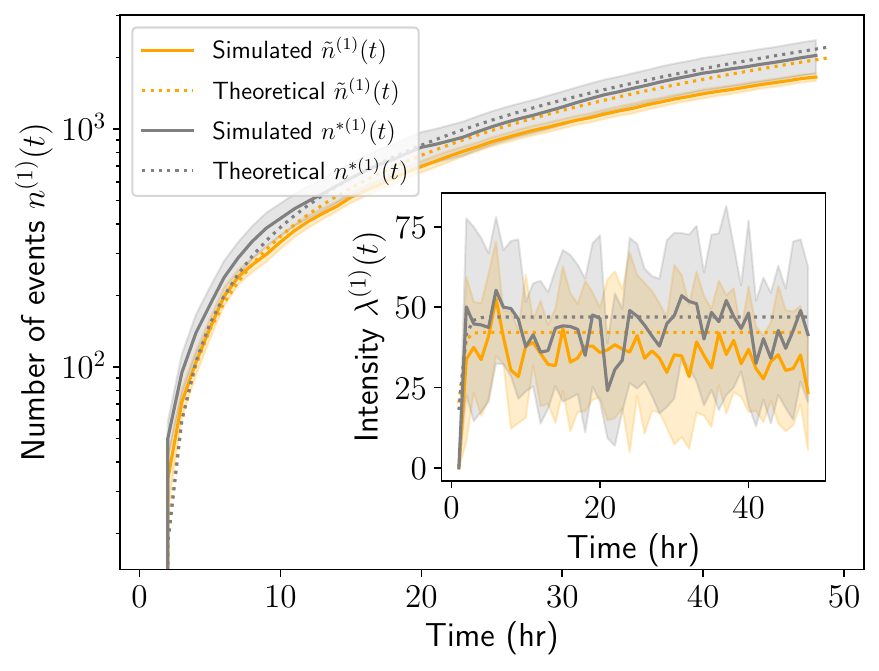}%
  \label{sub:int_1_grid_ex}}
  \subfloat[Cascade $3$]{\includegraphics[width=1.75in]{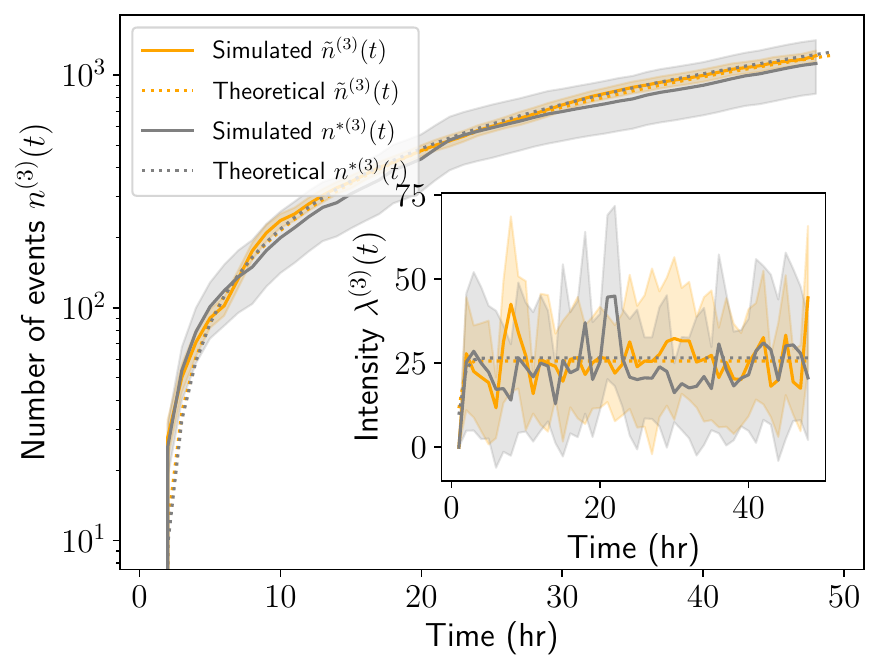}%
  \label{sub:int_3_grid_ex}}
  \caption{Number of events and intensity of cascades $c_1$ and $c_3$ over time, for the true $(n^{*(\cdot)}(t),\lambda^{*(\cdot)}(t))$, and the generated events given the learned \MIC model $(\Tilde{n}^{(\cdot)}(t),\Tilde{\lambda}^{(\cdot)}(t))$. Error bars correspond to the simulation on $10$ event generations. Dotted lines are the theoretical expected quantities for the \MIC model, computed using both the true and the inferred parameter values. The event dataset has been generated with the following parameterization: $\beta=33.37$ and $\mathbf{\Sigma} =\big[\begin{smallmatrix}1 & 0 & 0\\.71 & .29 & 0\\ 0 & 0 & 1\end{smallmatrix}\big]$.}
  \label{fig:int_eve_grid_ex}
\end{figure}

\begin{figure*}[!t]
  \centering
  \subfloat[Test log-likelihood]{\raisebox{0.33cm}{\rotatebox{90}{\hspace{0em}\colorbox{gray!15}{\scriptsize\hspace{4em}\texttt{music2}\hspace{4em}}}\hspace{1.3em}%
  \includegraphics[width=1.7in, valign=b]{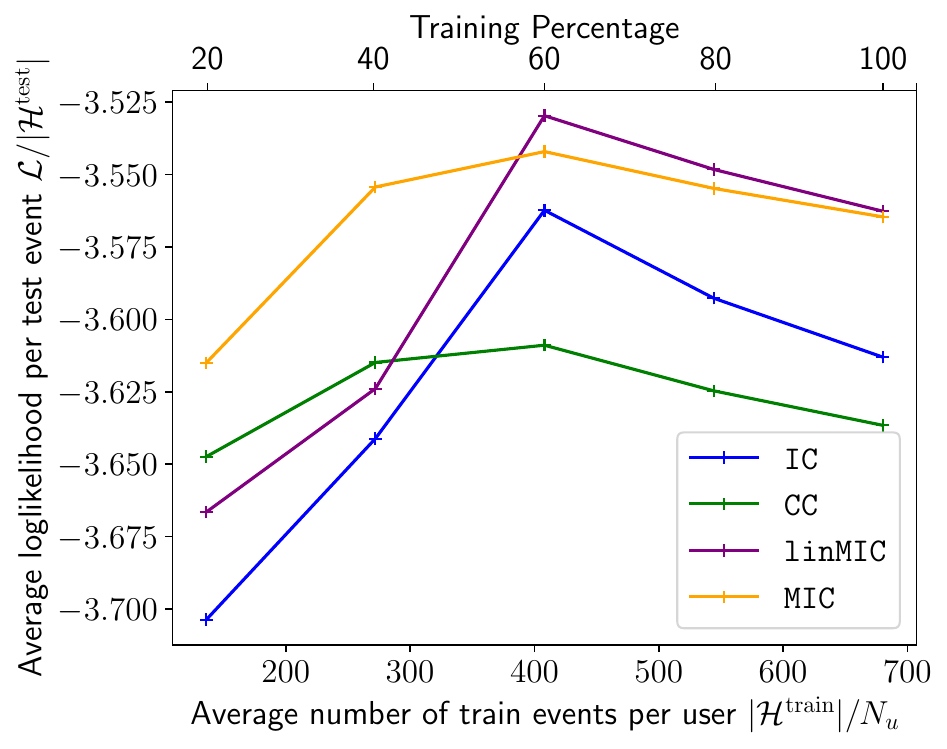}}%
  \label{sub:lglk_music2}}
  \hspace{2.6em}
  \subfloat[Inverse $l_1$-distance]{\includegraphics[width=1.7in, valign=b]{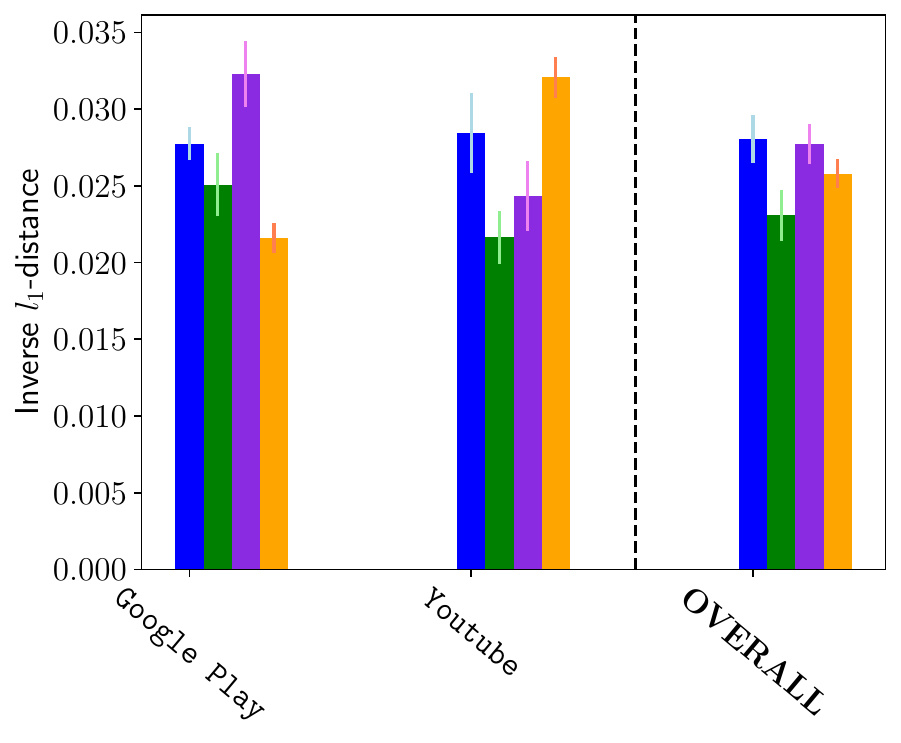}%
  \label{sub:inv_l1_music2}}
  \hspace{0.2em}
  \subfloat[Pearson correlation]{\includegraphics[width=1.7in, valign=b]{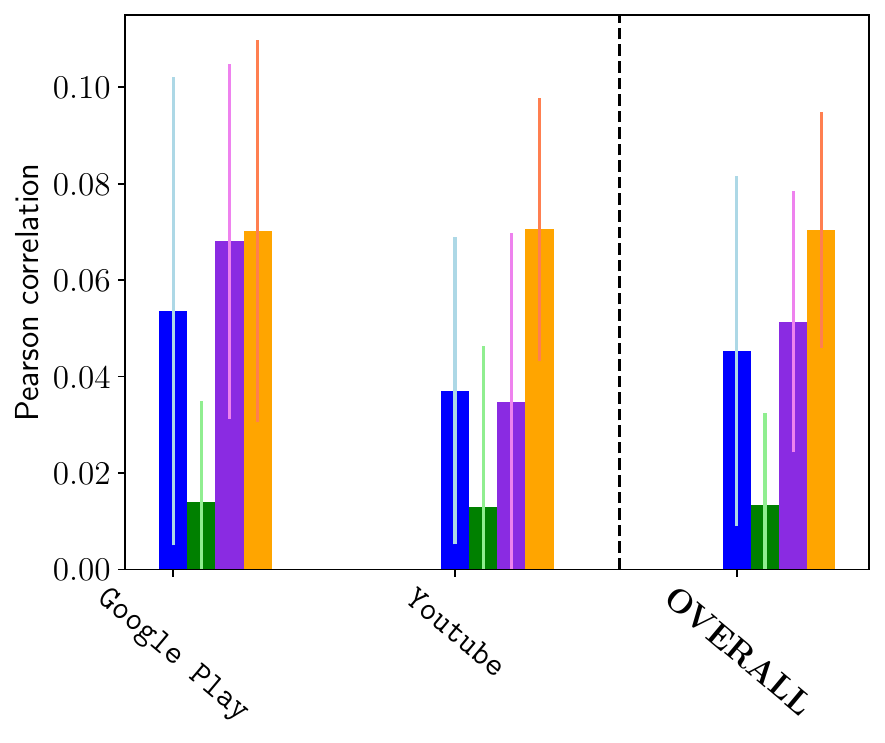}%
  \label{sub:corr_2_music2}}
	\\[-1em]
  \hspace{0.2em}\subfloat[Test log-likelihood]{\raisebox{0.33cm}{\rotatebox{90}{\hspace{0em}\colorbox{gray!15}{\scriptsize\hspace{4em}\texttt{url}\hspace{4em}}}\hspace{1.3em}%
  \includegraphics[width=1.6in, valign=b]{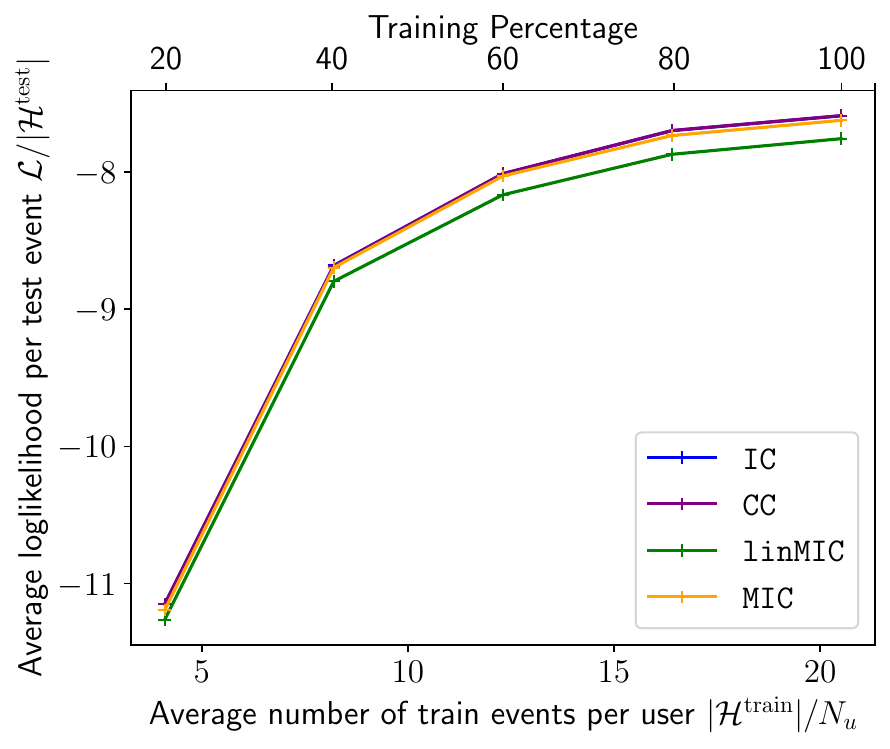}}%
  \label{sub:lglk_url}}
  \hspace{2.7em}
  \subfloat[Inverse $l_1$-distance]{\includegraphics[width=1.7in, valign=b]{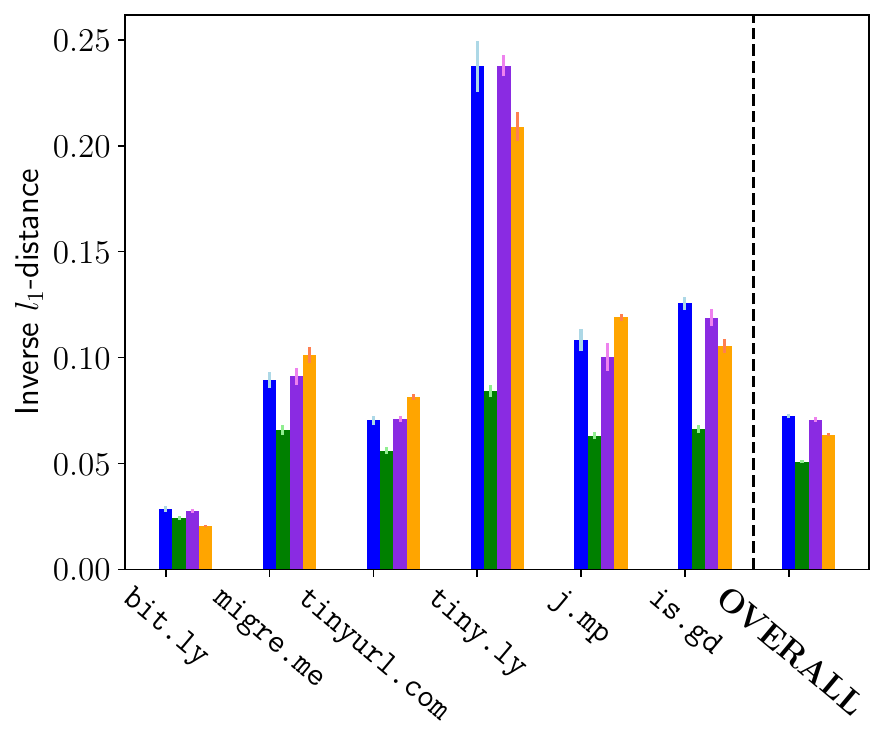}%
  \label{sub:inv_l1_url}}
  \hspace{0.2em}
  \subfloat[Pearson correlation]{\includegraphics[width=1.7in, valign=b]{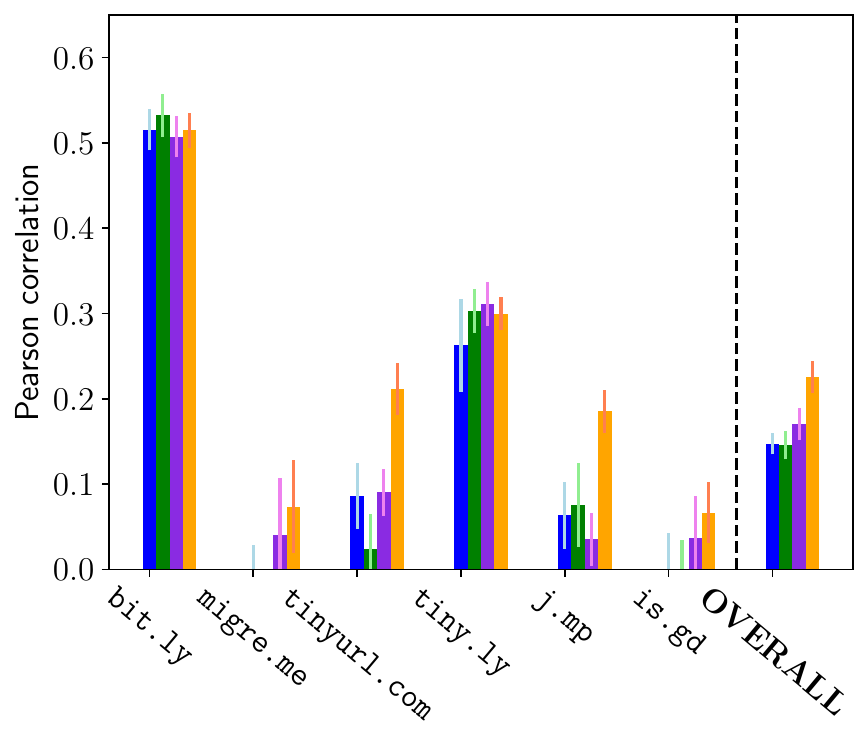}%
  \label{sub:corr_url}}
  \\[-1em]
	\subfloat[Test log-likelihood]{\raisebox{0.4cm}{\rotatebox{90}{\hspace{0em}\colorbox{gray!15}{\scriptsize\hspace{2.4em}\texttt{élysée2017}\hspace{2.4em}}}\hspace{1em}%
  \includegraphics[width=2.in, valign=b]{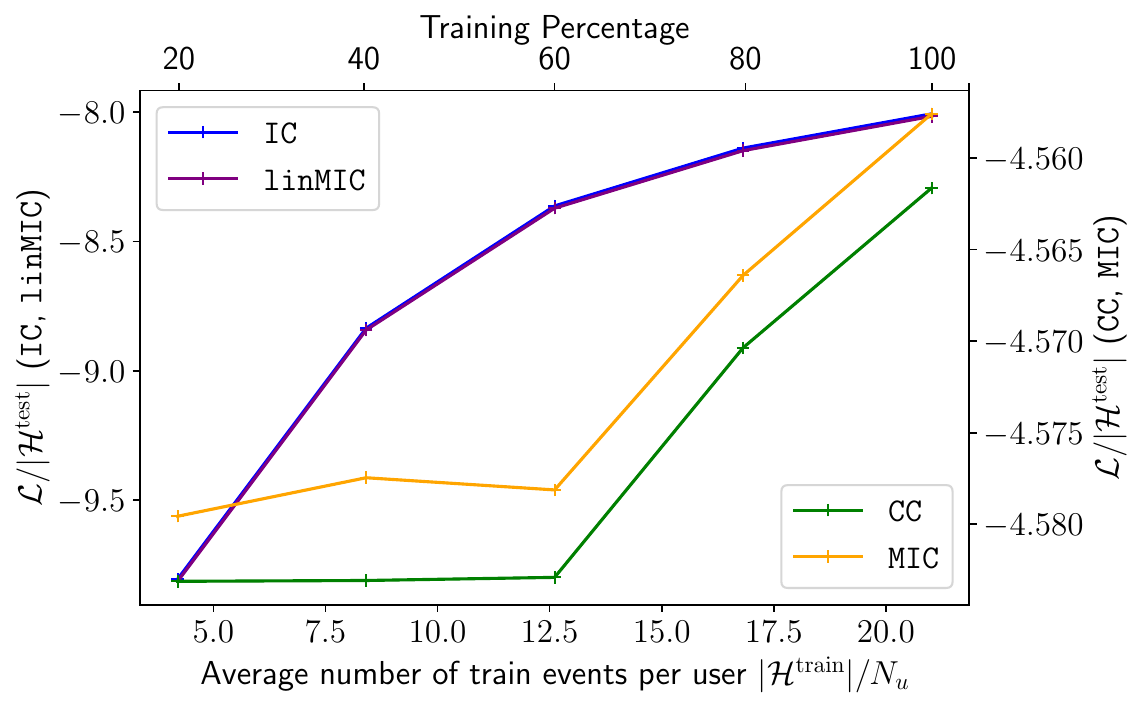}}%
  \label{sub:lglk_elysée2017}}
  \hspace{0.2em}
  \subfloat[Inverse $l_1$-distance]{\includegraphics[width=1.7in, valign=b]{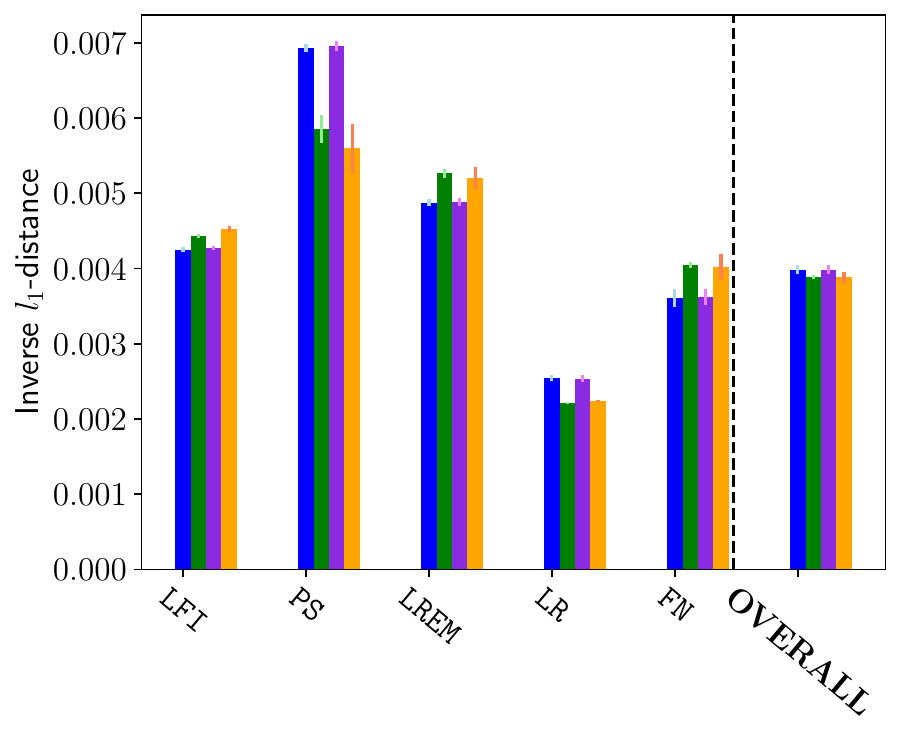}%
  \label{sub:inv_l1_elysée2017}}
  \hspace{0.2em}
  \subfloat[Pearson correlation]{\includegraphics[width=1.7in, valign=b]{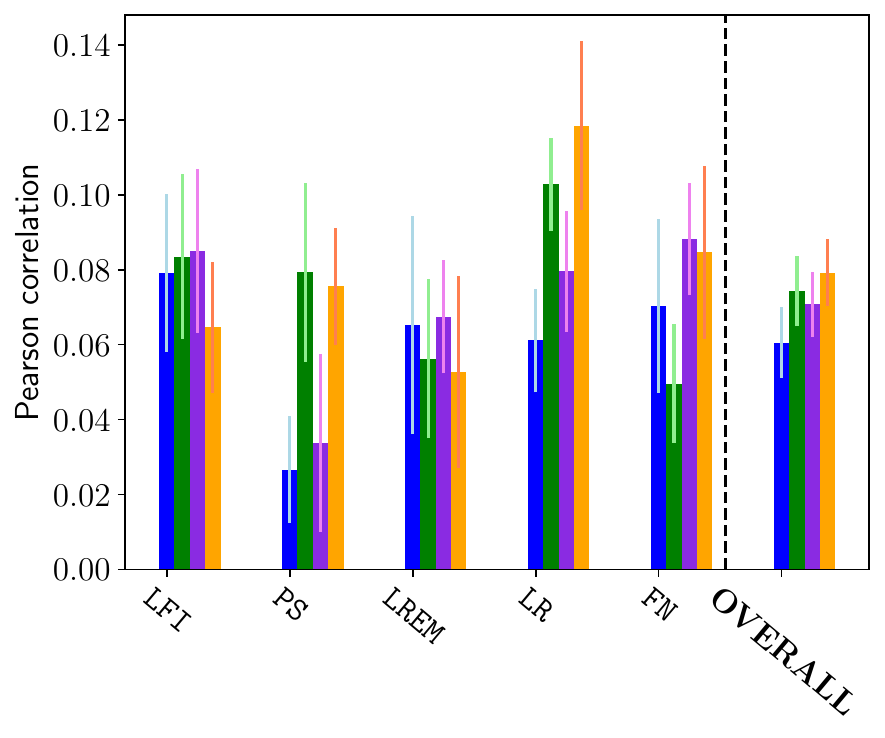}%
  \label{sub:corr_elysée2017}}
  \caption{Evaluation of the compared methods when applied on the \texttt{url} dataset (top row, $6$ cascades) and the \texttt{élysée2017} dataset (bottom row, $5$ cascades), using three measures: (a,d,g)~Test log-likelihood for a varying fraction of the initial train dataset (the bottom x-axis shows number of training events per user, and the top x-axis shows the correspondence to percentages). For \texttt{élysée2017}, the left y-axis corresponds to \IC, \MIClin, while the right y-axis corresponds to \CC and \MIC (note the difference in the scale between the left and right y-axes).
  (b,e,h)~Inverse $l_1$-distance for the adoption of each cascade, and for the overall intensity. (c,f,i)~Pearson correlation between the real and the simulated intensity of each cascade.}
  \label{fig:eval_url_ely2017}
\end{figure*}

\subsubsection{Comparing empirical intensities and number of events%
}

In this part we focus on the evaluation of the \MIC inference scheme (\Sec{sec:par-inferrence}), as well as the precision of the theoretical results (\Sec{sec:th_derivation}). We choose an instance of the previous experiment, namely the synthetic dataset generated by \MIC with $\beta = 33.37$, $\mathbf{\Sigma} =\big[\begin{smallmatrix}1 & 0 & 0\\.71 & .29 & 0\\ 0 & 0 & 1\end{smallmatrix}\big]$.
We treat $\beta$ and $\tau$ as known so as to assess independently the inference of the cascade interaction matrix (\Alg{alg:optimize_theta}), which in this case is recovered accurately:
    $\mathbf{\tilde{\Sigma}} = \big[\begin{smallmatrix}
      1 & 0 & 0\\
      .73 & .25 & .01\\
      0 & 0 & 1
    \end{smallmatrix}$\big].
Next, to evaluate the precision of the other inferred parameters ($\mathbf{\tilde{M}}, \mathbf{\tilde{\Weight}}$), we compare the number of events and intensity of the true $(n^{*(\cdot)}(t),\lambda^{*(\cdot)}(t))$, with the ones generated by simulations of the learned \MIC model $(\Tilde{n}^{(\cdot)}(t),\Tilde{\lambda}^{(\cdot)}(t))$. To validate our theoretical results, we also compute the analytical expectation of number of events and intensity for both ground truths and learned model.%

\Fig{fig:int_eve_grid_ex} shows the temporal evolution of both simulated and theoretical conditional event intensities and the realized number of events for each cascade in the generated data. It can be noticed that the inferred model is able to replicate the true event generating process%
, both in terms of average intensities and variance. In addition to that, the theoretical and simulated conditional intensities agree well for each cascade: the expected intensity converges to the stationary value after a negative exponential increase. The second cascade, less active, is however underestimated by simulations, which probably originates from the stationary state approximation made in \Eq{eq:approx_lambda}. Nevertheless, this experiment indicates that \MIC is not only statistically more adaptive than its competitors (as also shown earlier in \Fig{fig:grid_AUCs}), but also that the accurate parameter inference enables the model to provide accurate predictions for new events over time, both numerically and theoretically.

\subsection{Experiments on real data}\label{realexp}

\subsubsection{Data description}
We use three Twitter datasets and a one from a music platform, containing user activity logs while interacting with different content types. Social networks are characterized by heterogeneous and non-stationary user and cascade activity, \ie large differences in overall volume of events per user and per cascade, as well as temporal variations and bursts of the intensity of events. For the real datasets, the hyperparameters $\beta$ and $\tau$ are cross-validated (see details in the Appendix).%

\inlinetitle{\mySqBullet~\texttt{music2} dataset \textup{\cite{zarezadeCorrelatedCascadesCompete2017}}}{}~is a registry of Twitter users sharing URLs to Google Play and Youtube (taken as two cascades) of songs during one month of $2015$. The dataset is filtered so to keep the $\Nu = 93$ users with the highest adoption of these $\Nc=2$ cascades (more than $50$ tweets), which amounts to a total of $N_e = 79065$ events.

\inlinetitle{\mySqBullet~\texttt{url} dataset \textup{\cite{hodasSimpleRulesSocial2014}}}{}~contains tweets with URLs transformed by shortening services, commonly used when users share links on Twitter. We selected a (rather) stationary segment of the stream, and then we filtered the event log so as to keep $\Nu=637$ users related to the $\Nc=6$ cascades: \texttt{bit.ly}, \texttt{migre.me}, \texttt{tinyurl.com}, \texttt{tiny.ly}, \texttt{j.mp}, and \texttt{is.gd}, during $10$ days in $2010$, which amounts to a total of $N_e = 65321$ events.

\inlinetitle{\mySqBullet~\texttt{élysée2017} dataset \textup{\cite{fraisierElysee2017fr2017French2018}}}{}~contains Twitter activity around the $2017$ French presidential elections. We use a retweet network of $N_u=8003$ users involved in retweeting messages from accounts with $\Nc = 5$ political affiliations (\texttt{LREM} (center), \texttt{LR} (right), \texttt{PS} (left), \texttt{FN} (far right), and \texttt{FI} (far left)). The dataset spans over $T=1$ month around the election day, and contains $N_e = 841097$ retweet events.

\inlinetitle{\mySqBullet~\texttt{lastFM} dataset \textup{\cite{herradaMusicRecommendationDiscovery}}}{}~is a registry of user listenings to music from the LastFM platform and spans over $T=6$ months. To ease visualization, we keep the $N_c = 50$ most popular artists (taken as cascades), as well as $N_u = 559$ users with at least $50$ events each in those cascades. %
The final dataset is then composed of a total of $N_e = 56296$ listening events.

\begin{table}
\centering
\caption{Test log-likelihoods divided by the number of test events for all users, for the four datasets. The best scores are highlighted in gray (values closer to zero are better; draws imply differences of less than $10^{-2}$).}
\centering
\vspace{-1.5em}
$\begin{array}{r|c|c|cccc}
		\cmidrule[0.7pt]{4-7}
		\multicolumn{6}{r}{\textsc{Dataset}}\\
		\midrule
	  \textsc{Model}& \phi(x) & \Sigma & \texttt{url} & \texttt{lastfm}& \texttt{élysée2017}& \texttt{music2}\\
    \cmidrule[0.7pt]{1-7}
		\IC& x & \Id &\cellcolor{lightgray}{-7.59} & -20.11 & -8.01 &-3.61\\
    \CC& \exp(\beta x)& \Id  &-7.76  &-19.37 & \cellcolor{lightgray}{-4.56}&-3.64\\
    \MIClin& x & \mathbf{\tilde{\Sigma}} &\cellcolor{lightgray}{-7.59} & -19.73 & -8.02 &\cellcolor{lightgray}{-3.56}\\
    \MIC& \exp(\beta x)&\mathbf{\tilde{\Sigma}}  &-7.62 &\cellcolor{lightgray}{-19.31} & \cellcolor{lightgray}{-4.56}&\cellcolor{lightgray}{-3.56}\\

		\bottomrule
  \end{array}$
  \label{tab:full_lglk}
\end{table}

\begin{figure}
  \centering
  \subfloat[Test log-likelihood]{\raisebox{0.cm}{\rotatebox{90}{\hspace{0em}\colorbox{gray!15}{\scriptsize\hspace{4em}\texttt{lastfm}\hspace{4em}}}
  {\includegraphics[width=1.66in, valign=b]{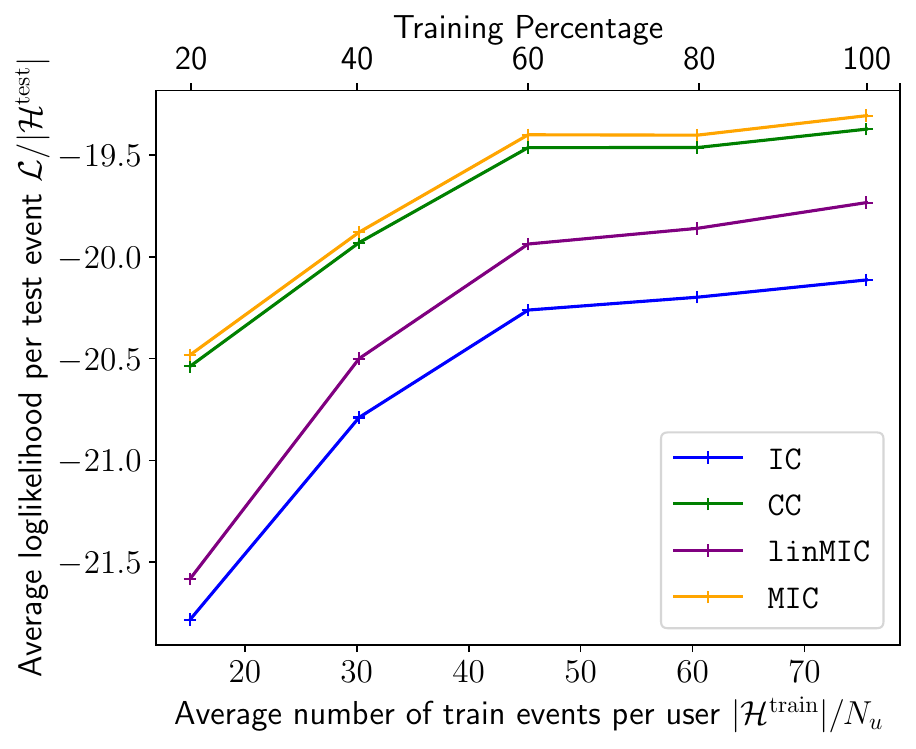}}}%
  \label{sub:lglk_lastfm}}
  \hfil
  \subfloat[Distrib. of cascade activity]{\raisebox{0cm}{\includegraphics[width=1.65in, valign=b]{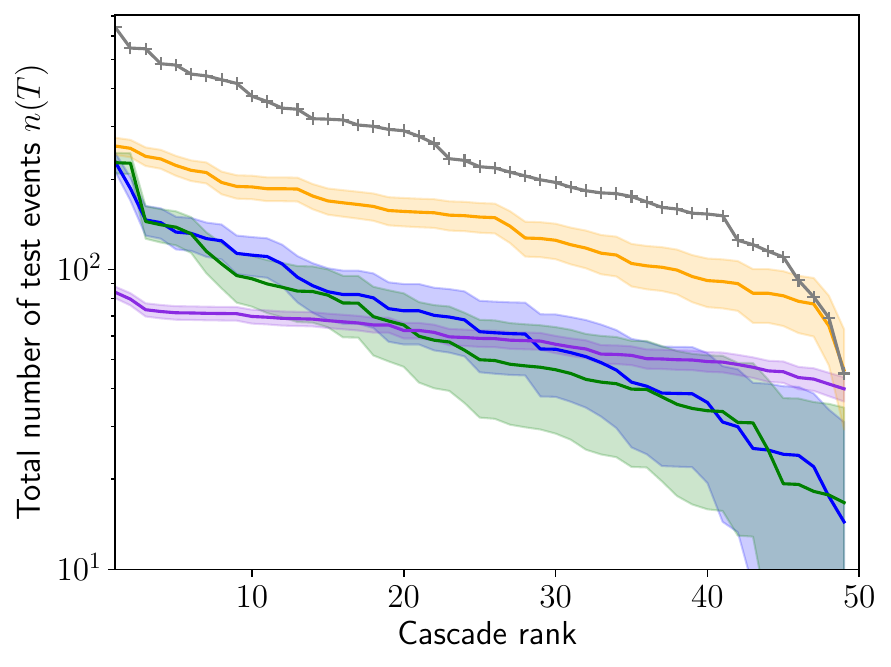}}%
  \label{sub:rank_c_lastfm}}
  \caption{Evaluation of the compared methods when applied on the \texttt{lastfm} dataset ($50$ cascades) using two measures: (a)~Test log-likelihood for a varying size of the initial train dataset (the bottom x-axis shows number of training events per user, and the top x-axis shows the correspondence to percentages).
  (b)~Ranked number of events for each cascade. Real data is compared to the generated events by each of the models.}
  \label{fig:eval_lastfm}
\end{figure}
\begin{figure}[!t]\footnotesize
  \centering
  \subfloat[\texttt{bit.ly}]{\includegraphics[width=1.76in]{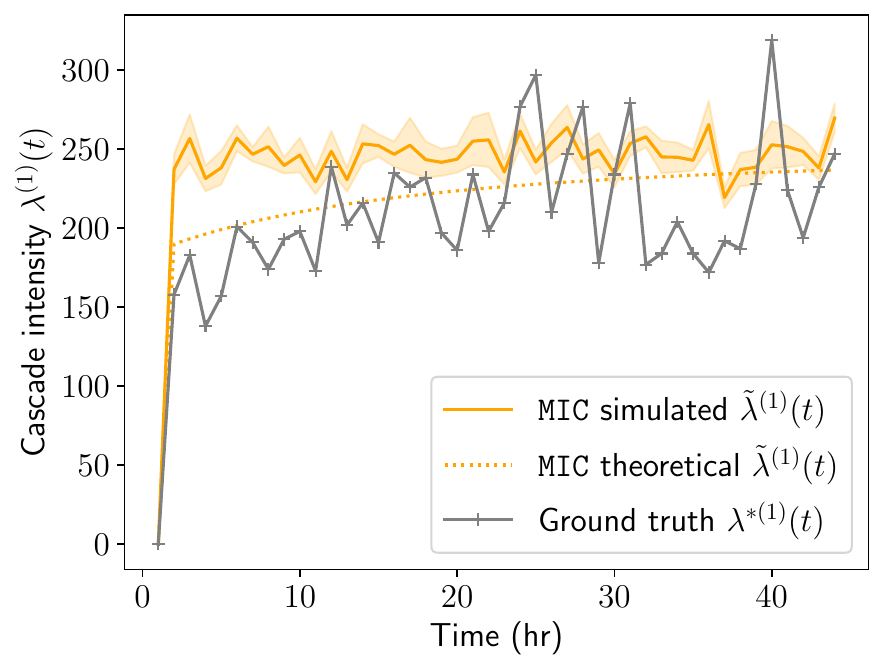}%
  \label{sub:ints_url_p1}}
  \hfil
  \subfloat[\texttt{j.mp}]{\includegraphics[width=1.74in]{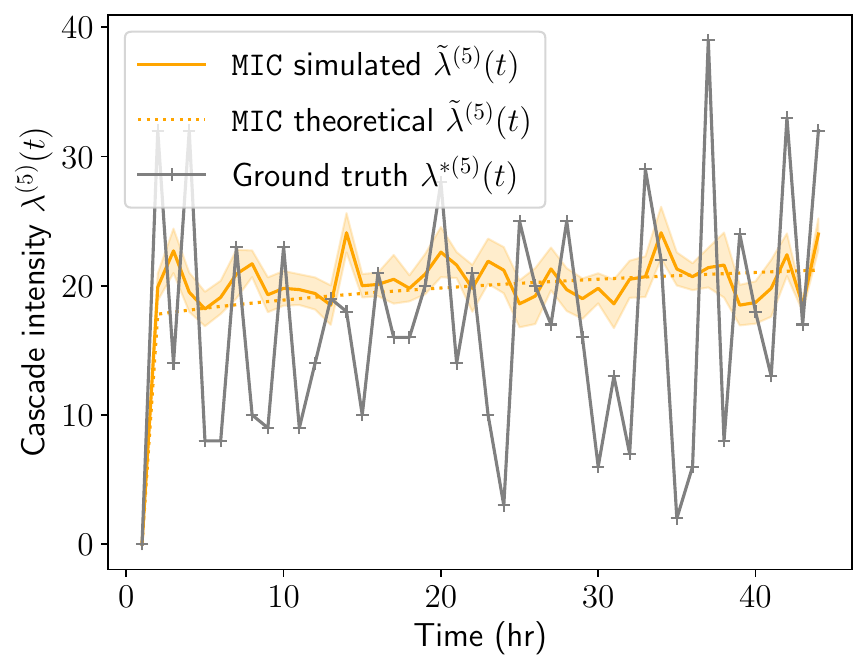}%
  \label{sub:ints_url_p5}}
  \caption{Temporal evolution of \MIC's conditional event intensities when applied to the \texttt{url} dataset. For each cascade, a signal is plotted for the learned \MIC and is compared to the associated theoretical expected intensity of the \MIC model (\Eq{eq:th_lambda}) and to test data.}
  \label{fig:ints_url}
\end{figure}

\subsubsection{Results}

Table \ref{tab:full_lglk} reports the test log-likelihood divided by the number of test events for all users, for the four real datasets. There, \MIC achieves the best or tied for the best performance in three datasets (and close second for the fourth one), confirming its ability to adapt to different real-world cascade dynamics. The table also illustrates the interest of considering both nonlinear cascade competition and interaction, by comparing \MIC to its ablated variants. A more detailed analysis follows, where we discuss the performance of the models on each dataset, and using various evaluation measures.%

\Fig{fig:eval_url_ely2017} shows results, arranged in rows, for three of the datasets that have a small number of cascades. The columns correspond to the three evaluation measures we use: test log-likelihood, inverse $l_1$-distance, and Pearson correlation. The first row refers to the smallest dataset, \texttt{music2}, where we see that \MIC starts with a much higher log-likelihood with less data and then competes with \MIClin. The other two measures --that do not encompass cascade interaction-- with \CC having the worst score. Despite the small size and simplicity of this dataset, we already see that accounting for cascade interactions is beneficial.
For the \texttt{url} dataset, the test log-likelihoods plotted in \Fig{sub:lglk_url} indicate that all models benefit as the training set grows (note: \IC is plotted exactly below \MIClin),
with \CC being slightly inferior. For the generation of events, \Fig{sub:inv_l1_url} indicates that even if the scores vary notably across the cascades, \MIC achieves a better approximation for $3$ out of the $6$ cascades, and is overall a little below linear models. Nevertheless, the Pearson correlation in \Fig{sub:corr_url} shows that, cascade-wise, \MIC is better or comparable to the other models, and overall exhibits a higher correlation to the real data. Note that \IC and \CC obtain negative scores for the $2$nd and the $6$th cascade.

In \Fig{fig:eval_url_ely2017}, the bottom row reports the results on the \texttt{élysée2017} dataset. The test log-likelihoods in \Fig{sub:lglk_elysée2017} indicate that \CC and \MIC  --which both consider Boltzmann mixing functions-- achieve a clearly higher performance than linear models, even with few training data. This suggests that the activity of a large set of users in this political context is better captured by nonlinear models. For the generation of events, the inverse $l_1$-distance in \Fig{sub:inv_l1_elysée2017} shows no clear superior accuracy, while the Pearson correlation in \Fig{sub:corr_elysée2017} confirm the overall superiority of \MIC in replicating the dynamics of the data generating process.

The results for the \texttt{lastfm} dataset are in \Fig{fig:eval_lastfm}. Similarly to the \texttt{élysée2017} case, the test log-likelihood in nonlinear models appear to have a superior test log-likelihood score than linear ones, with \MIC being the model with the best performance. Given the large number of cascades, we omit plotting for this dataset the inverse $l_1$-distance and Pearson correlation measures. Nevertheless, the larger diversity of this dataset can provide insights regarding the distribution of cascade popularity (\Fig{sub:rank_c_lastfm}) in log-scale. There, \MIC exhibits a more accurate distribution of events per cascade compared to the other models, both of number of events generated and in shape.
This plot suggests that none of the models gets sufficiently close to the test data, especially for the most popular cascades, hence seem unable to fully capture the complexity of this dataset. In fact, in this dataset, the largest event counts mostly originate from exogenous factors, such as the release of a new album or events like concerts.

\begin{figure}
  \centering
  \subfloat{%
	\rotatebox{90}{\hspace{0em}\colorbox{gray!15}{\tiny\hspace{1.5em}\textsc{all cascades}\hspace{1.5em}}}\hspace{0.2em}%
  {\includegraphics[width=1.1in,valign=b]{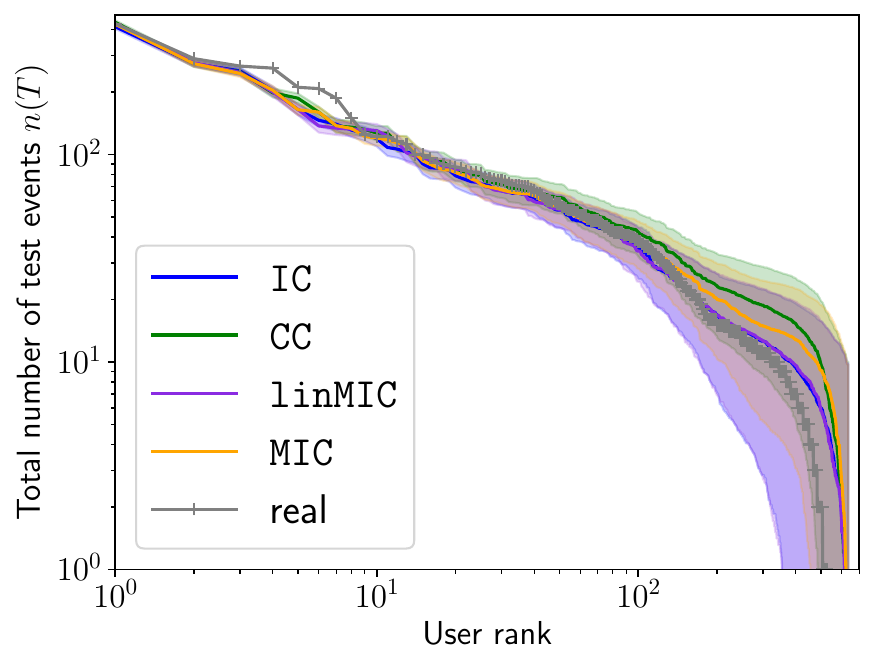}}%
  \label{sub:full_user_rank_url}}
  \hfil
  \subfloat{{\includegraphics[width=1.1in,valign=b]{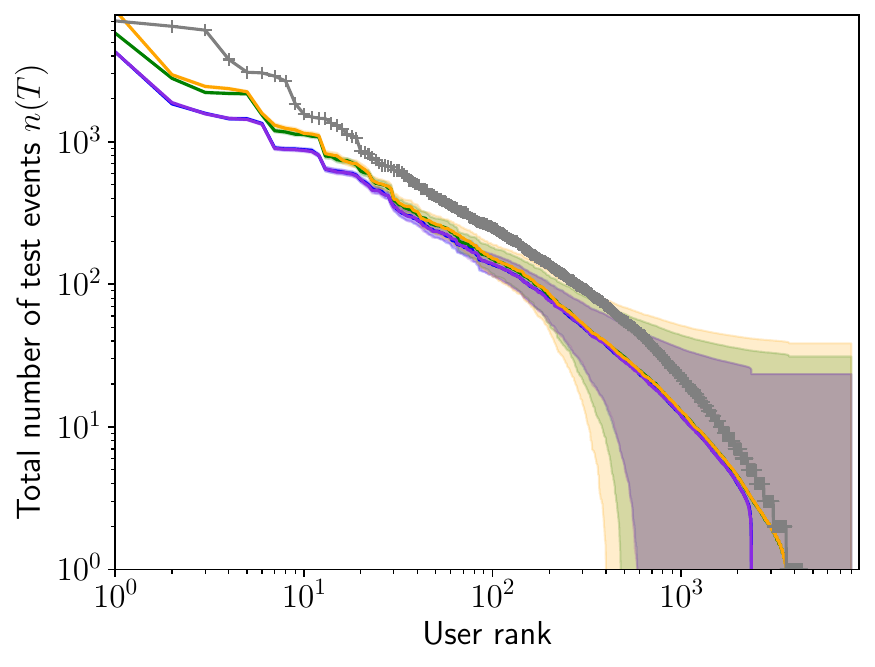}}%
  \label{sub:full_user_rank_ely2017_ter.pdf}}
  \hfil
  \subfloat{{\includegraphics[width=1.1in,valign=b]{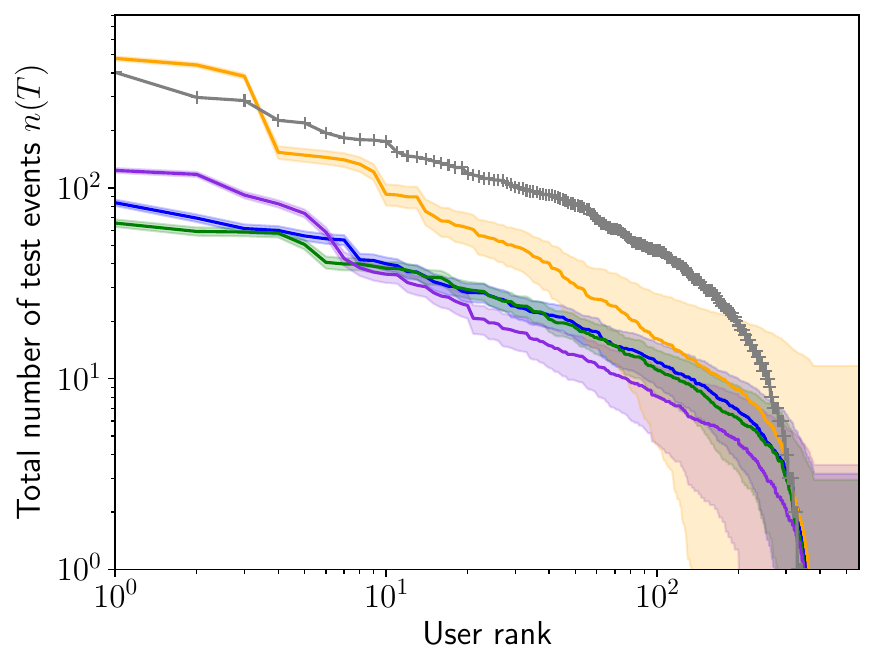}}%
  \label{sub:full_user_rank_lastfm}}\\
  \subfloat{%
	\rotatebox{90}{\hspace{0em}\colorbox{gray!15}{\tiny\hspace{.2em}\textsc{individual cascades}\hspace{.2em}}}\hspace{0.2em}%
{\includegraphics[width=1.1in,valign=b]{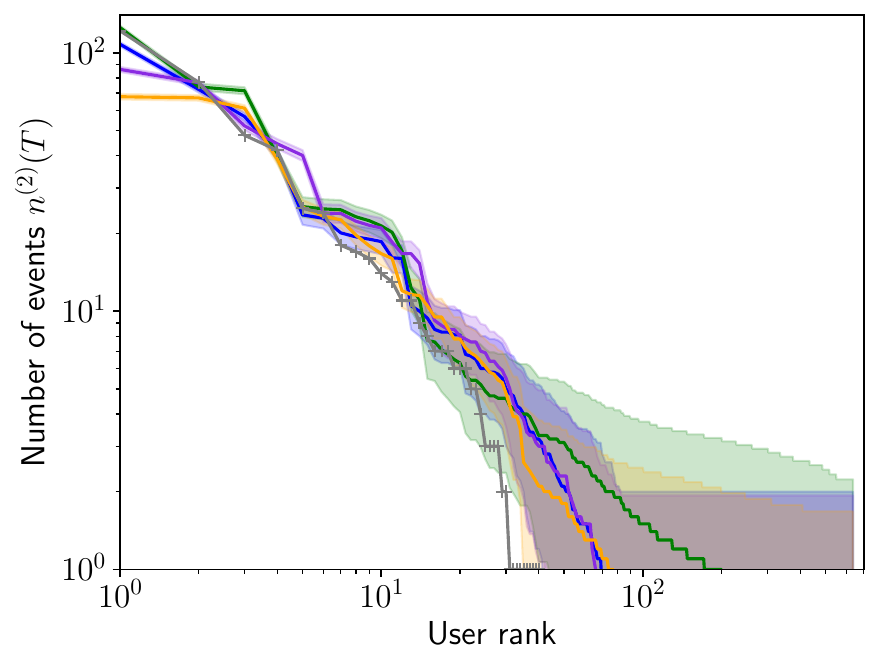}}%
  \label{sub:user_rank_url_c_1}}
  \hfil
  \subfloat{{\includegraphics[width=1.1in,valign=b]{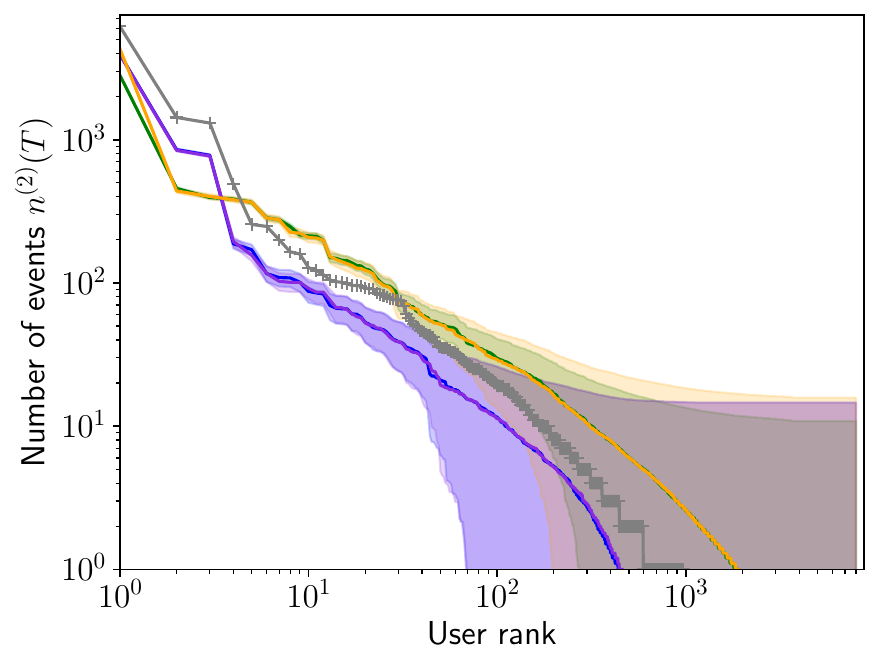}}%
  \hfil
  \label{sub:user_rank_ely2017_ter_c_1}}
  \subfloat{{\includegraphics[width=1.1in,valign=b]{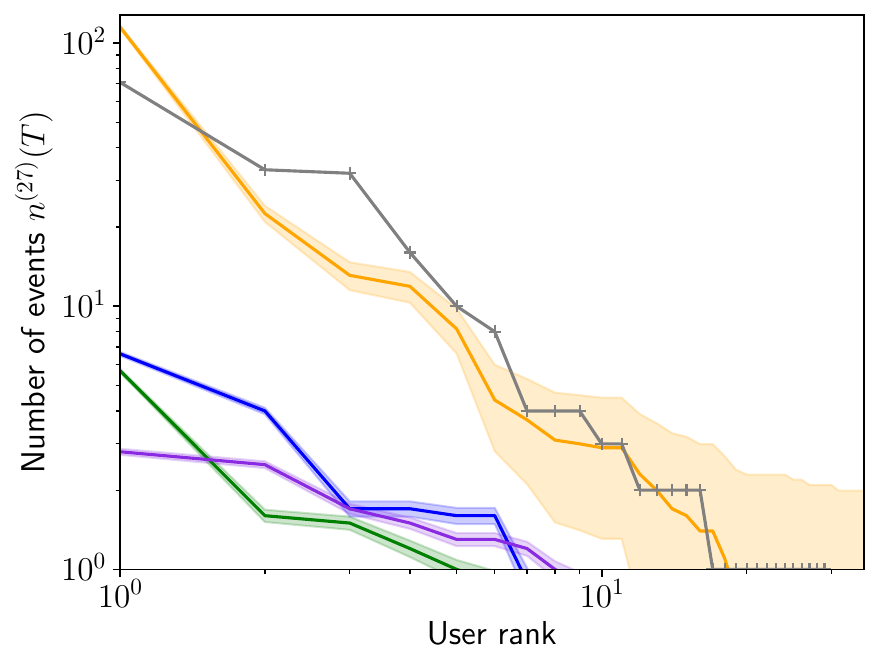}}%
  \label{sub:user_rank_lastfm_c_26}}\\
  \subfloat{%
	\rotatebox{90}{\hspace{0em}\colorbox{gray!15}{\tiny\hspace{.2em}\textsc{individual cascades}\hspace{.2em}}}\hspace{0.2em}%
{\includegraphics[width=1.1in,valign=b]{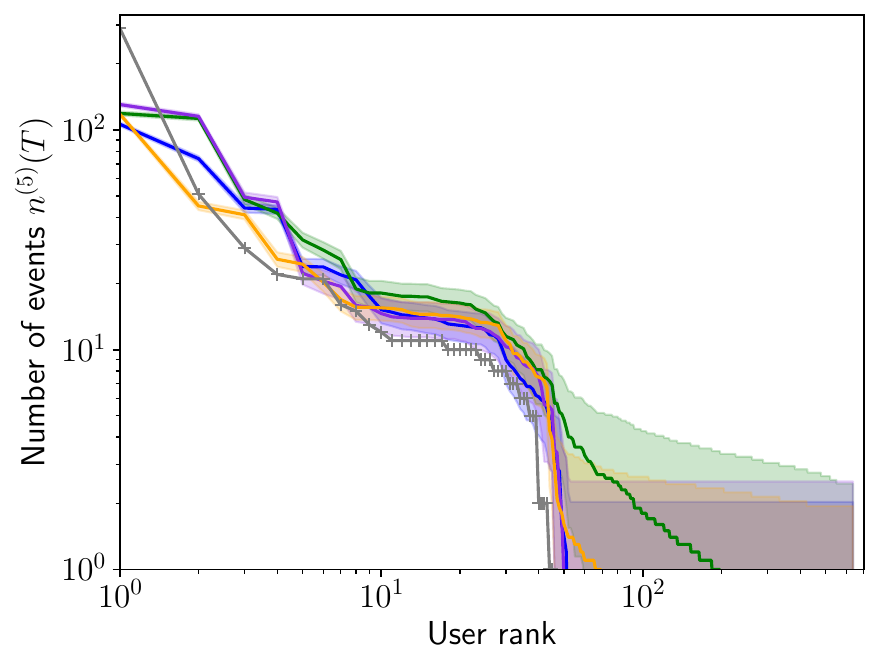}}%
  \label{sub:user_rank_url_c_4}}
  \hfil
  \subfloat{{\includegraphics[width=1.1in,valign=b]{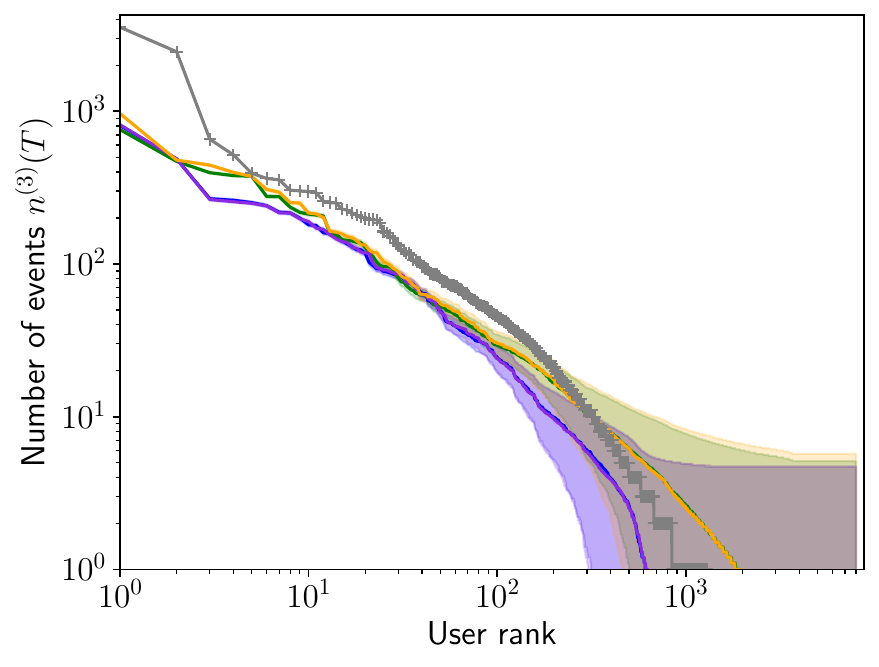}}%
  \label{sub:user_rank_ely2017_ter_c_2}}
  \hfil
  \subfloat{{\includegraphics[width=1.1in,valign=b]{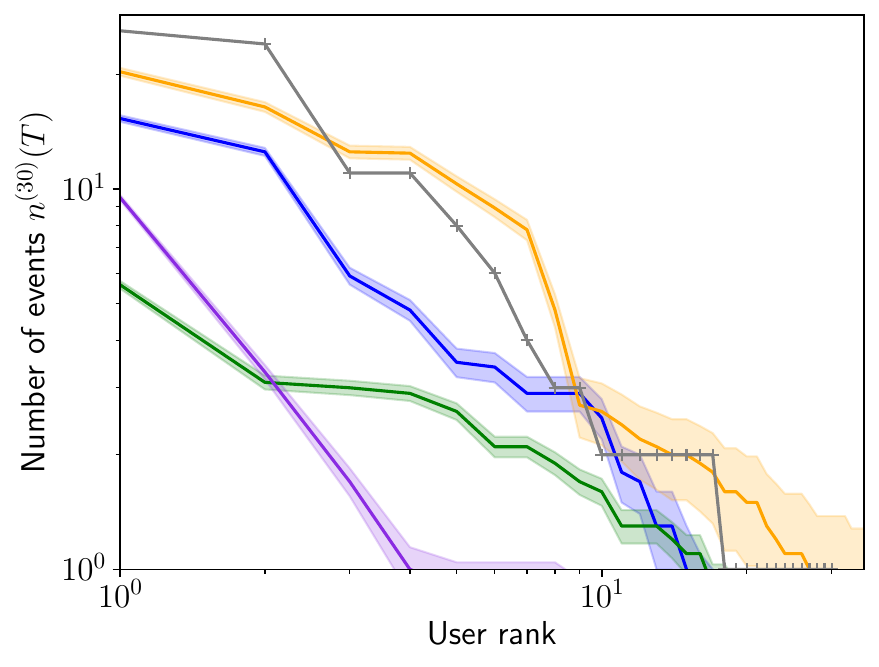}}%
  \label{sub:user_rank_lastfm_c_29}}\\
  \hfil
  \makebox[1in]{\ \quad\footnotesize (a)~\texttt{url}}
  \hfil
  \makebox[1in]{\quad\footnotesize (b)~\texttt{élysée2017}}
  \hfil
  \makebox[1in]{\quad\footnotesize (c)~\texttt{lastfm}}

  \caption{Ranked number of events per user of the compared methods when applied on the \texttt{url}, \texttt{élysée2017}, and \texttt{lastfm} datasets. The first row shows the  event distribution over all the cascades, which are however representative of the datasets. The error bars correspond to the standard deviation of the number of events per user.}
  \label{fig:user_rank_distributions}
\end{figure}%

Finally, we assess qualitatively \MIC in \Fig{fig:ints_url}, where we compare the intensity of test data to the one sampled from the learned \MIC models, as well as the theoretical expected intensity. The results show that the empirical conditional intensities over time of \MIC model are in agreement with the real test data, indicating an agreement between our numerical and analytical results in the context of the \texttt{url} dataset. %

\subsection{Heterogenous user activity in OSNs}\label{sec:heterogenous-activity}
We shift our focus to the behavior of the models at the level of individual users. %
We first compare in \Fig{fig:user_rank_distributions} user activity in the real test data to the activity generated by each trained model. The visual results validate the truncated power law behavior of social network activity, which was similarly observed cascade-wise in \Fig{fig:eval_lastfm}. For completeness we provide all the plots despite visible differences appear in only some of the cases. %
For instance, we notice in \Fig{sub:full_user_rank_url} that all methods generate roughly the same distribution. However, when we look at cascade-specific distributions from \texttt{url} and \texttt{élysée2017}, \MIC exhibits a greater capacity to replicate the distribution of user activity. This better capturing the heterogeneity of user activity is all the more visible when cascades are more in number and more diverse, as in \texttt{lastfm} (\Fig{sub:full_user_rank_lastfm}).

Interestingly, when comparing the test log-likelihoods at different quantiles of most active users for the four datasets in \Tab{tab:quant_lglk}, we can see that \MIC achieves a better score than the other models, except for \texttt{url} where it is close second. Moreover, the distance to the other models seems to grow as we focus on more active users, which are notably the users for which more can be learned by analyzing their higher activity.

\begin{table}[t]\footnotesize
\caption{Test log-likelihoods divided by the number of test events for all users, as well as the top $5\%$, $10\%$, and $25\%$ most active users for the three datasets. The best scores are highlighted in gray (values closer to zero are better; draws imply differences of less than $10^{-2}$).}
\centering
\vspace{-0.7em}
	$\begin{array}{crcccc}
		\cmidrule[0.7pt]{3-6}
		&&\multicolumn{4}{c}{\textsc{Dataset}}\\
		\midrule
	  \mathcal{L}_{\text{\texttt{<model>}}}/|\mathcal{H}^{\text{test}}|
		&\textsc{Model} & \texttt{music2} & \texttt{url} & \texttt{lastfm} & \texttt{élysée2017}\\
    \midrule
    \multirow{ 4}{*}{\text{top}~5\%~\text{users}}
		&\IC &-2.14 &\cellcolor{lightgray}{-1.98} & -20.65 & -8.35\\
    &\MIClin &-2.09 &\cellcolor{lightgray}{-1.98} & -20.17 & -8.36\\
    &\CC &-2.16 &-1.99 &-20.24 & -4.84\\
    &\MIC &\cellcolor{lightgray}{-2.05} &-1.99 &\cellcolor{lightgray}{-20.14} & \cellcolor{lightgray}{-4.83}\\
    \midrule
    \multirow{ 4}{*}{\text{top}~10\%~\text{users}}
		&\IC &-2.48 &\cellcolor{lightgray}{-3.09} & -20.05 & -9.06\\
    &\MIClin &-2.40 &\cellcolor{lightgray}{-3.09} & -19.57 & -8.39\\
    &\CC &-2.48 &-3.11 &-19.52 & \cellcolor{lightgray}{-4.86}\\
    &\MIC &\cellcolor{lightgray}{-2.38} &-3.10 &\cellcolor{lightgray}{-19.42} & \cellcolor{lightgray}{-4.86}\\
    \midrule
    \multirow{ 4}{*}{\text{top}~25\%~\text{users}}
		&\IC &-2.83 &\cellcolor{lightgray}{-5.14} & -20.4 & -8.39\\
    &\MIClin &-2.78 &\cellcolor{lightgray}{-5.14} & -20.00 & -8.02\\
    &\CC &-2.87 &-5.19 &-19.76 & -4.81\\
    &\MIC &\cellcolor{lightgray}{-2.77} &-5.16 &\cellcolor{lightgray}{-19.31} & \cellcolor{lightgray}{-4.80}\\
		\bottomrule
  \end{array}$
  \label{tab:quant_lglk}

\end{table}

These results emphasize the importance of user- and cascade-wise evaluation measures, and gives evidence on \MIC's superior performance in capturing user heterogeneity in social networks. %
This kind of evaluation constitutes an additional viewpoint on the performance of such models, typically overlooked in the literature. Nevertheless, evaluating further on this direction would require working with more diverse datasets, like \texttt{lastfm}.

\begin{figure*}[!t]
\vspace{-1em}
  \centering
  \subfloat[Visualization of real events (\texttt{lastfm})]{\rotatebox{90}{\hspace{0em}\colorbox{gray!15}{\footnotesize\hspace{5em}\textsc{layout for real data}\hspace{5em}}}\hspace{5em}
  \includegraphics[width=2.8in, valign=b]{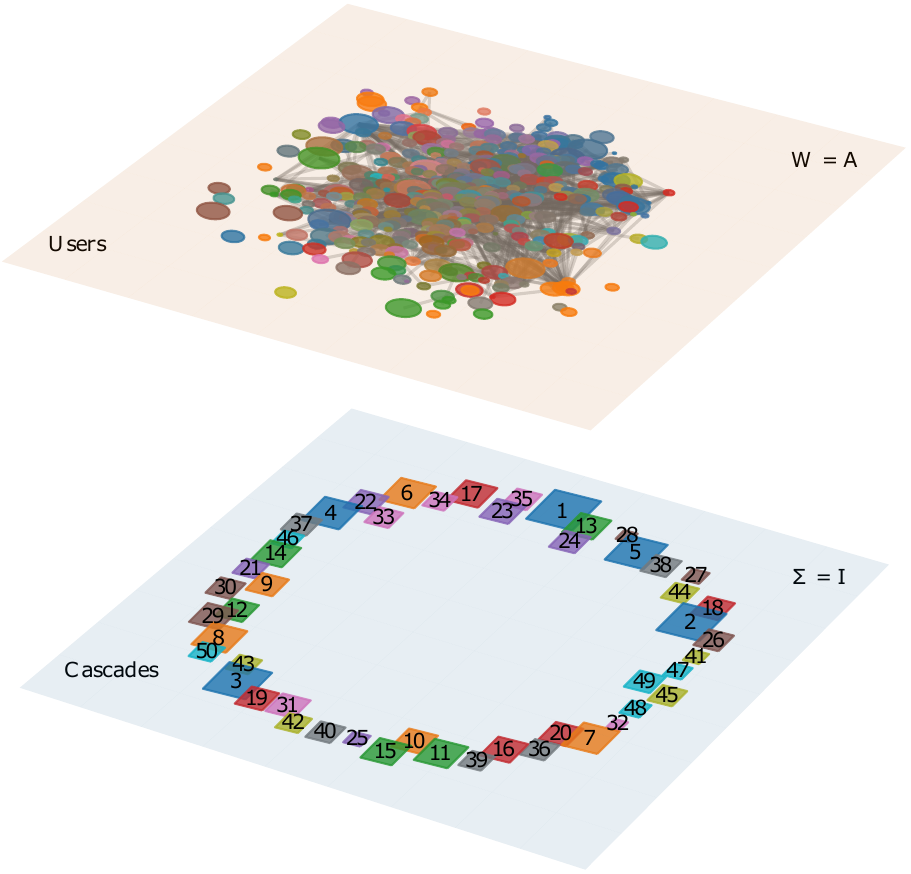}}%
  \label{sub:visu_real_lastfm}
  \hfil
  \subfloat[Visualization of real events (\texttt{élysée2017})]{%
	\includegraphics[width=2.8in, valign=b]{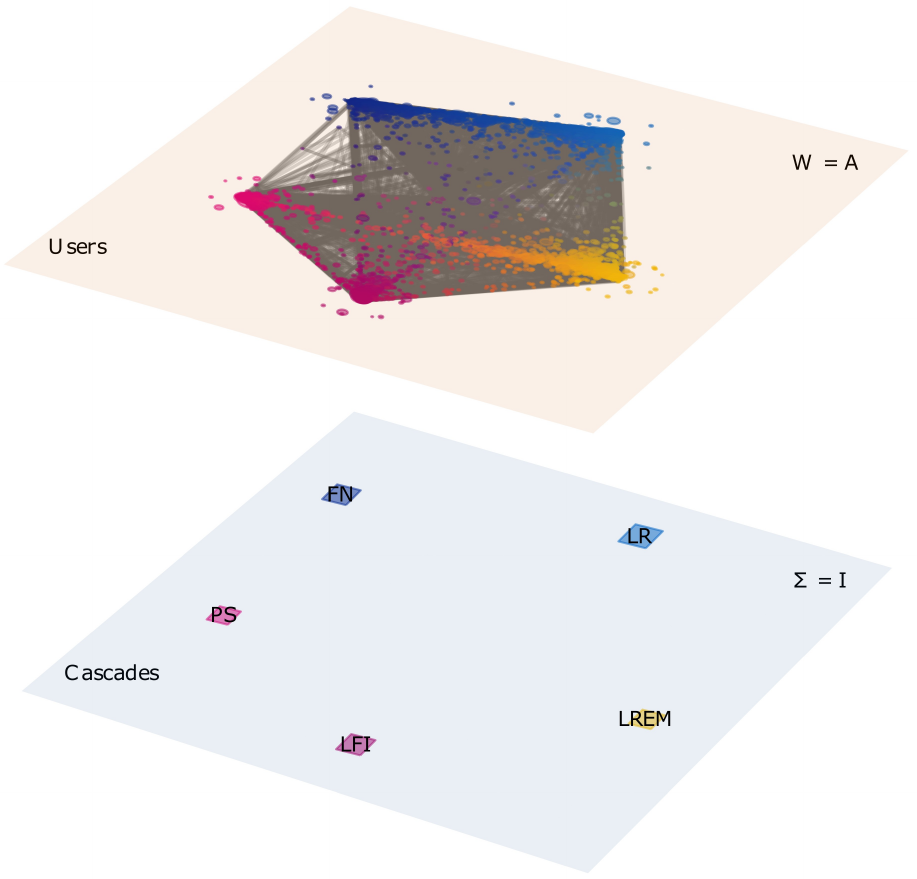}}%
  \label{sub:visu_real_ely2017}\\
  \subfloat[Visualization produced by \MIC (\texttt{lastfm})]{
  \rotatebox{90}{\hspace{0em}\colorbox{gray!15}{\footnotesize\hspace{3em}\textsc{layout with mic spatialization}\hspace{3em}}}\hspace{5em}%
  \includegraphics[width=2.8in, valign=b]{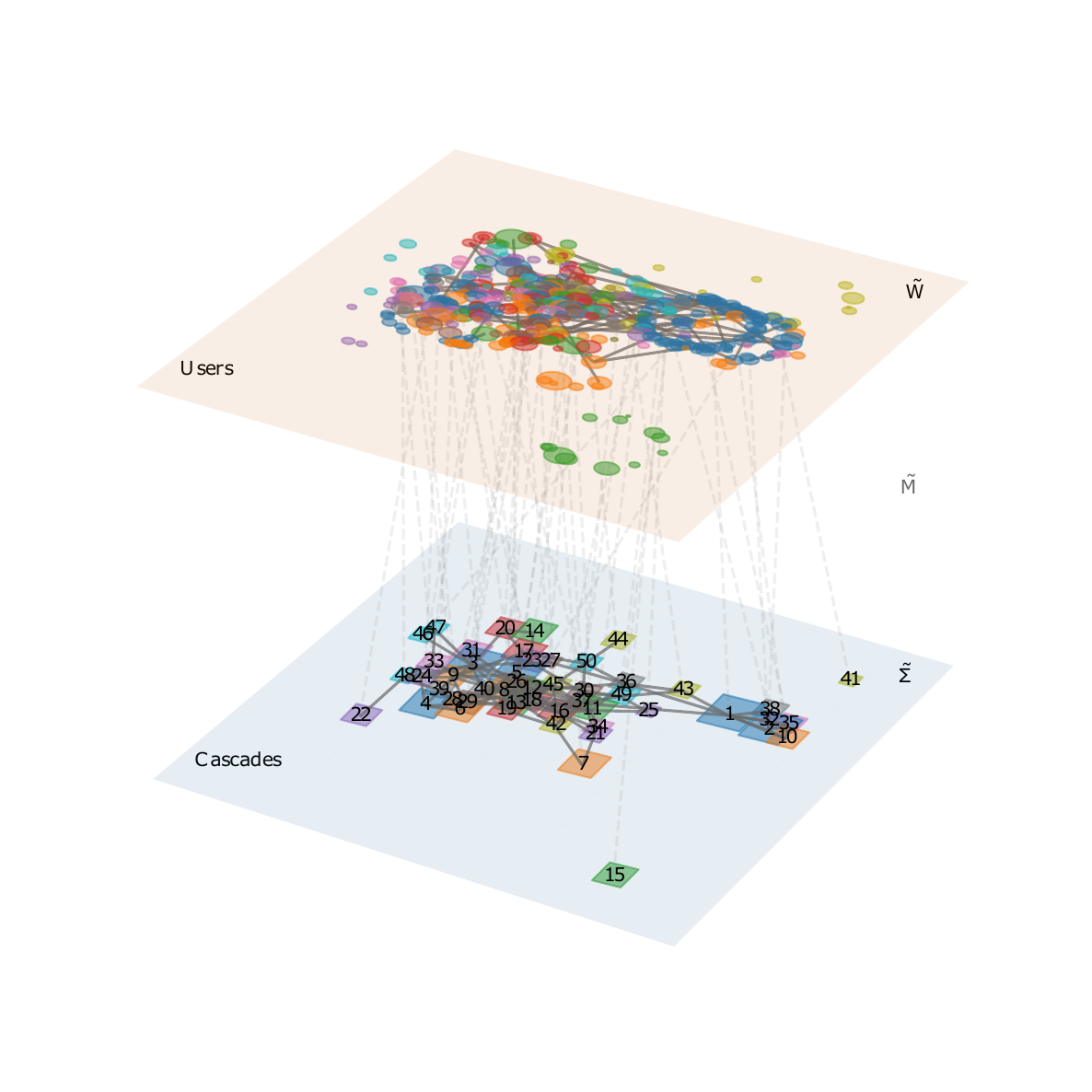}
  \label{sub:visu_a_lastfm}}
  \hfil
  \subfloat[Visualization produced by \MIC (\texttt{élysée2017})]{%
	\includegraphics[width=2.8in, valign=b]{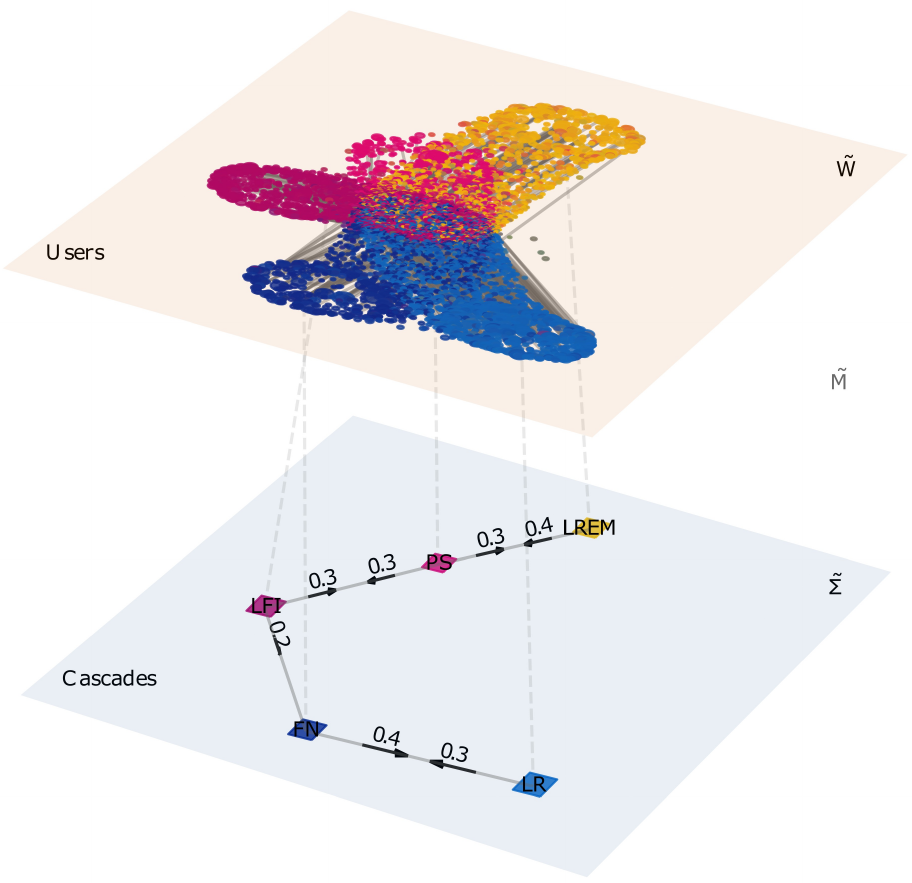}%
  \label{sub:visu_a_ely2017}}%
\\
   \caption{Bi-layered network visualization of the activity of the \texttt{élysée2017} (left) and \texttt{lastfm} (right) datasets. First row: the real data are visualized using a classic two-layer spring layout. Node size is proportional to the associated volume of events, and edges indicate the presence of influence/interactions. In the bottom layer, colors are defined to each cascade. In the top layer, user node color reflects the mixture cascades of activity. The unweighted adjacency matrix of the user graph is used for the top layers, while in the bottom layers there is no actual graph, so nodes are arranged in a circular layout aligned with the upper layer. Second row (c),(d): %
	a \MIC model is trained on the event log of each dataset, and the layout is based on the inferred parameters $\mathbf{\tilde{\Theta}}=(\mathbf{\tilde{\Sigma}}, \mathbf{\tilde{M}}, \mathbf{\tilde{\Weight}})$. Edges within the cascade (resp. user) graph are drawn given the cascade interaction matrix $\mathbf{\tilde{\Sigma}}$ (resp. user weight matrix $\mathbf{\tilde{\Weight}}$) and dashed edges between layers are drawn according to the highest values of user baseline intensities $\mathbf{\tilde{M}}$.}
  \label{fig:all_vis}

\end{figure*}

\subsection{Bi-layered visualization of social network activity}\label{sec:vizualization}

This section meets the promise %
of the schema in \Fig{fig:adoption_scheme}, by demonstrating that \MIC model can indeed produce such insightful visualizations for real data, using a \texttt{MultiLayerNetViz}, \cite{abelMultiLayerNetViz2026} a Python visualization module. Before discussing the results displayed in \Fig{fig:all_vis}, we should note that representing social network activity data as a bi-layered network, one layer for the users and another for the cascades, is intuitive, natural, and has been used in prior exploratory studies \cite{dedomenicoMuxVizToolMultilayer2015,loganSocialNetworkAnalysis2023}.  The contribution of our work in this aspect stems from the fact that \MIC parameters $\mathbf{\Theta}=(\mathbf{\Sigma}, \mathbf{M}, \mathbf{\Weight})$
can describe this structure, therefore such a bi-layered visualization can be computed numerically within our modeling framework by inferring the model parameters for real data. In these visualizations, node size reflects the associated volume of events, by users or cascades. An RGB color code is assigned to each cascade. User colors are computed using a weighted average of the cascade colors over the user activity distribution.

In the first row of \Fig{fig:all_vis}, the \texttt{élysée2017} and \texttt{lastfm} datasets are visualized using a typical bi-layer spring layout based of the real data. The adjacency matrix ($\mathbf{W} = \mathbf{A}$) gives the user graph at top layer, and the cascade activity is at the bottom layer, though lacking structure as this is not provided by the data ($\Sigma = \mathbf{\Id}$).
However, these visualizations are model-agnostic and do not capture the interactions between users and cascades, as well as the interplay between the two layers.

The second row (\Fig{sub:visu_a_lastfm},\,\ref{sub:visu_a_ely2017}) provides the bi-layered visualizations computed based on the \MIC model, specifically using the respective inferred parameters $\mathbf{\tilde{\Theta}}=(\mathbf{\tilde{\Sigma}}, \mathbf{\tilde{M}}, \mathbf{\tilde{\Weight}})$.  %
Edges within the cascade (resp. user) layer are drawn for %
values of $\mathbf{\tilde{\Sigma}}>0.1$ (resp. $\mathbf{\tilde{\Weight}}>10^{-5}$ for \texttt{lastfm}, $\mathbf{\tilde{\Weight}}>10^{-2}$ for \texttt{élysée2017})%
. Dashed edges between the two layers correspond to the highest value of the cascade-to-user coupling intensity $\mathbf{\tilde{M}}_{(\cdot)}^{(c)}$ for each cascade $c$. Cascade positions are computed using a spring layout given the inferred $\mathbf{\tilde{\Sigma}}$. The position of each user $u$ is computed based on his activity with respect to the cascades: the learned cascade mixing function $f_u(\cdot|t)$ maps user $u$'s activity to a probability distribution over the cascades, which is then projected into the cascade layout. To avoid cluttering, we use a convex combination of these coordinates with a spring layout given by $\mathbf{\tilde{W}}$ to obtain the final node positions.
This spatialization aims at uncovering how users are positioned according to their inferred cascade interests, compared to their activity in the social network.

For the \MIC-based visualization of the \texttt{lastfm} dataset (\Fig{sub:visu_a_lastfm}), the higher number of cascades leads to a dense network structure, with highly central artists (\eg $c_1$ and $c_3$). However, the inferred network has a typical center-periphery-isolated structure, as well as hubs connecting distinct communities (\eg $c_1$). The stronger mixing of cascades
is possibly due to the diversity of music tastes and cross-genre listening among users. Finally, in the \MIC-based visualization of the \texttt{élysée2017} dataset (\Fig{sub:visu_a_ely2017}), the cascade layer uncovers insightful %
interactions between political accounts. The inferred node chain \texttt{LREM-PS-LFI-FN-LR} does not reflect the left-right political spectrum, but rather %
the antagonism %
between parties during those %
elections. On the one hand, the central positions of \texttt{LFI} and \texttt{FN} cascades highlight their role as opposition parties against the mainstream ones, \texttt{LREM},\texttt{PS} and \texttt{LR}. On the other hand, the structure also reveals political alignments, with \texttt{LREM} and \texttt{PS} being close, as well as \texttt{LR} and \texttt{FN}. The user layer reveals interesting patterns, with distinct communities gravitating around specific political cascades, and the emergence of a group of indecisives, that are closer to \texttt{PS} and \texttt{LFI}.

\section{Conclusion}\label{sec:conclusion}
In this paper, we presented the \textit{\methodName} (\MIC) model for information diffusion in OSNs. \MIC is a marked Hawkes %
process providing an interpretable description of the coupling between cascades and user activity, %
and its parameter inference can be efficiently performed via Maximum Likelihood Estimation. We also derived differential equations for the moments of event counts and conditional intensity, providing analytical insights into diffusion dynamics.

The presented experiments provided a comprehensive evaluation of the \MIC model, using synthetic and real data, various complementary evaluation measures. %
The design of an ablation study, as well as the quantitative comparison with its competitors gives evidence for the different contributions of \MIC. Experiments on the more complex datasets \texttt{élysée2017} and \texttt{lastfm}, support the \MIC's proposition of considering both cascade interaction and nonlinear mixing functions. %
We have commented over the results by stressing that each evaluation measure provides a different perspective on the model's performance.
The results suggests that \MIC %
contributes in bridging the gap between theoretical and numerical approaches %
for modeling intricate information diffusion patterns. %

\inlinetitle{Perspectives}{.}~Further investigation can be concentrated on improving the accuracy of the model by refining modeling choices and parameter estimation, \eg by introducing time-dependent parameters ($\Theta_t$) and non-parametric triggering kernels \cite{zhouLearningTriggeringKernels2013,lemonnierMultivariateHawkesProcesses2017}. Such improvements can be easily implemented thanks to \MIC's modularity. Data preprocessing can mitigate the effects of non-stationarity and/or heterogeneity, \eg using existing methods to detect change-points in Hawkes processes \cite{zhangConjugateBayesianTwostep2024}, or deal with parameter learning given unbalanced data availability \cite{salehiLearningHawkesProcesses2019a}. %
The theoretical properties of the model could be elaborated to alleviate the current limitations regarding the decoupling assumptions between users and cascades.

\ifCLASSOPTIONcompsoc
  \section*{Acknowledgments}
\else
  \section*{Acknowledgment}
\fi

We are grateful to the anonymous reviewers for their insightful comments, which helped us improve the manuscript. G. Abel was funded by CDSN scholarship from ENS Paris-Saclay. A. Kalogeratos was supported by the Industrial Data Analytics and Machine Learning Chair hosted at ENS Paris-Saclay. J. Randon-Furling was funded by CPJ ModSHS at ENS Paris-Saclay.

\smallskip
\noindent \textbf{Contribution statement.} \mySqBullet~G.A.: Conceptualization, Methodology, Investigation, Formal analysis, Writing - original draft, Writing - review \& editing, Visualization, Validation, Software, Data curation; \mySqBullet~A.K.: Conceptualization, Methodology, Writing - original draft, Writing-review \& editing, Supervision, Validation, Methodology, Main scientific responsible. \mySqBullet~J.-P.N.: Writing-review \& editing, Supervision, Validation. \mySqBullet~J.R.-F.: Supervision, Validation.

\bibliographystyle{IEEEtran}
\bibliography{references}

@inproceedings{fraisierElysee2017fr2017French2018,
  title = {{The} 2017 {French} {Presidential} {Campaign} on {Twitter}},
  author = {Fraisier, Oph{\'e}lie and Cabanac, Guillaume and Pitarch, Yoann and Besan{\c c}on, Romaric and Boughanem, Mohand},
  year = {2018},
  booktitle = {AAAI \Conference on Web and Social Media},
  issn = {2334-0770},
  doi = {10.1609/icwsm.v12i1.14984},
  urldate = {2025-11-12},
  langid = {english}
}

@article{abelMultiLayerNetViz2026,
  title={{MultiLayerNetViz}: A Multilayered Visualization for Information Diffusion in Social Networks},
  author={Abel, Gaspard and Kalogeratos, Argyris and Nadal, Jean-Pierre and Randon-Furling, Julien},
  year={2026},
   url = {https://github.com/gas-abel/MultiLayerNetViz/blob/main/main.pdf},
  }

@book{aalenSurvivalEventHistory2008,
  title = {Survival and {{Event History Analysis}}: {{A Process Point}} of {{View}}},
  shorttitle = {Survival and {{Event History Analysis}}},
   author = {Aalen, Odd and Borgan, Ornulf and Gjessing, Hakon},
  year = 2008,
  publisher = {Springer},
  googlebooks = {2toprArSUMAC},
  isbn = {978-0-387-68560-1},
  langid = {english}
}

@article{aielloFriendshipPredictionHomophily2012,
  title = {Friendship Prediction and Homophily in Social Media},
  author = {Aiello, Luca Maria and Barrat, Alain and Schifanella, Rossano and Cattuto, Ciro and Markines, Benjamin and Menczer, Filippo},
  year = 2012,
  journal = {ACM \Transactions on the Web},
  volume = {6},
  number = {2},
  pages = {1--33},
  issn = {1559-1131, 1559-114X},
  doi = {10.1145/2180861.2180866},
  urldate = {2025-02-05},
  langid = {english}
}

@article{bacryModellingMicrostructureNoise2013,
  title = {Modelling Microstructure Noise with Mutually Exciting Point Processes},
  author = {Bacry, E. and Delattre, S. and Hoffmann, M. and Muzy, J. F.},
  year = 2013,
  journal = {Quantitative Finance},
  volume = {13},
  number = {1},
  pages = {65--77},
  publisher = {Routledge},
  issn = {1469-7688},
  doi = {10.1080/14697688.2011.647054},
  urldate = {2024-06-21}
}

@article{barberaTweetingLeftRight2015,
  title = {Tweeting {{From Left}} to {{Right}}: {{Is Online Political Communication More Than}} an {{Echo Chamber}}?},
  shorttitle = {Tweeting {{From Left}} to {{Right}}},
  author = {Barber{\'a}, Pablo and Jost, John T. and Nagler, Jonathan and Tucker, Joshua A. and Bonneau, Richard},
  year = 2015,
  journal = {Psychological Science},
  volume = {26},
  number = {10},
  pages = {1531--1542},
  publisher = {SAGE Publications Inc},
  issn = {0956-7976},
  doi = {10.1177/0956797615594620},
  urldate = {2025-10-27},
  langid = {english}
}

@article{baumannEmergencePolarizedIdeological2021,
  title = {Emergence of {{Polarized Ideological Opinions}} in {{Multidimensional Topic Spaces}}},
  author = {Baumann, Fabian and {Lorenz-Spreen}, Philipp and Sokolov, Igor M. and Starnini, Michele},
  year = 2021,
  journal = {Physical Review X},
  volume = {11},
  number = {1},
  pages = {011012},
  publisher = {American Physical Society},
  doi = {10.1103/PhysRevX.11.011012},
  urldate = {2023-09-08}
}

@book{boydConvexOptimization2004,
  title = {Convex {{Optimization}}},
  author = {Boyd, Stephen and Vandenberghe, Lieven},
  year = 2004,
  publisher = {Cambridge University Press},
  googlebooks = {IUZdAAAAQBAJ},
  isbn = {978-1-107-39400-1},
  langid = {english}
}

@inproceedings{caoDeepHawkesBridgingGap2017,
  title = {{DeepHawkes}: Bridging the Gap between Prediction and Understanding of Information Cascades},
  shorttitle = {{DeepHawkes}},
  booktitle = {{{ACM}} on {{\Conference}} on {{Information}} and {{Knowledge Management}}},
  author = {Cao, Qi and Shen, Huawei and Cen, Keting and Ouyang, Wentao and Cheng, Xueqi},
  year = 2017,
  pages = {1149--1158},
  doi = {10.1145/3132847.3132973},
  urldate = {2024-06-21},
  isbn = {978-1-4503-4918-5}
}

@article{chavalariasCanSingleLine2024,
  title = {Can a {{Single Line}} of {{Code Change Society}}? {{The Systemic Risks}} of {{Optimizing Engagement}} in {{Recommender Systems}} on {{Global Information Flow}}, {{Opinion Dynamics}} and {{Social Structures}}},
  shorttitle = {Can a {{Single Line}} of {{Code Change Society}}?},
  author = {Chavalarias, David and Bouchaud, Paul and Panahi, Maziyar},
  year = 2024,
  journal = {Journal of Artificial Societies and Social Simulation},
  volume = {27},
  number = {1},
  pages = {9},
  issn = {1460-7425}
}

@article{conjeaudDeGrootBasedOpinionFormation2024,
  title = {{{DeGroot-Based Opinion Formation Under}} a {{Global Steering Mechanism}}},
  author = {Conjeaud, Ivan and {Lorenz-Spreen}, Philipp and Kalogeratos, Argyris},
  year = 2024,
  journal = {IEEE \Transactions on Computational Social Systems},
  volume = {11},
  number = {3},
  pages = {4040--4057},
  issn = {2329-924X},
  doi = {10.1109/TCSS.2023.3330293},
  urldate = {2024-06-20}
}

@article{cuiElementaryDerivationMoments2020,
  title = {An Elementary Derivation of Moments of {{Hawkes}} Processes},
  author = {Cui, Lirong and Hawkes, Alan and Yi, He},
  year = 2020,
  journal = {Advances in Applied Probability},
  volume = {52},
  pages = {102--137},
  doi = {10.1017/apr.2019.53}
}

@book{daleyIntroductionTheoryPoint2003,
  title = {An {{Introduction}} to the {{Theory}} of {{Point Processes}}},
  author = {Daley, D.J. and {Vere-Jones}, D.},
  year = 2003,
  series = {Prob. and Its {{Applications}}},
  publisher = {Springer},
  doi = {10.1007/b97277},
  urldate = {2024-07-09},
  copyright = {http://www.springer.com/tdm},
  isbn = {978-0-387-95541-4},
  langid = {english}
}

@article{dedomenicoMuxVizToolMultilayer2015,
  title = {{{MuxViz}}: A Tool for Multilayer Analysis and Visualization of Networks},
  shorttitle = {{{MuxViz}}},
  author = {De Domenico, Manlio and Porter, Mason A. and Arenas, Alex},
  year = 2015,
  journal = {Journal of Complex Networks},
  volume = {3},
  number = {2},
  pages = {159--176},
  issn = {2051-1310},
  doi = {10.1093/comnet/cnu038},
  urldate = {2025-05-02}
}

@article{deffuantMixingBeliefsInteracting2000,
  title = {Mixing Beliefs among Interacting Agents},
  author = {Deffuant, Guillaume and Neau, David and Amblard, Frederic and Weisbuch, G{\'e}rard},
  year = 2000,
  journal = {Advances in Complex Systems},
  volume = {03},
  number = {01n04},
  pages = {87--98},
  publisher = {World Scientific Publishing Co.},
  issn = {0219-5259},
  doi = {10.1142/S0219525900000078},
  urldate = {2024-07-08}
}

@article{degrootReachingConsensus1974,
  title = {Reaching a {{Consensus}}},
  author = {Degroot, Morris H.},
  year = 1974,
  journal = {Journal of the American Statistical Association},
  volume = {69},
  number = {345},
  pages = {118--121},
  publisher = {Taylor \& Francis},
  issn = {0162-1459},
  doi = {10.1080/01621459.1974.10480137},
  urldate = {2024-06-24}
}

@inproceedings{deLearningForecastingOpinion2016,
  title = {Learning and {{Forecasting Opinion Dynamics}} in {{Social Networks}}},
  booktitle = {{Advances in Neural Information Processing Systems}},
  author = {De, Abir and Valera, Isabel and Ganguly, Niloy and Bhattacharya, Sourangshu and Gomez Rodriguez, Manuel},
  year = 2016,
  urldate = {2024-07-01}
}

@article{farajtabarCOEVOLVEJointPoint2017,
  title = {{{COEVOLVE}}: {{A Joint Point Process Model}} for {{Information Diffusion}} and {{Network Evolution}}},
  shorttitle = {{{COEVOLVE}}},
   author = {Farajtabar, Mehrdad and Wang, Yichen and {Gomez-Rodriguez}, Manuel and Li, Shuang and Zha, Hongyuan and Song, Le},
  year = 2017,
  journal = {Journal of Machine Learning Research},
  volume = {18},
  number = {41},
  pages = {1--49},
  issn = {1533-7928},
  urldate = {2024-06-21}
}

@article{goldenbergTalkNetworkComplex,
  title = {Talk of the {{Network}}: {{A Complex Systems Look}} at the {{Underlying Process}} of {{Word-of-Mouth}}},
  author = {Goldenberg, Jacob and Libai, Barak and Muller, Eitan},
	journal = {Marketing Letters},
	volume = 12,
	issue = 3,
	pages = {211–223},
  langid = {english}
}

@article{hawkesSpectraSelfexcitingMutually1971,
  title = {Spectra of Some Self-Exciting and Mutually Exciting Point Processes},
  author = {Hawkes, Alan},
  year = 1971,
  journal = {Biometrika},
  volume = {58},
  number = {1},
  pages = {83--90},
  issn = {0006-3444},
  doi = {10.1093/biomet/58.1.83},
  urldate = {2024-06-21}
}

@book{herradaMusicRecommendationDiscovery,
  title = {Music {{Recommendation}} and {{Discovery}} in the Long Tail},
  author = {Herrada, Oscar Celma},
  langid = {english},
	publisher = {Springer},
	year = {2010},
}

@article{hethcoteMathematicsInfectiousDiseases,
  title = {The {{Mathematics}} of {{Infectious Diseases}}},
  author = {Hethcote, Herbert},
  journal = {SIAM Review},
	volume = {42},
	number = {4},
	pages = {599--653},
	year = {2000},
}

@article{hodasSimpleRulesSocial2014,
  title = {The {{Simple Rules}} of {{Social Contagion}}},
  author = {Hodas, Nathan O. and Lerman, Kristina},
  year = 2014,
  journal = {Scientific Reports},
  volume = {4},
  number = {1},
  pages = {4343},
  publisher = {Nature Publishing Group},
  issn = {2045-2322},
  doi = {10.1038/srep04343},
  urldate = {2024-06-21},
  copyright = {2014 The Author(s)},
  langid = {english}
}

@inproceedings{hosseiniRecurrentPoissonFactorization2017,
  title = {Recurrent {{Poisson Factorization}} for {{Temporal Recommendation}}},
  booktitle = {{{ACM SIGKDD \International \Conference}} on {{Knowledge Discovery}} and {{Data Mining}}},
  author = {Hosseini, Seyed Abbas and Alizadeh, Keivan and Khodadadi, Ali and Arabzadeh, Ali and Farajtabar, Mehrdad and Zha, Hongyuan and Rabiee, Hamid R.},
  year = 2017,
  pages = {847--855},
  doi = {10.1145/3097983.3098197},
  urldate = {2023-09-15},
  isbn = {978-1-4503-4887-4}
}

@InProceedings{isikFlexibleTriggeringKernels2022,
  title = 	 {Hawkes Process with Flexible Triggering Kernels},
  author =       {Isik, Yamac and Chapfuwa, Paidamoyo and Davis, Connor and Henao, Ricardo},
  booktitle = 	 {Machine Learning for Healthcare \Conference},
  pages = 	 {308--320},
  year = 	 {2023},
}

@inproceedings{iwataDiscoveringLatentInfluence2013,
  title = {Discovering Latent Influence in Online Social Activities via Shared Cascade Poisson Processes},
  booktitle = {{{ACM SIGKDD}} \International \Conference on {{Knowledge}} Discovery and Data Mining},
  author = {Iwata, Tomoharu and Shah, Amar and Ghahramani, Zoubin},
  year = 2013,
  pages = {266--274},
  doi = {10.1145/2487575.2487624},
  urldate = {2025-02-19},
  isbn = {978-1-4503-2174-7},
  langid = {english}
}

@inproceedings{lemonnierMultivariateHawkesProcesses2017,
  title = {Multivariate {{Hawkes Processes}} for {{Large-Scale Inference}}},
  author = {Lemonnier, R{\'e}mi and Scaman, Kevin and Kalogeratos, Argyris},
  year = 2017,
  booktitle = {AAAI \Conference on Artificial Intelligence},
  issn = {2374-3468},
  doi = {10.1609/aaai.v31i1.10846},
  urldate = {2023-09-21},
  copyright = {Copyright (c)},
  langid = {english}
}

@inproceedings{lindermanDiscoveringLatentNetwork2014,
  title = {Discovering {{Latent Network Structure}} in {{Point Process Data}}},
  booktitle = {{{\International \Conference}} on {{Machine Learning}}},
  author = {Linderman, Scott and Adams, Ryan},
  year = 2014,
  pages = {1413--1421},
  issn = {1938-7228},
  urldate = {2025-02-19},
  langid = {english}
}

@inproceedings{liPublicOpinionField2025,
  title = {Public {{Opinion Field Effect}} and {{Hawkes Process Join Hands}} for {{Information Popularity Prediction}}},
  author = {Li, Junliang and Yang, Yajun and Zhang, Yujia and Hu, Qinghua and Zhao, Alan and Gao, Hong},
  year = 2025,
  booktitle = {AAAI \Conference on Artificial Intelligence},
  pages = {12076--12083},
  issn = {2374-3468},
  doi = {10.1609/aaai.v39i11.33315},
  urldate = {2025-10-27},
  copyright = {Copyright (c) 2025 Association for the Advancement of Artificial Intelligence},
  langid = {english}
}

@article{liuCTLISLearning2019a,
  title = {{CT LIS}: Learning Influences and Susceptibilities through Temporal Behaviors},
  author = {Liu, Shenghua and Shen, Huawei and Zheng, Houdong and Cheng, Xueqi and Liao, Xiangwen},
  year = {2019},
  journal = {ACM \Transactions on Knowledge Discovery from Data},
  volume = {13},
  number = {6},
  pages = {1--21},
  langid = {english}
}

@article{loganSocialNetworkAnalysis2023,
  title = {Social Network Analysis of {{Twitter}} Interactions: A Directed Multilayer Network Approach},
  shorttitle = {Social Network Analysis of {{Twitter}} Interactions},
  author = {Logan, Austin P. and LaCasse, Phillip M. and Lunday, Brian J.},
  year = 2023,
  journal = {Social Network Analysis and Mining},
  volume = {13},
  number = {1},
  pages = {65},
  issn = {1869-5469},
  doi = {10.1007/s13278-023-01063-2},
  urldate = {2025-05-02},
  langid = {english}
}

@inproceedings{luceriUnmaskingWebDeceit2024,
  title = {Unmasking the {{Web}} of {{Deceit}}: {{Uncovering Coordinated Activity}} to {{Expose Information Operations}} on {{Twitter}}},
  shorttitle = {Unmasking the {{Web}} of {{Deceit}}},
  booktitle = {ACM Web \Conference},
  author = {Luceri, Luca and Pant{\`e}, Valeria and Burghardt, Keith and Ferrara, Emilio},
  year = 2024,
  pages = {2530--2541},
  doi = {10.1145/3589334.3645529},
  urldate = {2025-10-27},
  isbn = {979-8-4007-0171-9}
}

@inproceedings{meiNeuralHawkesProcess2017,
  title = {The {{Neural Hawkes Process}}: {{A Neurally Self-Modulating Multivariate Point Process}}},
  shorttitle = {The {{Neural Hawkes Process}}},
  booktitle = {Advances in Neural Information Processing Systems},
  author = {Mei, Hongyuan and Eisner, Jason M},
  year = 2017,
  urldate = {2023-09-19}
}

@inproceedings{meiTransformerEmbeddingsIrregularly2021,
  title = {Transformer {{Embeddings}} of {{Irregularly Spaced Events}} and {{Their Participants}}},
  booktitle = {\International {{\Conference}} on {{Learning Representations}}},
  author = {Mei, Hongyuan and Yang, Chenghao and Eisner, Jason},
  year = 2021,
  urldate = {2025-10-27},
  langid = {english}
}

@article{muchnikOriginsPowerlawDegree2013,
  title = {Origins of Power-Law Degree Distribution in the Heterogeneity of Human Activity in Social Networks},
  author = {Muchnik, Lev and Pei, Sen and Parra, Lucas C. and Reis, Saulo D. S. and Andrade Jr, Jos{\'e} S. and Havlin, Shlomo and Makse, Hern{\'a}n A.},
  year = {2013},
  journal = {Scientific Reports},
  volume = {3},
  number = {1},
  pages = {1783},
  publisher = {Nature Publishing Group},
  issn = {2045-2322},
  doi = {10.1038/srep01783},
  urldate = {2025-02-21},
  copyright = {2013 The Author(s)},
  langid = {english}
}

@inproceedings{myersBurstyDynamicsTwitter2014,
  title = {The Bursty Dynamics of the {{Twitter}} Information Network},
  booktitle = {\International \Conference on {{World}} Wide Web},
  author = {Myers, Seth A. and Leskovec, Jure},
  year = {2014},
  pages = {913--924},
  doi = {10.1145/2566486.2568043},
  urldate = {2024-07-02},
  isbn = {978-1-4503-2744-2}
}

@inproceedings{myersClashContagionsCooperation2012,
  title = {Clash of the Contagions: Cooperation and Competition in Information Diffusion},
  shorttitle = {Clash of the {{Contagions}}},
  booktitle = {{IEEE} {{\International \Conference}} on {{Data Mining}}},
  author = {Myers, Seth A. and Leskovec, Jure},
  year = {2012},
  pages = {539--548},
  urldate = {2024-07-08}
}

@inproceedings{nayakSmartInformationSpreading2019,
  title = {Smart Information Spreading for Opinion Maximization in Social Networks},
  booktitle = {IEEE \Conference on Computer Communications},
  author = {Nayak, Anuj and Hosseinalipour, Seyyedali and Dai, Huaiyu},
  year = {2019},
  pages = {2251--2259}
}

@article{nianInfluenceOpinionLeaders2025,
  title = {The {{Influence}} of {{Opinion Leaders}} on {{Public Opinion Spread}} and {{Control Strategies}} in {{Online Social Networks}}},
  author = {Nian, Fuzhong and Zhang, Zhen},
  year = 2025,
  journal = {IEEE \Transactions on Computational Social Systems},
  volume = {12},
  number = {5},
  pages = {2494--2507},
  issn = {2329-924X},
  doi = {10.1109/TCSS.2025.3537701},
  urldate = {2025-11-20}
}

@article{notarmuziUniversalityCriticalityComplexity2022,
  title = {Universality, Criticality and Complexity of Information Propagation in Social Media},
  author = {Notarmuzi, Daniele and Castellano, Claudio and Flammini, Alessandro and Mazzilli, Dario and Radicchi, Filippo},
  year = {2022},
  journal = {Nature Communications},
  volume = {13},
  number = {1},
  pages = {1308},
  publisher = {Nature Publishing Group},
  issn = {2041-1723},
  doi = {10.1038/s41467-022-28964-8},
  urldate = {2023-09-06},
  copyright = {2022 The Author(s)},
  langid = {english}
}

@article{ogataLewisSimulationMethod1981,
  title = {On {{Lewis}}' Simulation Method for Point Processes},
  author = {Ogata, Y.},
  year = {1981},
  journal = {IEEE \Transactions on Inform. Theory},
  volume = {27},
  number = {1},
  pages = {23--31},
  issn = {1557-9654},
  doi = {10.1109/TIT.1981.1056305},
  urldate = {2024-06-21}
}

@inproceedings{poux-medardMultivariatePoweredDirichletHawkes2023,
  title = {Multivariate Powered {Dirichlet-Hawkes} Process},
  booktitle = {Advances in {{Information Retrieval}}},
  author = {{Poux-M{\'e}dard}, Ga{\"e}l and Velcin, Julien and Loudcher, Sabine},
  editor = {Kamps, Jaap and Goeuriot, Lorraine and Crestani, Fabio and Maistro, Maria and Joho, Hideo and Davis, Brian and Gurrin, Cathal and Kruschwitz, Udo and Caputo, Annalina},
  year = {2023},
  pages = {47--61},
  publisher = {Springer Nature Switzerland},
}

@article{rainerOpinionDynamicsBounded2002,
  title = {Opinion {{Dynamics}} and {{Bounded Confidence}}: {{Models}}, {{Analysis}} and {{Simulation}}},
  shorttitle = {Opinion {{Dynamics}} and {{Bounded Confidence}}},
  author = {Rainer, Hegselmann and Krause, Ulrich},
  year = {2002},
  journal = {Journal of Artificial Societies and Social Simulation},
  volume = {5},
  number = {3}
}

@inproceedings{rodriguezUncoveringTemporalDynamics2011,
author = {Gomez-Rodriguez, Manuel and Balduzzi, David and Sch\"{o}lkopf, Bernhard},
title = {Uncovering the temporal dynamics of diffusion networks},
year = {2011},
booktitle = {\International \Conference on Machine Learning},
pages = {561–568},
numpages = {8},
}

@article{rothSociosemanticConfigurationOnline2023a,
  title = {Socio-Semantic Configuration of an Online Conversation Space: {{The}} Case of {{Twitter}} Users Discussing the \#{{IPCC}} Reports},
  shorttitle = {Socio-Semantic Configuration of an Online Conversation Space},
  author = {Roth, Camille and Hellsten, Iina},
  year = 2023,
  journal = {Social Networks},
  series = {Social {{Networks}} and {{Anthropogenic Climate Change}}},
  volume = {75},
  pages = {186--196},
  issn = {0378-8733},
  doi = {10.1016/j.socnet.2022.06.007},
  urldate = {2025-10-27}
}

@article{royHatefulSentimentDetection2024,
  title = {Hateful {{Sentiment Detection}} in {{Real-Time Tweets}}: {{An LSTM-Based Comparative Approach}}},
  shorttitle = {Hateful {{Sentiment Detection}} in {{Real-Time Tweets}}},
  author = {Roy, Sanjiban Sekhar and Roy, Akash and Samui, Pijush and Gandomi, Mostafa and Gandomi, Amir H.},
  year = 2024,
  journal = {IEEE \Transactions on Computational Social Systems},
  volume = {11},
  number = {4},
  pages = {5028--5037},
  issn = {2329-924X},
  doi = {10.1109/TCSS.2023.3260217},
  urldate = {2025-11-20}
}

@article{sadriExploringNetworkProperties2020,
  title = {Exploring Network Properties of Social Media Interactions and Activities during {{Hurricane Sandy}}},
  author = {Sadri, Arif Mohaimin and Hasan, Samiul and Ukkusuri, Satish V. and Cebrian, Manuel},
  year = {2020},
  journal = {Transportation Research Interdisciplinary Perspectives},
  volume = {6},
  pages = {100143},
  issn = {2590-1982},
  doi = {10.1016/j.trip.2020.100143},
  urldate = {2025-02-21}
}

@inproceedings{salehiLearningHawkesProcesses2019a,
  title = {Learning {Hawkes} processes from a Handful of Events},
  booktitle = {Advances in Neural Information Processing Systems},
  author = {Salehi, Farnood and Trouleau, William and Grossglauser, Matthias and Thiran, Patrick},
  year = {2019},
  urldate = {2024-10-18}
}

@inproceedings{tanUserlevelSentimentAnalysis2011,
  title = {User-Level Sentiment Analysis Incorporating Social Networks},
  booktitle = {{{ACM SIGKDD}} \International \Conference on {{Knowledge}} Discovery and Data Mining},
  author = {Tan, Chenhao and Lee, Lillian and Tang, Jie and Jiang, Long and Zhou, Ming and Li, Ping},
  year = {2011},
  pages = {1397--1405},
}

@phdthesis{thomaslinigerMultivariateHawkesProcesses2009,
  title = {Multivariate {{Hawkes processes}}},
  author = {T. Liniger},
  year = {2009},
  school = {ETH Zurich}
}

@inproceedings{valeraModelingAdoptionUsage2015,
  title = {Modeling Adoption and Usage of Competing Products},
  booktitle = {IEEE \International \Conference on {{Data Mining}}},
  author = {Valera, Isabel and {Gomez-Rodriguez}, Manuel},
  year = {2015},
  pages = {409--418},
  issn = {1550-4786},
  doi = {10.1109/ICDM.2015.40}
}

@article{vendevilleVoterModelCan2025,
  title = {Voter Model Can Accurately Predict Individual Opinions in Online Populations},
  author = {Vendeville, Antoine},
  year = 2025,
  journal = {Physical Review E},
  volume = {111},
  number = {6},
  pages = {064310},
  publisher = {American Physical Society}
}

@article{vermaImpactCompetingZealots2014,
  title = {The Impact of Competing Zealots on Opinion Dynamics},
  author = {Verma, Gunjan and Swami, Ananthram and Chan, Kevin},
  year = {2014},
  journal = {Physica A: Statistical Mechanics and its Applications},
  volume = {395},
  pages = {310--331},
  issn = {0378-4371},
  doi = {10.1016/j.physa.2013.09.045},
  urldate = {2024-06-28}
}

@inproceedings{wangStochasticDifferentialEquation,
  title = {A Stochastic Differential Equation Framework for Guiding Online User Activities in Closed Loop},
  author = {Wang, Yichen and Theodorou, Evangelos and Verma, Apurv and Song, Le},
  langid = {english},
  pages = 	 {1077--1086},
  year = 	 {2018},
  booktitle = 	 {\International \Conference on Artificial Intelligence and Statistics},
}

@article{wengCompetitionMemesWorld2012,
  title = {Competition among Memes in a World with Limited Attention},
  author = {Weng, L. and Flammini, A. and Vespignani, A. and Menczer, F.},
  year = {2012},
  journal = {Scientific Reports},
  volume = {2},
  number = {1},
  pages = {335},
  publisher = {Nature Publishing Group},
  issn = {2045-2322},
  doi = {10.1038/srep00335},
  urldate = {2024-06-21},
  copyright = {2012 The Author(s)},
  langid = {english}
}

@article{westExploitingSocialNetwork2014,
  title = {Exploiting Social Network Structure for Person-to-Person Sentiment Analysis},
  author = {West, Robert and Paskov, Hristo S. and Leskovec, Jure and Potts, Christopher},
  year = {2014},
  journal = {\Transactions of the Association for Computational Linguistics},
  volume = {2},
  pages = {297--310},
}

@inproceedings{xueEasyTPPOpenBenchmarking2023a,
  title = {{{EasyTPP}}: {{Towards Open Benchmarking Temporal Point Processes}}},
  shorttitle = {{{EasyTPP}}},
  booktitle = {{{\International \Conference}} on {{Learning Representations}}},
  author = {Xue, Siqiao and Shi, Xiaoming and Chu, Zhixuan and Wang, Yan and Hao, Hongyan and Zhou, Fan and Jiang, Caigao and Pan, Chen and Zhang, James Y. and Wen, Qingsong and Zhou, Jun and Mei, Hongyuan},
  year = 2023,
  urldate = {2025-10-27},
  langid = {english}
}

@inproceedings{xuLearningGrangerCausality2016a,
  title = {Learning {Granger} causality for {Hawkes} processes},
  booktitle = {\International \Conference on Machine Learning},
  author = {Xu, Hongteng and Farajtabar, Mehrdad and Zha, Hongyuan},
  year = {2016},
  pages = {1717--1726},
  langid = {english}
}

@article{yangBiVirusCompetingSpreading2018,
  title = {A Bi-Virus Competing Spreading Model with Generic Infection Rates},
  author = {Yang, Lu-Xing and Yang, Xiaofan and Tang, Yuan Yan},
  year = {2018},
  journal = {IEEE \Transactions on Network Science and Engineering},
  volume = {5},
  number = {1},
  pages = {2--13},
  issn = {2327-4697},
  doi = {10.1109/TNSE.2017.2734075},
  urldate = {2024-06-24}
}

@inproceedings{yangMixtureMutuallyExciting2013,
  title = {Mixture of Mutually Exciting Processes for Viral Diffusion},
  booktitle = {{{\International \Conference}} on {{Machine Learning}}},
  author = {Yang, Shuang-Hong and Zha, Hongyuan},
  year = {2013},
  pages = {1--9},
  urldate = {2024-06-21},
  langid = {english}
}

@inproceedings{zarezadeCorrelatedCascadesCompete2017,
  title = {Correlated {Cascades}: Compete or Cooperate},
  author = {Zarezade, Ali and Khodadadi, Ali and Farajtabar, Mehrdad and Rabiee, Hamid and Zha, Hongyuan},
  year = {2017},
  booktitle = {AAAI \Conference on Artificial Intelligence},
  urldate = {2023-09-15},
  copyright = {Copyright (c)},
  langid = {english},
	pages = {238--244},
}

@inproceedings{zhangConjugateBayesianTwostep2024,
  title = {Conjugate {{Bayesian Two-step Change Point Detection}} for {{Hawkes Process}}},
  author = {Zhang, Zeyue and Lu, Xiaoling and Zhou, Feng},
  year = 2024,
  booktitle = {Advances in Neural Information Processing Systems},
  langid = {english}
}

@article{zhangEfficientNonparametricBayesian2022,
  title = {Efficient Non-parametric {Bayesian} {Hawkes} Processes},
  author = {Zhang, Rui and Walder, Christian and Rizoiu, Marian-Andrei and Xie, Lexing},
  year = {2022},
  journal = {Preprint arXiv:1810.03730},
  urldate = {2023-09-29},
  archiveprefix = {arXiv}
}

@inproceedings{zhouLearningTriggeringKernels2013,
  title = {Learning Triggering Kernels for Multi-dimensional {Hawkes} Processes},
  booktitle = {\International \Conference on Machine Learning},
  author = {Zhou, Ke and Zha, Hongyuan and Song, Le},
  year = {2013},
  pages = {1301--1309},
  issn = {1938-7228},
  urldate = {2023-09-18},
  langid = {english}
}

@article{zhouSurveyInformationCascade2022,
  title = {A Survey of Information Cascade Analysis: Models, Predictions, and Recent Advances},
  author = {Zhou, Fan and Xu, Xovee and Trajcevski, Goce and Zhang, Kunpeng},
  year = {2022},
  journal = {ACM Computing Surveys},
  volume = {54},
  number = {2},
  pages = {1--36},
  issn = {0360-0300, 1557-7341},
  doi = {10.1145/3433000},
  urldate = {2024-07-10},
  langid = {english}
}

@article{zhuSpatiotemporaltextualPointProcesses2022,
  title = {Spatiotemporal-Textual Point Processes for Crime Linkage Detection},
  author = {Zhu, Shixiang and Xie, Yao},
  year = {2022},
  journal = {The Annals of Applied Statistics},
  volume = {16},
  number = {2},
  issn = {1932-6157},
  doi = {10.1214/21-AOAS1538},
  urldate = {2024-04-18},
  langid = {english}
}

@incollection{Limnios2025,
title = {Selective review of penalized learning methods for event processes},
editor = {Ioannis S. Triantafyllou and Sonia Malefaki and Alex Karagrigoriou},
booktitle = {Stochastic Modeling and Statistical Methods},
publisher = {Academic Press},
pages = {159-189},
year = {2025},
series = {Advances in Reliability Science},
isbn = {978-0-443-31694-4},
doi = {https://doi.org/10.1016/B978-0-44-331694-4.00014-1},
author = {Myrto Limnios and Niels R. Hansen}
}

\begin{figure*}[!t]
    \centering
    \hspace{-0.3em}\subfloat{\raisebox{-1.4cm}{\rotatebox{90}{\hspace{0em}\colorbox{gray!15}{\scriptsize\hspace{4em}\texttt{music2}\hspace{4em}}}}\includegraphics[width=1.8in, valign=c]{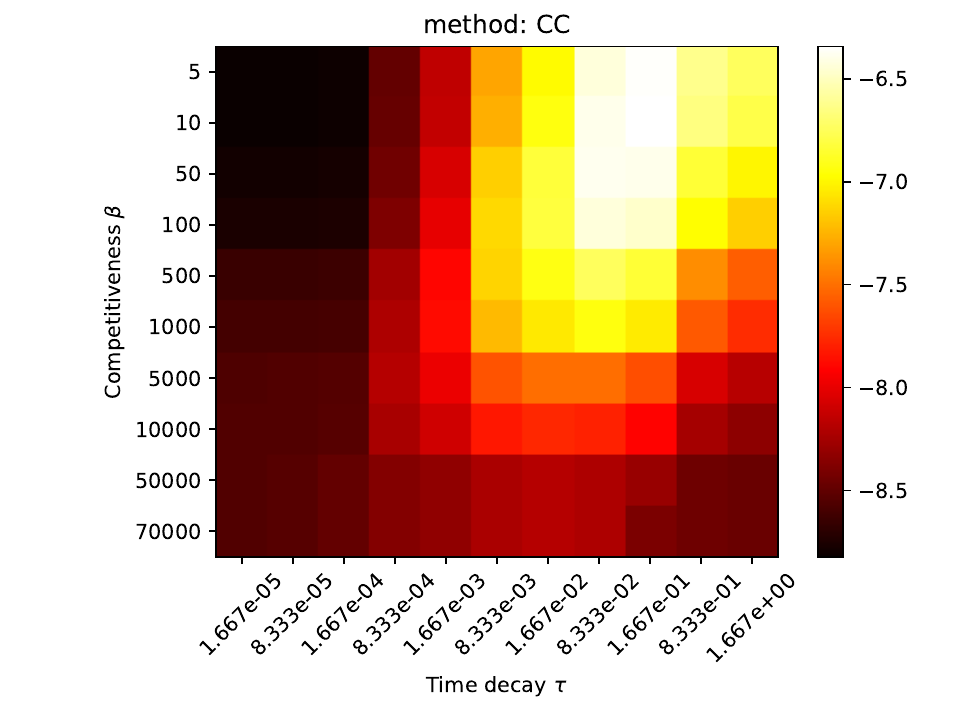}}
    \subfloat{\includegraphics[width=1.8in, valign=c]{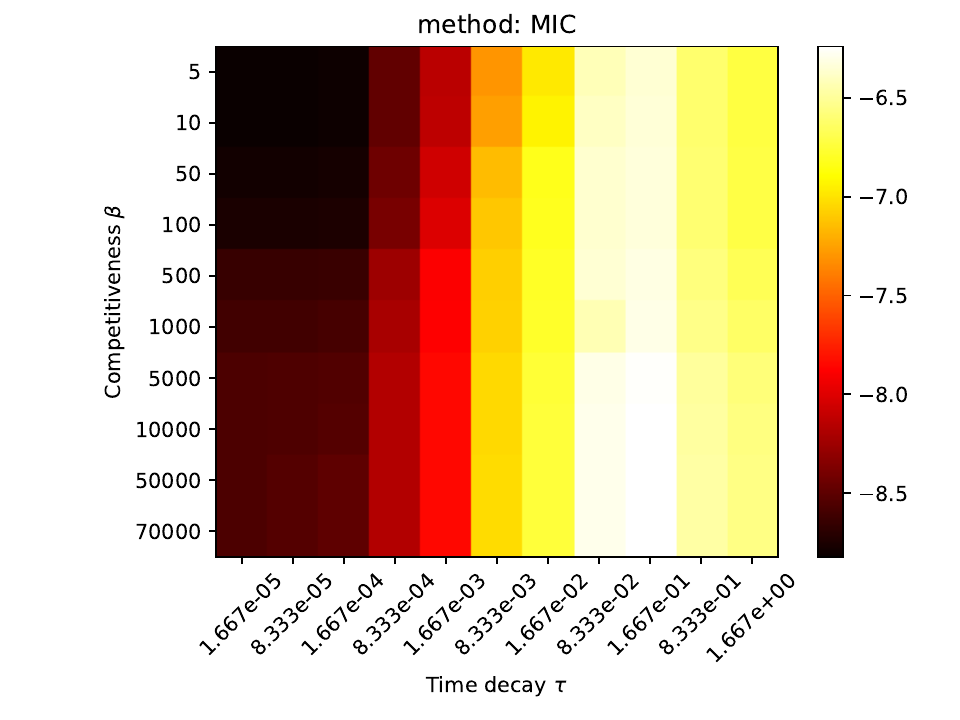}}
    \subfloat{\includegraphics[width=1.6in, valign=c]{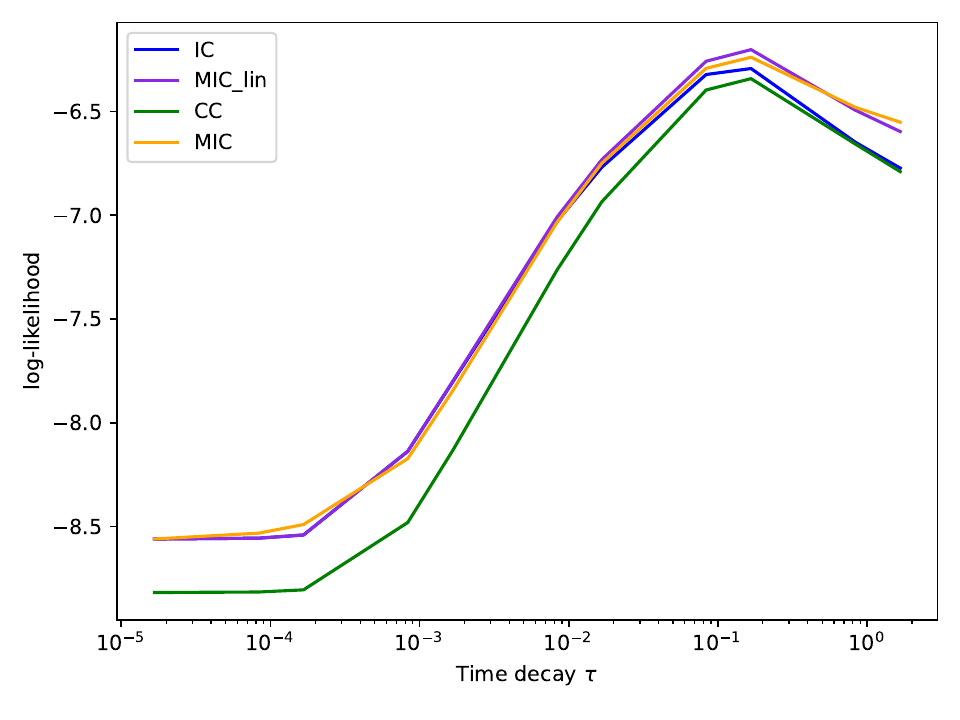}}\\
\vspace{-1em}
  \hspace{-0.3em}\subfloat{\raisebox{-1cm}{\rotatebox{90}{\hspace{0em}\colorbox{gray!15}{\scriptsize\hspace{4em}\texttt{url}\hspace{4em}}}}\includegraphics[width=1.8in, valign=c]{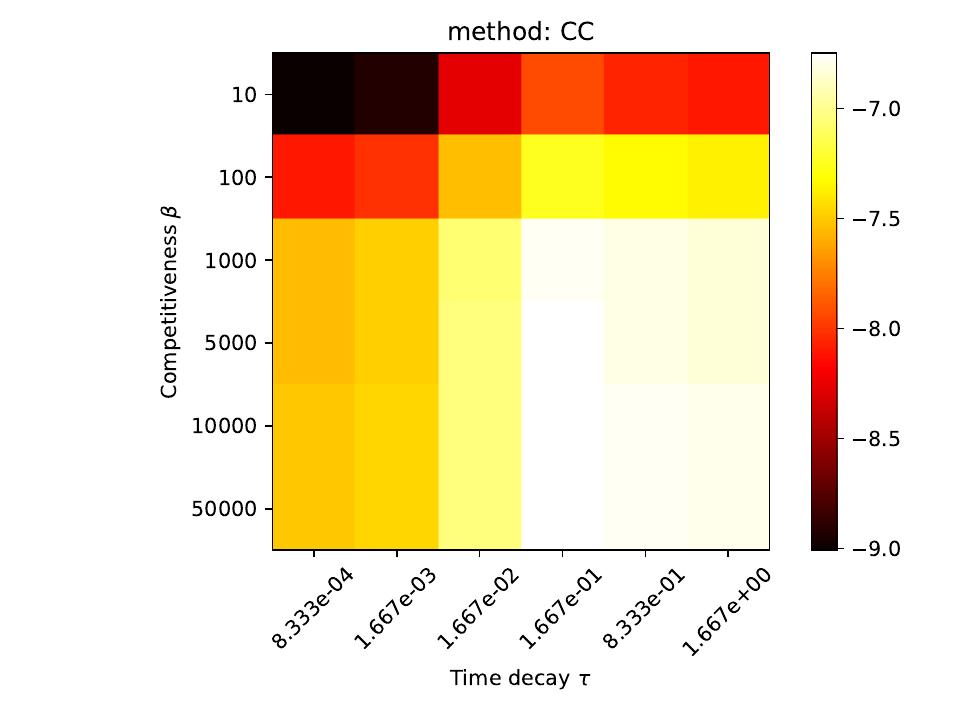}}
  \subfloat{\includegraphics[width=1.8in, valign=c]{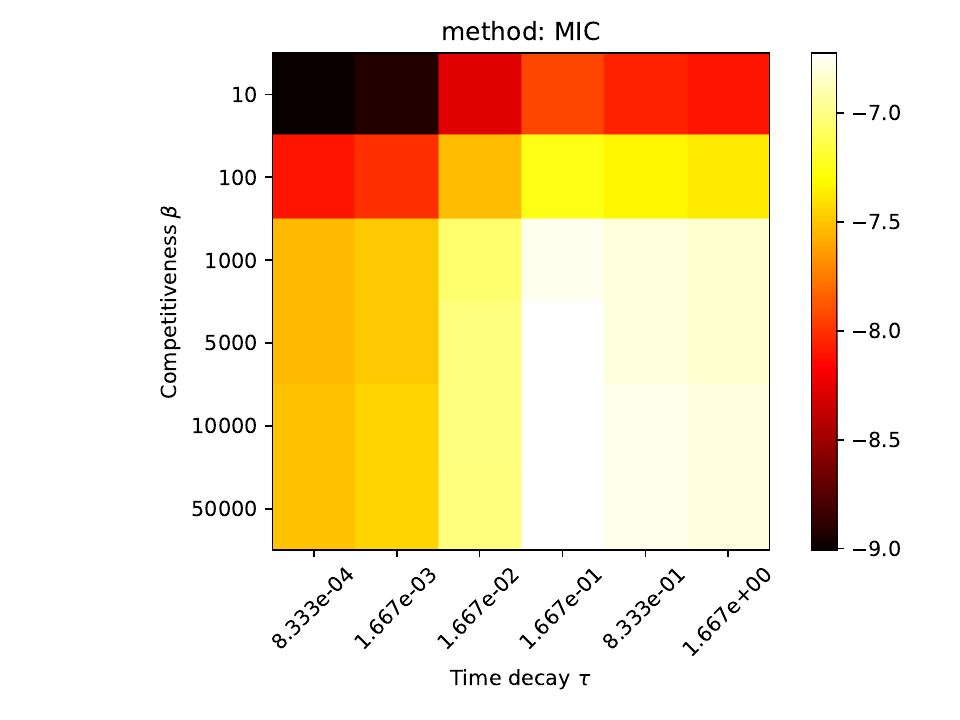}}
  \subfloat{\includegraphics[width=1.6in, valign=c]{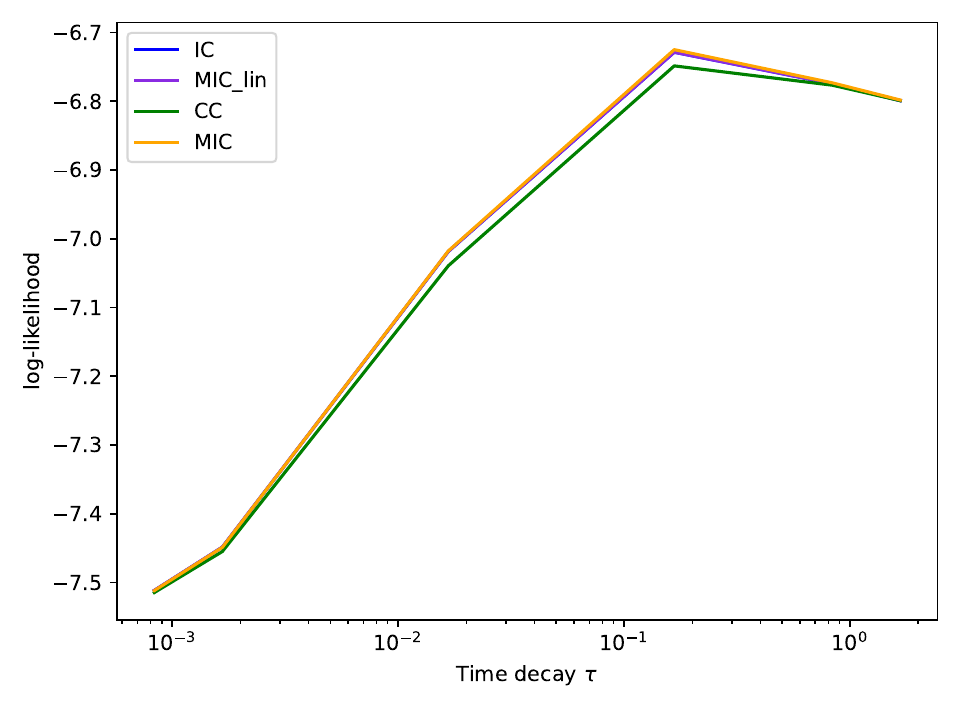}}%
	\vspace{-1em}
  \caption{Left (resp. middle): Cross-validation heatmaps of hyperparameters $\beta$ and $\tau$ on the \texttt{music2} and \texttt{url} datasets for the \CC (resp. \MIC) models. Right: Cross-validation of the time-decay hyperparameter $\tau$ on the \texttt{music2} and \texttt{url} datasets for all models.}
  \label{fig:crossval_music2_url}
\end{figure*}


\appendix

\begin{figure}[h!]
  \centering
  \hspace{-0.3em}\subfloat{\includegraphics[width=1.8in]{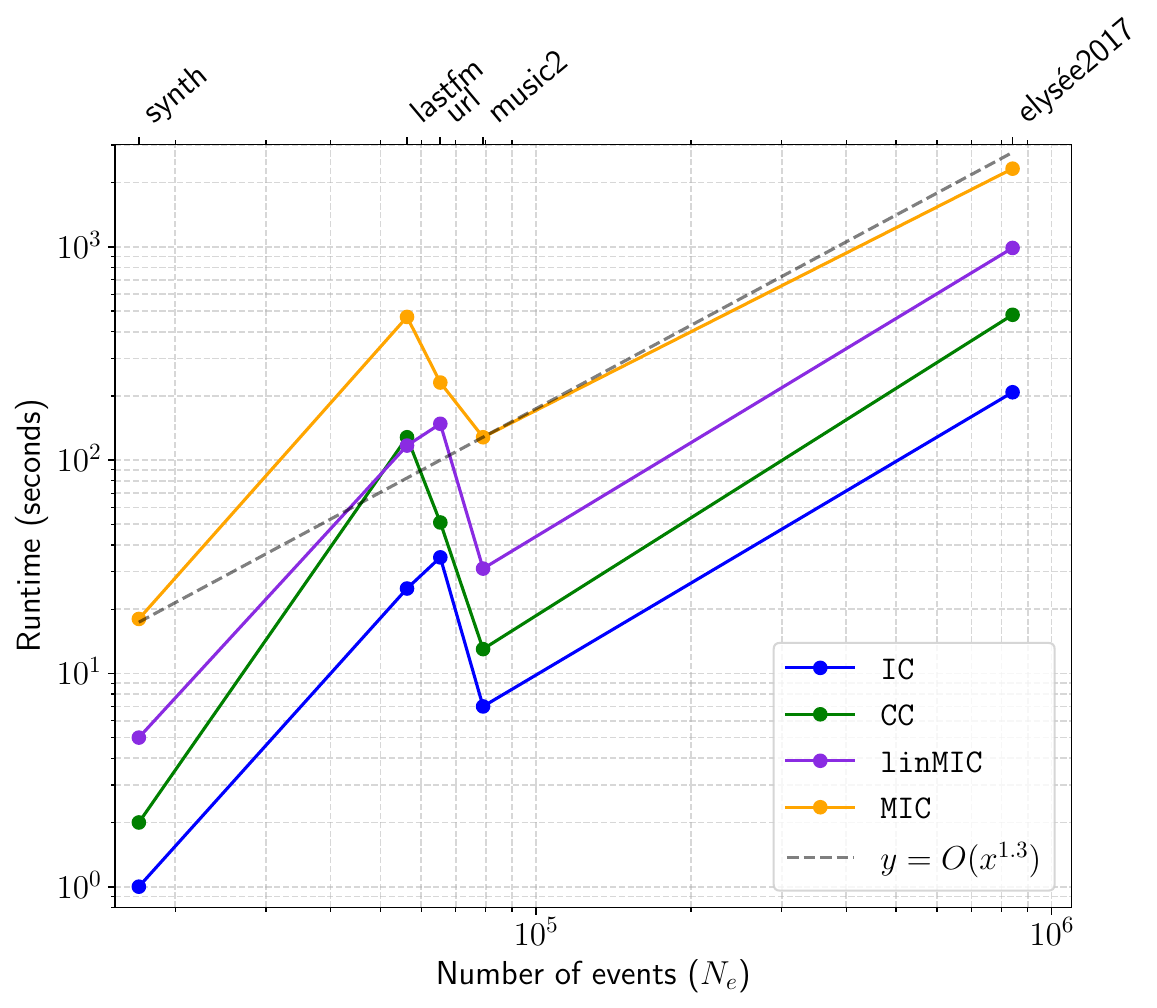}}
  \subfloat{\includegraphics[width=1.8in]{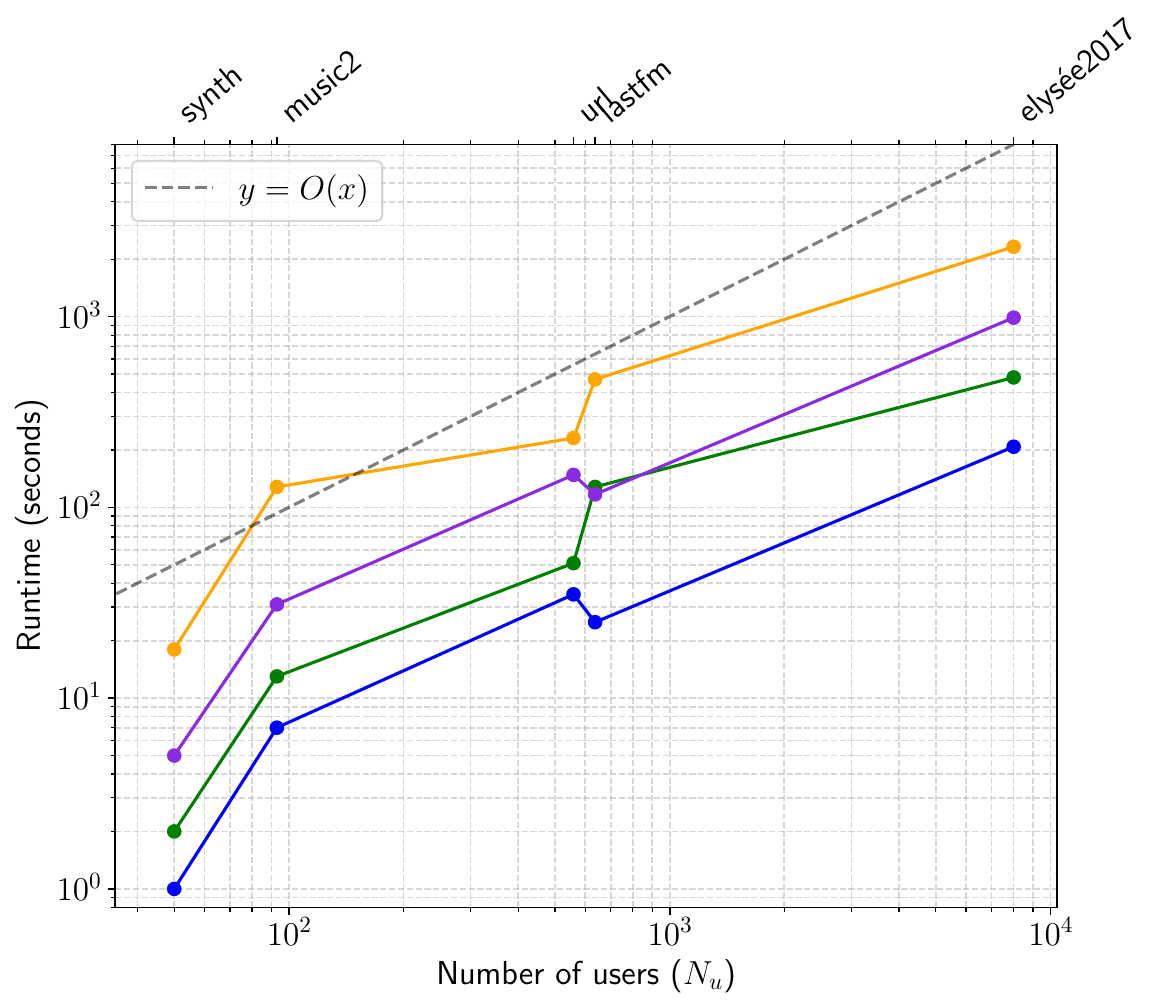}}%
	\vspace{-2em}
  \caption{Runtime (seconds) for parameter inference versus number of events (left) and number of users (right) for each dataset (top x-axis).}
  \label{fig:runtime}
\end{figure}

\mySqBullet~\Tab{tab:cross-vals} reports the optimal hyperparameter values obtained through cross-validation for each dataset and model. These values were used in all experiments reported in the main paper. \Fig{fig:crossval_music2_url} shows the full cross-validation heatmaps for the \texttt{music2} and \texttt{url} datasets, illustrating the sensitivity of model performance to the choice of hyperparameters $\beta$ and $\tau$.\\
\mySqBullet~\Fig{fig:runtime}: the empirical running times of our parameter inference algorithm (\Alg{alg:optimize_theta}) as a function of the number of events and users across the different datasets. The results show that our inference algorithm scales reasonably well with both dataset size and network complexity, with runtime growing sub-quadratically (appears to be bounded by $\Ne^{1.3}$).%
\\
\mySqBullet~\Tab{tab:symbols}: a comprehensive reference that lists the mathematical notation and symbols used throughout this paper. It includes notations for the network structure, event sequences, model parameters, and derived quantities.
\begin{table}[h!]\footnotesize
  \caption{The cross-validated values for the hyperparameters $\beta$ and $\tau$, for each model when used on the different real datasets.}
  \vspace{-1em}
  \centering
  \scalebox{0.9}{%
  $\begin{array}{crcccc}
    \cmidrule[0.7pt]{3-6}
    &&\multicolumn{4}{c}{\textsc{\footnotesize Dataset}}\\
    \midrule
    \textsc{\scriptsize Hyperparameter}
    &\textsc{\scriptsize Model}& \texttt{music2} & \texttt{url} & \texttt{lastfm} & \texttt{élysée2017} \\
    \midrule
    \multirow{1}{*}{\scriptsize Time-decay $\tau$}
    & all &6 \min & 6 \min & 10\mathrm{h} & 6 \min\\
    \midrule
    \multirow{ 2}{*}{\scriptsize Competitiveness $\beta$}
    &\CC &10 &5\times 10^3 &10^4 & 10^2\\
    &\MIC &5\times 10^3 &10^4 &10^4 & 10^2\\
    \bottomrule
  \end{array}$
  }
    \label{tab:cross-vals}
\end{table}%
\vspace{-2.5em}

\begin{table}[t!]
  \caption{Table of symbols.}
  \centering
  \begin{tabular}{p{1.5cm}p{6.7cm}}
  \toprule
  \textbf{Symbol} & \textbf{Description} \\
	\midrule
  $\mathcal{G}\left(\mathcal{V},E \right)$ & Directed network representing the OSN \\
  $\mathcal{V}$ & Set of $\Nu$ nodes (users) \\
  $\mathbf{A}$ & Graph adjacency matrix (with binary edge weights) \\
  $E$ & Set of edges \\
  $\mathcal{F}_u$ & Set of followers of user $u$ \\
  $\mathcal{C}$ & Set of $\Nc$ information cascades \\
  $\mathbf{n}(t)$ & Counting process vector \\
  $\mathcal{H}$ & Sequence of $\Ne$ events \\
  $e_i\!=\!(u_i,t_i,c_i)$\!\!\!\!\! & Event $i$ by user $u_i$, at time $t_i$, for cascade $c_i$ \\
  $\mathcal{H}(t)$ & Set of events occurring in $[0,t]$ \\
  $\mathcal{H}_u(t)$ & Subset of events of user $u$ \\
  $\mathcal{H}^{(c)}(t)$ & Subset of events of cascade $c$ \\
  $\mathcal{H}_u^{(c)}(t)$ & Subset of events of user $u$ and cascade $c$ \\
  $\mathcal{H}_{u}(t,t')$ & Subset of events of user $u$  in the interval $(t,t']$ \\
  $\boldsymbol{\lambda}(t)$ & Conditional intensity vector \\
  $\mathbf{M}$ & Set of user baseline intensities \\
  $\mu_u$ & Background intensity of user $u$ \\
  $\mathbf{\Weight}$ & User weight matrix of social influence\\
  $\weight_{vu}$ & Social influence kernel from $v$ to $u$ \\
  $\kappa(t)$ & Temporal kernel \\
  $\nu_u^{(c)}(t)$ & Independent intensity of user $u$ for cascade $c$ \\
  $\lambda_u^{(c)}(t)$ & Conditional marked intensity of user $u$ for cascade $c$ \\
  $f_u(c \vbar t)$ & Conditional mark probability function \\
  $n_u^{(c)}(t)$ & Number of events of user $v$ for cascade $c$ at time $t$\\
	$\mathbf{\Sigma}$ & Cascade interaction matrix \\
  $\sigma_{sc}$ & Influence of cascade $s$ on cascade $c$ \\
	$\mathbf{\Theta}$ & Set of \MIC parameters, $\mathbf{\Theta} = \left( \mathbf{M} , \mathbf{\Sigma}, \mathbf{\Weight} \right)$\\
	$\tilde{\mathbf{\Theta}}$ & Tilde for any inferred model parameters\\
	$\mathbf{\Theta}^*$  & Star for any true model parameters used to generate synthetic data\\
  $\SF{t'}{t}$ & Survival function between $t$ and $t'$\\
  $Z_u(t)$ & Partition function of the conditional mark probability \\
  $\beta$ & Hyperparameter modulating cascade competition \\
  $\mathcal{L}$ & Log-likelihood \\
  $\mathcal{L}_u$ & Partial log-likelihood for user $u$ \\
  $K(T,t_i)$ & Integral of the temporal kernel \\
  $\mathbf{\Id}$ & Identity matrix \\
  $\mathbb{E}[\cdot]$ & Expectation operator \\
  $\mathbb{P}(\cdot)$ & Probability of an event\\
	\bottomrule
  \end{tabular}
	\label{tab:symbols}
\end{table}

\end{document}